\RequirePackage[tbtags,fleqn]{amsmath}
\documentclass{aa}
\usepackage{amsmath,graphicx,amssymb}
\usepackage{txfonts}
\usepackage{natbib}
\usepackage{color}
\usepackage[T1]{fontenc}
\usepackage{url}
\definecolor{DarkBlue}{rgb}{0.0,0.1,0.4}
\definecolor{Red}{rgb}{0.6,0.2,0.1}
\definecolor{DarkRed}{rgb}{0.5,0.0,0.5}
\definecolor{Blue}{rgb}{0.0,0.0,1.0}
\definecolor{Zelinkava}{rgb}{0.2,0.5,0.2}
\definecolor{Pink}{rgb}{0.0,0.7,0.7}
\definecolor{White}{rgb}{1.0,1.0,1.0}
\usepackage{hyperref}
\bibpunct{(}{)}{;}{a}{}{,}

\hypersetup{
    colorlinks=true,
    linkcolor=Blue,
    citecolor=Blue,
    filecolor=Blue,     
    urlcolor=Blue,
}

\renewcommand{\d}{\text{d}}
\newcommand{\dd}{\,\text{d}}

\begin{document}

\title{Two-dimensional modeling of density and thermal structure of dense circumstellar outflowing disks}

\author{P.~Kurf\"urst\inst{1} \and A.~Feldmeier\inst{2} \and J.~Krti\v{c}ka\inst{1}}

\institute{Department of Theoretical Physics and Astrophysics,
           Masaryk University, Kotl\'a\v rsk\' a 2, CZ-611\,37 Brno, Czech Republic
           \and
           Institut f\"ur Physik und Astronomie, Universit\"at Potsdam, Karl-Liebknecht-Stra\ss e 24/25, 
           14476 Potsdam-Golm, Germany}

\date{Received}

\abstract 
{Evolution of massive stars is affected by a significant loss of mass either via (nearly)
spherically symmetric stellar winds or by aspherical mass-loss mechanisms, namely the outflowing equatorial disks. However, 
the scenario that leads to the formation of a disk or rings of gas and dust around massive stars is still under debate. It is also unclear how 
various forming physical mechanisms of the circumstellar environment affect its shape and density, as well as its kinematic and thermal structure.} 
{We study the hydrodynamic and thermal structure of optically thick, dense parts of outflowing circumstellar disks that may be formed around various types of 
critically rotating
massive stars, for example, Be stars, 
B[e] supergiant (sgB[e]) stars or Pop III
stars. We calculate self-consistent time-dependent models of temperature and density structure in the 
disk's inner dense region that is strongly affected by irradiation from a rotationally oblate central star and by
viscous heating.}
{Using the method of short characteristics, we specify the optical depth of the disk along the line-of-sight from stellar poles. Within the optically 
thick dense region with an optical depth of $\tau>2/3$ we calculate the vertical disk thermal structure using the diffusion approximation
while for the optically thin outer layers we assume a local thermodynamic equilibrium with the impinging stellar irradiation.
For time-dependent hydrodynamic modeling, we use
two of our own types of hydrodynamic codes: two-dimensional operator-split numerical code based on 
an explicit Eulerian finite volume scheme on a staggered grid, and unsplit code based on the Roe's method, both including 
full second-order Navier-Stokes shear viscosity.}
{Our models show the geometric distribution and contribution of viscous heating that begins to dominate in the central 
part of the disk for mass-loss rates higher than $\dot{M}\gtrsim 10^{-10}\,M_{\odot}\,\text{yr}^{-1}$.
In the models of dense viscous disks with $\dot{M}>
10^{-8}\,M_\odot\,\text{yr}^{-1}$, the viscosity increases the central
temperature up to several tens of thousands of Kelvins, however the temperature
rapidly drops with radius and with distance from the disk midplane.
The high mass-loss rates and high viscosity lead to instabilities with significant waves 
or bumps in density and temperature in the very inner disk region.}
{The two-dimensional radial-vertical models of dense outflowing disks including the full Navier-Stokes viscosity terms 
show very high temperatures that are however limited to only the central disk 
cores inside the optically thick area, while near the edge of the optically 
thick region the temperature may be 
low enough for the existence of neutral hydrogen, for example.}


\titlerunning{Two-dimensional modeling of density and thermal structure of dense circumstellar outflowing disks}

\authorrunning{P.~Kurf\"urst et al.}
\maketitle

\section{Introduction}
\label{intro}
Various types of stars are associated with equatorial disks. Notably, some of these disks can be classified
as dense equatorial Keplerian outflowing disks
which are typical for the classical Be stars or Pop III stars 
\citep{1991MNRAS.250..432L,2011A&A...527A..84K,2014A&A...569A..23K}. 
The disk-like equatorial flows may also be formed
around various classes of B[e] type stars \citep[e.g.,][]{2006ASPC..355...39H}, 
luminous blue variables (LBVs) \citep[e.g.,][]{1994ApJ...429..846S,2005A&A...439.1107D}, 
post asymptotic giant branch (post-AGB) stars \citep[e.g.,][]{1998A&A...334..210H} or
around the accretion inflows of, for example, young HAeB[e] stars \citep[e.g.,][]{1994A&AS..104..315T}. 
The possible formation channels of these disks include near-critical rotation,
the wind confinement due to the magnetic field, or the effect of binarity.

In Keplerian outflowing disks, we assume that it is 
viscosity that plays a key role in the outward transport of mass and angular momentum \citep{1991MNRAS.250..432L} 
and governs the process of the disk 
creation and its further feeding and evolution. The
disks are supposed to be rotationally supported and very thin in the region close to
the star. Assuming a vertically isothermal gas, the disks are in vertical hydrostatic equilibrium with the Gaussian profiles of the density and pressure
\citep[e.g.,][]{1991PASJ...43...75O}.
The vertical disk thickness (and
also the disk vertical opening angle) grows with radius leading to the ``flaring'' shape of the disk. 
Also, observational evidence supports the idea that these gaseous disks are Keplerian, at least in the inner region, that is, approximately
below the disk sonic point \citep{2008ApJ...684.1374C}. 
Classical Be stars are on average the fastest rotators among all other
(nondegenerate) types of stars. Their equatorial rotation rate is close to the critical limit \citep{2013A&ARv..21...69R}.
Whether the rotation of Be stars is (at least for a significant fraction of them) exactly critical, 
or mostly remains more or less subcritical, is still an open question that is
intensively discussed. The determination of the observable projected rotational velocity $\varv\sin\,i$
is, in the case of rapid rotators, strongly affected by the stellar gravity darkening which makes the fast rotating equatorial area hardly observable 
\citep[see, e.g.,][]{2004MNRAS.350..189T}.

The existence of outflowing disks stems from the same principle
as in the case of accretion disks, that is, the need for angular momentum transport. Evolutionary models of fast-rotating
stars show that the stellar rotational velocity may approach the
critical velocity \citep{2006A&A...447..623M,2008A&A...478..467E}, above
which no stellar spin up is possible. The further evolution of
critically rotating stars may require loss of angular momentum,
which is carried out by an outflowing decretion disk \citep{2011A&A...527A..84K}. The disk may in principle extend up to several hundreds of
stellar radii \citep{2014A&A...569A..23K} or even more. The disk angular momentum transport requires some kind of anomalous
viscosity, which we regard as being presumably connected with magnetorotational
instability \citep{2015A&A...573A..20K}.

The disk behavior, namely the distribution of density and the direction of the radial flow, 
may be far more complicated in models with very high values of density or viscosity, or
in the case of, for example, a central star with  subcritical rotation \citep[see][]{2017ASPC..508...17K}. 
In such models, the occurrence of drops in radial velocity in the inner disk region 
indicates the material
infall that may periodically increase the angular momentum of the inner disk, creating more or less regular
waves of the density. 

Apart from Be and Pop III stars, there may be another type of rapidly rotating star surrounded by a dense gaseous outflowing circumstellar disk:
 B[e] supergiants (sgB[e]s) \citep[for a review see, e.g.,][]{1998ASSL..233....1Z}.
These stars have a disk or a system of rings of high-density material containing
gas and dust \citep{2013A&A...549A..28K}.
The origin of the B[e] phenomenon in evolved massive stars
like sgB[e]s is still an open question. However, at least two
possible fast rotating sgB[e] stars
with rotation velocity reaching a substantial proportion (at least 70\%) of
critical velocity have been identified: the Small Magellanic Cloud (SMC)
stars LHA 115-S 23 \citep{2008A&A...487..697K} and LHA 115-S 65 \citep{2000ASPC..214...26Z,2010A&A...517A..30K}. 
Another example is the Galactic eccentric binary system GG Car with a circumbinary ring,
where two scenarios of ring origin have been discussed: either the nonconservative Roche
lobe overflow or the possibility that, during the
previous evolution of the system, the primary component underwent the classical Be star phase \citep{2013A&A...549A..28K}. A particular subgroup of
B[e] stars can also be identified with rapid rotators on a blue loop in Hertzprung-Russell (H-R)
diagram \citep[see][for details]{1998A&A...334..210H}. 
Their disks could be formed by the spin-up mechanism induced by stellar
contraction during the transition phase from red supergiant to the blue region on H-R diagram.

Most observable characteristics of the disks originate from regions at a
distance up to ten stellar radii from the central star
\citep{2011IAUS..272..325C}. Therefore, the regions close to the star are key
for understanding the observational behavior of stars with outflowing disks.
As a first step towards more realistic multidimensional models,
we provide detailed two-dimensional (2D) hydrodynamical simulations of central
parts of the disk that include the radiative irradiation and viscous heating.

\section{Basic equations of outflowing disk radial-vertical structure}
\label{baseeqs}
Using our 2D models, we study the profiles of basic hydrodynamic and thermal characteristics in the radial-vertical plane in 
the inner dense Keplerian regions of the disks up to distance, which roughly corresponds to the sonic point radius of 
outflowing disks \citep{2014A&A...569A..23K}.
We examine the distribution of main characteristics of disks for various values of mass-loss rate, 
$\dot{M}=\text{const.}$ \citep{2014A&A...569A..23K}, which is determined by 
the angular-momentum-loss rate needed to keep the star at or near critical rotation 
\citep[see also \citealt{2012IAUS..282..257K}, \citealt{2012ASPC..464..223K}]{2011A&A...527A..84K,2014A&A...569A..23K}.
The basic scenario follows the model of the viscous decretion disk proposed by 
\citet[see also \citealt{2001PASJ...53..119O}]{1991MNRAS.250..432L}. 
The main uncertainty is the viscous coupling, namely the value and the spatial dependence of viscosity parameter $\alpha$ \citep{1973A&A....24..337S}. 
Despite some recent
models indicating a constant viscosity throughout 
the disk \citep{2013MNRAS.428.2255P}, 
we have investigated the cases with variable $\alpha$ decreasing outward as a power law 
and examined various values of 
the disk base viscosity parameter $\alpha_0$.

The disk temperature is principally determined
by the irradiation from the central star \citep{1991MNRAS.250..432L,2008ApJ...684.1374C}. The impinging stellar irradiation
in the regions close to the star strongly depends 
on the geometry and other properties (equatorial and limb darkening) of the rotationally 
oblate central star. The stellar irradiation however does not deeply penetrate the central optically thick core of the disk \citep[cf.][]{1991MNRAS.250..432L}. 
If the disk mass-loss rate is relatively high, then the heating produced by viscous effects begins to
dominate near the disk midplane.
We study the distribution of temperature and 
the hydrodynamic structure of the dense disks 
corresponding to a final stationary state of self-consistent evolution 
of our 2D models, taking into account all the above-mentioned effects. 

\subsection{Hydrodynamic equations with vertical hydrostatic equilibrium}
\label{basehydroeqs}
We study the self-consistently calculated radial-vertical structure of dense viscous
outflowing disks assuming vertical hydrostatic equilibrium, that is, the vertical component of the disk gas velocity $V_z=0$. 
The basic hydrodynamic equations and the parameterization of one-dimensional (1D), radial, thin viscous disk structures are fully described 
in \citet{2014A&A...569A..23K}; see also, for example,~\citet{2001PASJ...53..119O,2011A&A...527A..84K}.
The cylindrical mass conservation (continuity) equation in the 2D axisymmetric 
$(\partial/\partial\phi=0,$ $\text{where}$ $\phi$ $\text{is the azimuthal angle})$ case is
\begin{align}
\label{base2}
\frac{\partial\rho}{\partial t}+\frac{1}{R}\frac{\partial}{\partial R}\left(R\rho V_{{R}}\right)=0,
\end{align}
where $\rho$ is the density and $V_R(R,z)$ is the velocity of a radial 
outflow of the disk matter. 

The corresponding radial momentum conservation equation is
\begin{align}
\label{base3}
\frac{\partial V_R}{\partial t}+V_R\frac{\partial V_R}{\partial R}=
\frac{V_{\phi}^2}{R}-\frac{1}{\rho}\,\frac{\partial\left(a^2\rho\right)}{\partial R}-\frac{GM_{\star}R}{\left(R^2+z^2\right)^{3/2}},
\end{align}
where $V_\phi(R,z)$ is the disk azimuthal velocity, $a(R,z)$ is the speed of sound,
$G$ is the gravitational constant, $M_\star$ is mass of the central star, and $z$ is the vertical coordinate. The second and third terms on the 
right-hand side express
the radial pressure gradient and radial component of gravitational acceleration, respectively. 
The explicit 2D form of the conservation equation of the angular momentum 
(cf. 1D Eq.~(4) in \citet{2014A&A...569A..23K} and Eq.~(7.3) in
\citet{kurfurst2015thesis}) is
\begin{align}
\label{base4}
\frac{\partial}{\partial t}\left(R\rho V_{\phi}\right)+\frac{1}{R}\frac{\partial}{\partial R}\left(R^2\rho V_RV_{\phi}\right)
=f_{\text{visc}}^{(2)},
\end{align}
where the form of the full second-order Navier-Stokes viscosity term $f_{\text{visc}}^{(2)}$ (cf. 1D Eq.~(9) in \citet{2014A&A...569A..23K}) is
\begin{align}
\label{base5}
f_{\text{visc}}^{(2)}=-\frac{1}{R}\frac{\partial}{\partial R}\left[\alpha a^2R^2\rho\left(1-\frac{\partial\ln V_{\phi}}{\partial\ln R}\right)\right].
\end{align}
The term in the inner bracket is equal to $3/2$ in the case of Keplerian azimuthal velocity, $V_\phi\sim R^{-1/2}$, and it is equal to $2$ in the case of angular momentum conserving azimuthal velocity,
$V_\phi\sim R^{-1}$. The contribution of the second-order viscous term is thus of significant importance and cannot be omitted, as most authors do. 
It also prevents the outer regions of the disk from switching to a physically implausible (negative) backward rotational motion \citep[see][]{2014A&A...569A..23K}.

We assume radial variations
of viscosity parameter $\alpha$ introduced in \citet{2014A&A...569A..23K},
\begin{align}
\label{base1}
\alpha=\alpha_0\left(R_{\text{eq}}/R\right)^n,
\end{align}
where $\alpha_0$ is the disk base viscosity, 
$R$ is the cylindrical radial distance, $R_{\text{eq}}$ is the stellar equatorial radius and
$n$ is a free parameter describing the radial viscosity dependence, $n>0$. We employ in our models
the values of the disk base viscosity parameter $\alpha_0 = 0.025$
\citep[cf.][]{2013MNRAS.428.2255P}, $\alpha_0 = 0.1$, and $\alpha_0 = 1$. 
We regard the vertical profile of the viscosity as constant.

We do not employ the energy conservation equation in our calculations 
(neither do other authors who calculate the global structure of outflowing disks \citep{2001PASJ...53..119O,2007ASPC..361..230O,2007ApJ...668..481S}).
Since the speed of sound plays a key role in gas dynamics, the flow characteristics are basically determined 
by subsonic or supersonic nature of the flow \citep[see, e.g.,][]{1967pswh.book.....Z}.
In astrophysics, the temperature of the gas is very often determined by thermal balance
between heat sources and radiative cooling. Radiative heating and cooling timescales are in this case quite short compared 
to other timescales in the problem, namely timescale of sound wave propagation \citep[e.g.,][]{1992pavi.book.....S}. 
It is therefore adequate to use the thermal balance in the present situation (Sect.~\ref{tempstruct}), where
the gas temperature is mainly determined by
external processes (by irradiation of external sources, etc.).

\subsection{The disk irradiation by the central star}\label{diskirrad}
\begin{figure} [t]
\centering\resizebox{1.0\hsize}{!}{\includegraphics{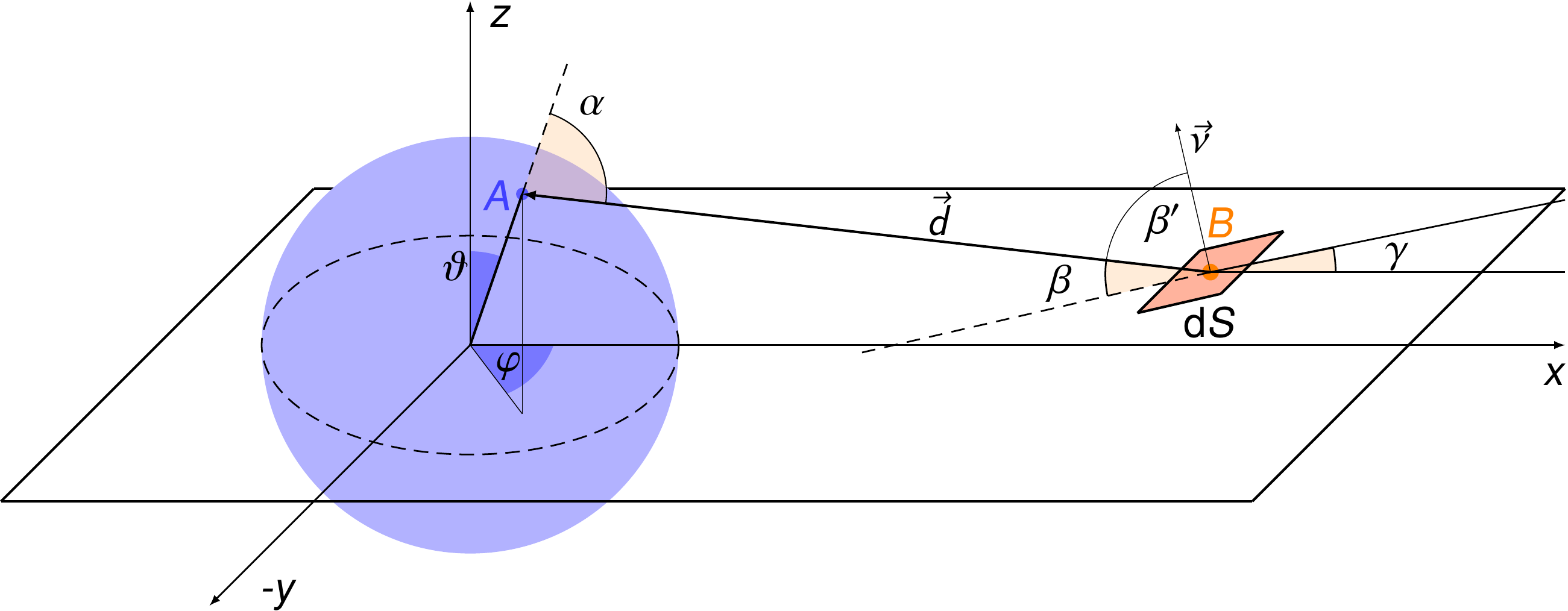}}
\caption{Schema of the geometry of the disk irradiation by a central star. The disk ``surface''
element $\d S$ (with central point $B$) is impinged by the stellar radiation coming along the line-of-sight
vector $\vec{d}$ from the stellar surface element $\d S_{\star}$ (with central point $A$). 
Here $\alpha$ is the angle between
the position vector of the point $A$ and the vector $\vec{d}$,
$\beta^\prime$ is the angle between normal vector $\vec{\nu}$ of the surface
element $\d S$ and the line-of-sight vector $\vec{d}$,
$\beta=\pi/2-\beta^\prime$, and 
$\gamma$ is the angle between normals of the surface
element $\d S$ and of the equatorial plane.
The idea is adapted from \citet{1989AcA....39..201S}.}
\label{fig1}
\end{figure}
\begin{figure} [t]
\centering\resizebox{0.45\hsize}{!}{\includegraphics{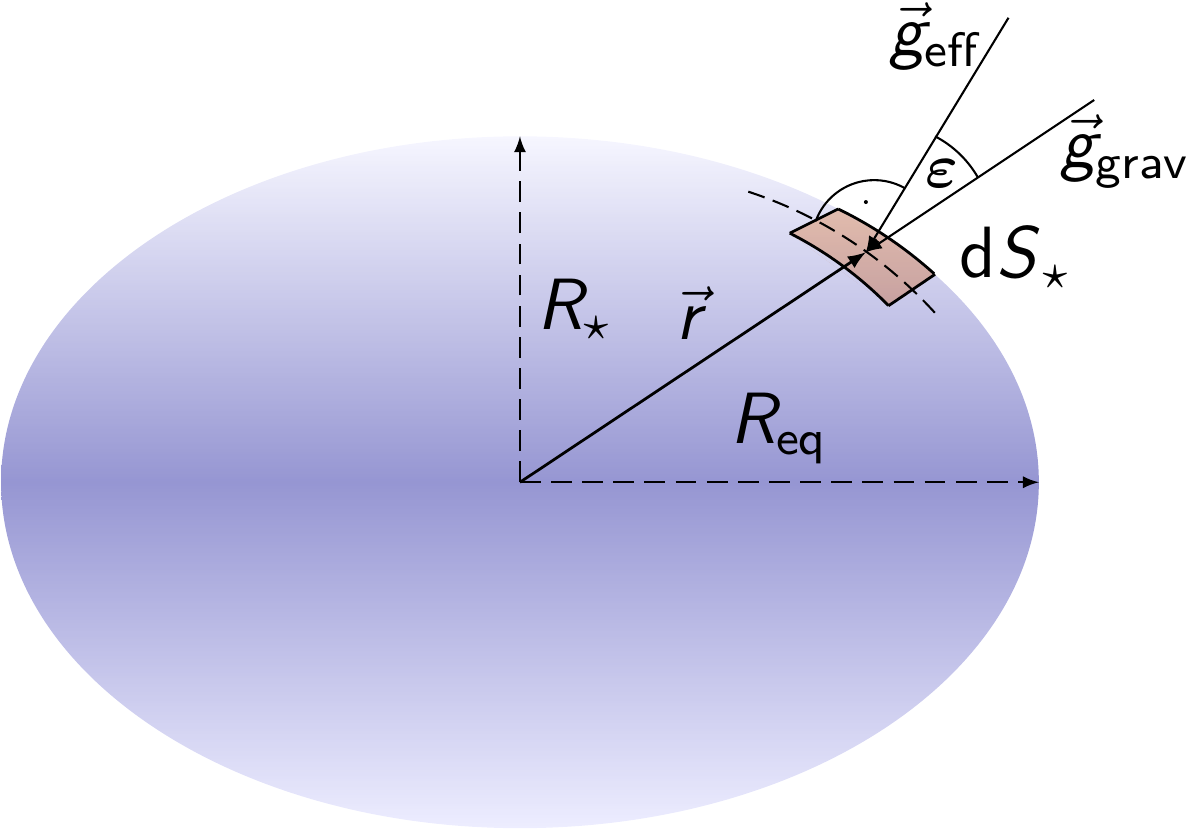}}
\caption{Schema of the rotationally oblate star with polar radius $R_\star$,
equatorial radius $R_\text{eq}$, and
stellar surface element $\d S_{\star}$ whose position vector is $\vec{r}$ (with magnitude $r$).
The vector $\vec{g}_\text{eff}$ is
normal to the stellar surface element $\d S_{\star}$. The angle $\epsilon$ denotes the deviation between the vectors $\vec{g}_\text{grav}$ 
(which is antiparallel to vector $\vec{r}$) and
$\vec{g}_\text{eff}$.}
\label{fig2}
\end{figure}
The total radiative (frequency integrated) flux $\mathcal{F}_\text{irr}
$ from the whole ``visible'' stellar surface, intercepted by the surface element 
$\d S$ of the disk (see the notation in Fig.~\ref{fig1}) is \citep{1978stat.book.....M}
\begin{align}\label{diskirrad1}
\mathcal{F}_\text{irr}=\oint I(\mu)\,\mu\dd\omega,
\end{align}
where $I$ is the intensity of stellar radiation, $\mu=\cos\alpha$, and $\d\omega$ is the solid angle ``filled'' with the element $\d S$
as it is ``seen'' from the center of the stellar surface element $\d S_{\star}$ (from the point $A$). From equalities
$I\,\mu\dd\omega=I\cos\beta^\prime\dd\omega^\prime=I\sin\beta\dd\omega^\prime$,
where $\d\omega^\prime$ is the solid angle ``filled'' with the stellar surface element
$\d S_{\star}$ as it is ``seen'' from the point $B$, we obtain $\d\omega^\prime=\d S_{\star}\,\mu/d^2$, where
$d$ is the magnitude of the line-of-sight vector $\vec{d}$.

We calculate the stellar equatorial (gravity)
darkening by expressing the radiative flux $F_{\star}$ from rotationally oblate
stellar surface in dependence on the stellar angular velocity $\Omega$
and stellar colatitude $\vartheta$
as (von Zeipel theorem, \citealt{1924MNRAS..84..665V,2009pfer.book.....M}, page 72)
\begin{align}\label{vonzeipel}
\vec{F}_{\star}(\Omega,\vartheta)=-\frac{L_{\star}}{4\pi GM_{\star}\left(1-\dfrac{\Omega^2}{2\pi G\langle\rho\rangle}\right)}
\vec{g}_\text{eff}(\Omega,\vartheta),
\end{align}
where $L_{\star}$ is the total stellar luminosity, $\langle\rho\rangle$ is the
mean density of the stellar body, and 
$\vec{g}_\text{eff}$ is the vector of effective gravity that is normal to the stellar surface,
\begin{align}\label{effegrav}
\vec{g}_\text{eff}=\left(-\frac{GM_{\star}}{r^2}+\Omega^2r\sin^2\vartheta,\Omega^2r\sin\vartheta\cos\vartheta,0\right), 
\end{align}
for stellar coordinate directions $r,\vartheta,\varphi$,
respectively (we 
hereafter distinguish between the spherical radial coordinate $r$
and cylindrical radial coordinate $R$). In
the case of critical rotation, the term in parentheses in the denominator of Eq.~\eqref{vonzeipel} is approximately 0.639 \citep[see the Table~(4.2) in][]{2009pfer.book.....M}.
The von Zeipel theorem in the case of a differentially rotating star differs only very slightly 
from Eq.~\eqref{vonzeipel} \citep{1999A&A...347..185M}. We approximate the total stellar luminosity $L_{\star}$ 
as the luminosity for selected effective temperature of the oblate star with average stellar radius $\langle r\rangle$ 
given at the colatitude $\sin^2\vartheta=2/3$ which corresponds to the radius at the root $P_2\left(\cos\vartheta\right)=0$
of the second Legendre polynomial \citep[cf.][page 80]{2009pfer.book.....M}. For example, in the case of critical rotation,
$\langle r\rangle\approx 1.1503\,R_{\star}$ (where $R_{\star}$ is the stellar polar radius). 

For calculation of the limb darkening of the central star, we involve only the basic linear limb-darkening law.
The specific intensity $I$ of stellar irradiation is in this case modified as \citep[see, e.g.,][]{1985A&AS...60..471W,2007ApJ...656..483H}
\begin{align}\label{limbdark1}
I(\mu)=I(1)[1-u(1-\mu)], 
\end{align}
with $I$ being the intensity of the radiation for  
a given point on the stellar surface and $u$ is the coefficient of the
limb darkening that determines the shape of the limb-darkening profile. Integrating Eq.~\eqref{limbdark1}
over, for example, the ``upper'' hemisphere of the star \citep[cf., e.g.,][]{1978stat.book.....M,1989AcA....39..201S}, 
we obtain the radiative flux from one half of the stellar surface 
(impinging ``one side'' of the disk),
corresponding to a given point on the stellar surface as
\begin{align}\label{diskirrad2}
F_{\star}(\Omega,\vartheta)=\int^{2\pi}_0\dd\varphi\int_0^1 I(\mu)\,\mu\dd\mu=\pi\,I(1)\left(1-\frac{u}{3}\right).
\end{align}
We adopt in our calculations the values of the linear limb-darkening coefficient $u$ from Table~1
in \citet{1993AJ....106.2096V}.

The area of the stellar surface element $\d S_{\star}$ of the rotationally oblate star (see Fig.~\ref{fig2}) in spherical coordinates is
\begin{align}\label{diskirrad3}
\d S_{\star}=\frac{r^2\sin\vartheta\dd\vartheta\dd\varphi}{\cos\epsilon},
\end{align}
where $\cos\epsilon$ is cosine of 
the angle between the position vector $\vec{r}$ of a point
on stellar surface and the vector $\vec{g}_\text{eff}$ of effective gravity \citep[cf.][page 27]{2009pfer.book.....M},
\begin{align}\label{diskirrad4}
\cos\epsilon=\left[\left(\frac{R_{\star}}{r}\right)^2-\frac{8}{27}\left(\frac{\Omega}{\Omega_\text{crit}}\right)^2\frac{r}{R_{\star}}\sin^2\vartheta\right] 
\frac{g_\text{pole}}{g_\text{eff}},
\end{align}
where $\Omega_\text{crit}$ is the angular velocity of the critically rotating star,
$g_\text{pole}=GM_{\star}/R_{\star}^2$ is the magnitude of the vector $\vec{g}_\text{grav}$ at the stellar pole,
and $g_\text{eff}$ is the magnitude of the vector 
$\vec{g}_\text{eff}$. 

The magnitude $d$ of the line-of-sight vector $\vec{d}$, where we note that the position vectors of the points $A$ and $B$ (see Fig.~\ref{fig1}) are
$\vec{r}_A=\vec{r},\,\vec{r}_B=(R,0,z)$, is 
\begin{align}\label{diskirrad5}
d^2=R^2+z^2+r^2-2r\left(R\sin\vartheta\cos\varphi+z\cos\vartheta\right). 
\end{align}
From Eq.~\eqref{effegrav} the cosine of the angle between vectors $\vec{g}_\text{eff}$ and $\vec{d}$ is
\citep[cf.][]{1989AcA....39..201S,2011ApJ...743..111M}
\begin{align}\label{diskirrad6}
\mu=\cos\alpha=\frac{A\left(R\cos\varphi-r\sin\vartheta\right)\sin\vartheta+B\cos\vartheta}{\left[A^2\sin^2\vartheta+C^2\right]^{1/2}d}\geq 0,
\end{align}
where $A={1}/{r^2}-{\Omega^2r}/{GM_{\star}}$, $B=z/r^2-\cos\vartheta/r$ and $C=\cos\vartheta/r^2$. 
The inequality $\cos\alpha\geq 0$ eliminates the line-of-sight vectors that connect the selected disk points with the ``invisible'' 
points on the stellar surface. 

Following the simplified linear interpolation approximation 
corresponding to the ``flaring disk'' (see~Figs.~\ref{fig1} and \ref{fig3}), $\tan\gamma=(z/H)\Delta H/\Delta R$,
where $H=H(R)$ is the disk vertical scale height 
(see Eq.~\eqref{diskh}) 
and where $\Delta H$ and $\Delta R$ are the differences of these quantities
within the radially neighboring computational cells (see the details of
numerical grid described in Sect.~\ref{numapproach}).
From this we obtain the normal vector $\vec{\nu}$
to each disk surface element $\d S$ in the arbitrary disk point $B(R,0,z)$
\begin{align}\label{diskirrad7}
\vec{\nu}=\left[-z\left(\frac{\Delta a}{a}+\frac{3}{2}\frac{\Delta R}{R}\right),0,\Delta R\right], 
\end{align}
where $a$ is the speed of sound
and $\Delta a$ is its difference between the radially neighboring computational cells. 

Calculation of the angle $\beta$
between this interpolated ``disk surface'' plane at arbitrary point $B$ and the line-of-sight vector $\vec{d}$ gives the following relation,
\begin{align}\label{diskirrad8}
\sin\beta=\frac{z\left[\left(a+\right)\left(R+\right)^{3/2}-1\right]\left(R-\right)+
\Delta R\left(z-\right)}
{\left\{z^2\left[\left(a+\right)\left(R+\right)^{3/2}-1\right]^2+(\Delta R)^2\right\}^{1/2}d}\geq 0,
\end{align}
where $\left(a+\right)=1+\Delta a/a$, $\left(R+\right)=1+\Delta R/R$, $\left(R-\right)=R-r\sin\vartheta\cos\varphi$ and 
$\left(z-\right)=r\cos\vartheta-z$.
In the isothermal case the term $\left(a+\right)$ is obviously equal to $1$. The
inequality $\sin\beta\geq 0$ eliminates the line-of-sight vectors that come from below the local
``disk surface'' plane. This condition is further enforced by the stellar surface ``visible
domain'' condition, $\cos\vartheta\geq\left(z/H\right)H(R_\text{eq})/r$, where $r$ is the spherical radial coordinate of the point on stellar
surface with height $z(R_\text{eq})$.
This condition eliminates the stellar irradiation that comes to point $B(R,z)$ from the star's equatorial belt
that lies below the vertical level $z(R_\text{eq})/H(R_\text{eq})\leq z/H(R)$.
Following Eqs.~\eqref{diskirrad1}-\eqref{diskirrad8} together with the described conditions, we integrate the (frequency integrated) 
stellar irradiative flux impinging each point $B$ in the disk 
over the whole ``visible'' domain of the stellar surface. We obtain the following relation (cf.~Eq.~\eqref{diskirrad1}),
\begin{align}\label{centerflux}
\mathcal{F}_{\text{irr}}=\frac{1}{\pi}\!\!\iint_{\vartheta,\varphi}\!\! F_{\star}(\Omega,\vartheta)\,\d S_{\star}
\frac{\left[1-u(1-\mu)\right]\,\mu\sin\beta}{\left(1-u/3\right)\,d^2},
\end{align}
where the radiative flux $F_{\star}(\Omega,\vartheta)$ that emerges from the stellar surface is given by Eq.~\eqref{vonzeipel}.

\subsection{Vertical temperature structure}\label{tempstruct}

Following \citet{1981ARA&A..19..137P,2002apa..book.....F}, 
we express the viscous heat (or thermal) flux $F_\text{vis}(R)$ 
per unit area of the disk in the 1D approach 
(taking into account two sides of the disk) as
\begin{align}\label{temp1}
F_\text{vis}(R)=\frac{1}{4\pi R}\mathcal{G}(R)\frac{\d\Omega(R)}{\d R},
\end{align}
where $\mathcal{G}$ is the viscous torque exerted by the outer disk ring on the inner disk ring,
\begin{align}\label{temp2}
\mathcal{G}(R)=2\pi R\nu(R)\Sigma(R)R^2\frac{\d\Omega(R)}{\d R},
\end{align}
where $\Sigma(R)$ is the vertically integrated (surface) density
\begin{align}\label{sigi}
 \Sigma(R)=\int_{-\infty}^\infty \rho\,\d z.
\end{align}
The kinematic viscosity $\nu=\alpha aH$ \citep{1973A&A....24..337S} can be, in the 2D case 
(neglecting vertical variations of angular velocity which is legitimate in the inner parts of the disk),
expressed in the form
\begin{align}\label{temp3}
\nu(R,z)=\frac{\alpha a^2(R,z)}{\Omega(R)}=\frac{\alpha a^2(R,z)\,R}{V_\phi(R)}.
\end{align}
From Eq.~\eqref{temp3}, assuming hydrostatic and thermal equilibrium
in the vertical direction \citep[cf.][]{1991MNRAS.250..432L},
which applies in the optically thick equatorial disk region that is not too far from the central star,
we write the viscous heat flux in height $z$ as (see Eqs.~(3.18) and (3.28) in \citet{kurfurst2015thesis})
\begin{align}\label{temp4}
F_\text{vis}(R,z)=\frac{1}{2}\frac{\alpha a^2(R,z)}{\Omega(R)}\left[R\frac{\d\Omega(R)}{\d R}\right]^2\int_0^z\rho(R,z^\prime)\dd z^\prime.
\end{align}
Integration from $-\infty$ to $\infty$ in Eq.~\eqref{temp4} gives
$F_\text{vis}(R)$ in Eq.~\eqref{temp1}.
From Eq.~\eqref{temp4}, it follows that there is zero net flux in
the disk midplane, $F_\text{vis}(R,0)=0$, which results from symmetry.

Using the equation for pressure, $P(R,z)=a^2(R,z)\,\rho(R,z),$ and neglecting
the vertical temperature gradient, we obtain the vertical slope of the viscous heat flux in the form
\begin{align}\label{temp5}
\frac{\d F_\text{vis}(R,z)}{\d z}=\frac{1}{2}\frac{\alpha P(R,z)}{\Omega(R)}\left[R\frac{\d\Omega(R)}{\d R}\right]^2.
\end{align}
From vertical hydrostatic equilibrium (involving Eq.~\eqref{zgrav}) the vertical pressure gradient simultaneously follows: 
\begin{align}\label{temp6}
\frac{\d P(R,z)}{\d z}=-\rho(R,z)\,\Omega^2(R,z)\,z,
\end{align}
where we explicitly include the possibility of vertical variations of disk angular
velocity (see Sect.~\ref{numapproach} for details).

Assuming local thermodynamic and radiative
equilibrium in the optically thick regions and omitting now the external radiative
flux $\mathcal{F}_\text{irr}$ 
(that we assume does not penetrate into optically thick layers),
the vertical gradient of the temperature \citep[e.g.,][]{1991MNRAS.250..432L}
generated by the viscous heating is
\begin{align}\label{temp7}
\frac{\d T(R,z)}{\d z}=\frac{3F_\text{vis}(R,z)}{16\sigma T^3(R,z)}\frac{\d\tau(R,z)}{\d z},
\end{align}
where $\sigma$ is the Stefan-Boltzmann constant. The vertical gradient of the disk optical depth $\tau$ in Eq.~\eqref{temp7} is
\begin{align}\label{temp8}
\frac{\d\tau(R,z)}{\d z}=-\kappa(\rho,T)\,\rho(R,z),
\end{align}
where $\kappa$ is the opacity of the gas in the disk \citep[e.g.,][]{1978stat.book.....M,1984oup..book.....M}.
Equation \eqref{temp7} holds only in the case where the radiative gradient $\nabla_{\!\text{rad}}$
does not exceed the adiabatic gradient, that is, $\nabla_{\!\text{rad}}<\nabla_{\!\text{ad}}$ \citep[see, e.g., Sect.~5.1.3 of][for details]{2009pfer.book.....M}.
We therefore examined also the regions where convection may develop. However, because
the adiabatic gradient for monoatomic perfect gas 
$\nabla_{\!\text{ad}}=2/5$,
our calculations show that the convective zones occur only in the models with the highest densities,
that is, with $\dot{M}\ge 10^{-7}\,M_{\odot}\,\textup{yr}^{-1}$
(see Sect.~\ref{massiveresults} and \citealt{1991MNRAS.250..432L}).

We include electron scattering, bound-free, and free-free opacities in our model.
For the calculation of Thomson (electron) opacity of partly ionized hydrogen (where for fully ionized
hydrogen $\kappa_\text{es}\approx 0.04\,\text{m}^2\text{kg}^{-1}$ in SI units)
we use the relation \citep{1958ses..book.....S,1978stat.book.....M,2009pfer.book.....M}
\begin{align}\label{temp9}
\kappa_\text{es}=\frac{n_\text{e}\,\sigma_\text{es}}{\rho}=\frac{n_\text{e}\,\sigma_\text{es}}{n_\text{N}\,m_\text{u}}, 
\end{align}
where $n_\text{e}$ is the number density of free electrons, $\rho$ is the gas density,
$\sigma_\text{es}$ is the Thomson scattering cross-section,
$\sigma_\text{es}\approx 6.65\times 10^{-29}\,\text{m}^2$, 
$n_\text{N}$ is the number
density of all particles, that is, the sum of free electrons (which equals the number of hydrogen ions) and neutral hydrogen
atoms, and $m_\text{u}$ is the atomic mass unit. 
The ratio $n_\text{e}/n_\text{N}$ is obtained from the Saha
equation. We involve only hydrogen atoms for calculation of $\kappa_\text{es}$ for simplicity.

We involve also the mean Roseland opacity for continuum, that is,~$\kappa_{\text{bf}}$ for bound-free and $\kappa_{\text{ff}}$ for free-free absorption according to 
assumed chemical composition of the disk material.
In the case of only hydrogenic composition (like in pop III stars' disks)
we involve only the mean Roseland opacity for the free-free absorption.
The bound-free and free-free
opacities are roughly given by a Kramers law (cf. Eqs.~(8.24) and (8.29) in \citet{2009pfer.book.....M}, cf. also, e.g.,~\citet{1958ses..book.....S}) in the form
(also taking into account the rapid decrease of concentration of free electrons in the 
region with low temperatures, analogously to Eq.~\eqref{temp9}),
\begin{align}\label{temp10}
\kappa_\text{bf}\cong\kappa_{0,\text{bf}}\frac{n_\text{e}}{n_\text{N}}Z(1+X)\,\rho T^{-7/2},\quad
\kappa_\text{ff}\cong\kappa_{0,\text{ff}}\frac{n_\text{e}}{n_\text{N}}\rho T^{-7/2}, 
\end{align}
where $X$ and $Z$ are the mass fractions of hydrogen and metals, respectively, $\rho$ is the gas density and $T$ is temperature.
The coefficients $\kappa_{0,\text{bf}}\approx 4.3\times 10^{21}\,\text{m}^5\,\text{kg}^{-2}\,\text{K}^{7/2}$ and 
$\kappa_{0,\text{ff}}\approx 7.4\times 10^{18}\,\text{m}^5\,\text{kg}^{-2}\,\text{K}^{7/2}$ \citep{2009pfer.book.....M}. 
In the case of standard (solar)
composition, bound-free opacity dominates over free-free opacity (e.g., in the case of solar composition with $Z\approx 0.02$ the
bound-free opacity by about an order of magnitude exceeds the free-free opacity),
however, because the free-free opacity does not depend on metallicity, it dominates
in very low-$Z$ objects (e.g., in Pop III stars). Finally, the overall optical depth $\kappa$ in 
Eq.~\eqref{temp8} is the sum $\kappa=\kappa_\text{es}+\kappa_\text{bf}+\kappa_\text{ff}$.
For other details see 
the description of numerical approach in Sect.~\ref{numapproach} and in Appendix~\ref{optdepth}. 

For inclusion of the effect of stellar irradiation we calculate the limit where the disk optical
depth $\tau=2/3$ (we denote it $\tau_{2/3}$), regarding the line-of-sight from stellar pole (see more details in Sect.~\ref{numapproach}) 
and calculating self-consistently the hydrodynamic and thermal properties of the gas. 
We calculate the temperature $T_{2/3}$ at the optical depth limit $\tau_{2/3}$ simply via the frequency integrated Eddington approximation 
\citep{1978stat.book.....M},
\begin{align}\label{temp11}
\sigma T_{2/3}^4=\frac{3}{4}\left(\tau_{2/3}+\frac{2}{3}\right)\left(\mathcal{F}_\text{irr}+F_\text{vis}\right), 
\end{align}
where $\mathcal{F}_\text{irr}$ is given by Eq.~\eqref{centerflux} and $F_\text{vis}$ is determined
by Eqs.~\eqref{temp4} and \eqref{temp5}.
The temperature is then calculated as follows:
the temperature $T_{2/3}$ is used as the starting point for calculation of
the disk thermal and density structure in the optically thick domain with
$\tau\geq\tau_{2/3}$, using the diffusion
approximation (excluding in the optically thick domain
the effects of $\mathcal{F}_\text{irr}$) from thermal and hydrostatic equilibrium equations \eqref{temp6}-\eqref{temp8}.
The computation is performed provided that the disk equatorial plane boundary condition
$F_\text{vis}(z=0)=0$ is satisfied. In the optically thin domain 
($\tau<\tau_{2/3}$) 
we calculate the temperature from a local thermodynamic
and radiative (gray)
equilibrium of the 
gas with the impinging external flux of the stellar irradiation 
(using Eq.~\eqref{temp11} with $\tau<\tau_{2/3}$ and omitting therein the contribution of the viscous heating $F_\text{vis}$).
This approximation is relevant in this case since
the temperature structure of the disk material is maintained primarily by an external source of energy, we therefore regard
the hydrodynamic equation of state as isothermal (in principle, however, 
the solution of the optically thin environment is not the main point of these models).

We also involve the effect of radiative loss of energy, $Q_T\approx\rho^2P(T)$, where $P(T)$ 
is a positive function that describes the temperature variations of the radiative loss
in the optically thin approximation,
which however plays a significant role only above $T\approx 15\,000\,\text{K}$ \citep{1978ApJ...220..643R,2011A&A...530A.124L}
(at lower temperatures the radiative energy is assumed to be mostly absorbed in continua of neutral helium and neutral hydrogen where it contributes to radiative heating and thus
to internal energy of the outflowing gas \citep{2012A&A...539A..39C}).
For the calculation we adopt the values of function $P(T)$ for 
various ranges of temperature in the optically thin domain from Eq.~A.1
in \citet{1978ApJ...220..643R}. The approximative boundary between the optically thick and 
optically thin domain is for the radiative loss calculated in the vertical direction $z$ (unlike the heating, where the boundary of the optically thick region is 
calculated along the line-of-sight from the stellar pole).
The decrease in temperature is then calculated similarly as in the case of the heating contribution, that is, with use of modified 
Eq.~\eqref{temp11},
where the last term in parentheses is replaced by the radiative cooling $F_\text{cool}$, determined by
\begin{align}\label{temp12}
Q_T=-\frac{\d F_\text{cool}}{\d z}=-n_\text{e}n_\text{H}P(T),
\end{align}
where $n_\text{H}$ is the number density of hydrogen particles \citep[cf.][]{2012A&A...539A..39C}.

\subsection{Initial state}\label{instate}
The initial conditions are derived assuming Keplerian rotating disk.
Integrating analytically the vertical hydrostatic equilibrium equation,
\begin{align} \label{vertgz}
\frac{\partial (a^2\rho)}{\partial z}=\rho g_z,
\end{align}
where the vertical component of gravitational acceleration, 
\begin{align} \label{zgrav}
g_z=-\frac{GM_{\star}z}{\left(R^2+z^2\right)^{3/2}},
\end{align}
and we can integrate the analytical expression for the initial density profile into the form
\begin{align} \label{base6}
\rho=\rho_\text{eq}\,\text{exp}\left[-\frac{GM_{\star}}{a^2}\left(\frac{1}{R}-\frac{1}{r}\right)\right],
\end{align}
where $\rho_\text{eq}=\rho(R,0)$ , which
is the density in the equatorial plane (disk midplane where $z=0$) and $r=\sqrt{R^2+z^2}$.
Eq.~\eqref{base6} provides 
a more realistic 2D analytical profile of a disk spatial density than the usually used Gaussian vertical
density distribution in a thin disk approximation.
This becomes relevant for 2D 
models, which
(due to a convergence necessity)
extend up to several hundreds of stellar radii, where the assumption $z\ll R$ does not hold.

To express the initial profile of the disk midplane density $\rho_\text{eq}$ analytically, we employ the 1D equation of the vertically integrated density 
\citep[e.g.,][]{2011A&A...527A..84K},
\begin{align} \label{base7}
\Sigma(R)=\sqrt{2\pi}\rho_{\text{eq}}(R)H(R),
\end{align}
where we adopt the approximation $\Sigma(R)\sim R^{-2}$ \citep[e.g.,][]{2001PASJ...53..119O}.
Denoting $H$ the vertical disk scale height, 
\begin{align}
\label{diskh}
H=\frac{a}{\Omega},
\end{align}
where $\Omega$ is the angular velocity,
we obtain for the Keplerian disk 
region $\rho_{\text{eq}}=\rho_0(R_\text{eq}/R)^{7/2}$ where $\rho_0$ is the disk base density 
(noting that $H\sim R^{3/2}$ in the disk Keplerian region).

To determine the disk base density value $\rho_0$, we first calculate the disk base surface density $\Sigma_0$ from a selected disk mass-loss rate $\dot{M}$. We use
the equation of mass conservation, $\dot{M}=2\pi R_\text{eq}\Sigma_0 V_R(R_\text{eq})$, where we adopt the already pre-calculated density independent 
value of the disk base radial velocity $V_R(R_\text{eq})$ from a 1D model \citep[see][for details]{2014A&A...569A..23K,kurfurst2015thesis}. 
We note that the value of the disk base integrated density $\Sigma_0$ (connected with the selected $\dot{M}$) 
and the viscosity parameters $\alpha_0$ and $n$ are the only parameters (except the parameters of central star) that 
stay fixed during the whole computational process.
We finally obtain the initial disk volume base density $\rho_0$ from Eq.~\eqref{base7} with use of Eq.~\eqref{diskh},
where we adopt an initial sound speed $a$ in accordance with an initially parameterized temperature profile, 
\begin{align}
\label{timeevol1}
T=T_0\left(R_{\mathrm{eq}}/R\right)^p,
\end{align}
where $T_0\approx 0.6\left\langle T_\text{eff}\right\rangle$ is the disk base temperature \citep[cf.][]{2008ApJ...684.1374C} 
and $p$ is a slope parameter (where in the isothermal case $p=0$).

Combining Eqs.~\eqref{base6} -- \eqref{diskh}, we write the initial density profile in the explicit form
\begin{align}\label{numappr4}
\rho(R,z)=\Sigma_0\left(\frac{R_\text{eq}}{R}\right)^2\sqrt{\frac{GM_\star}{2\pi a^2R^3}}\exp\left[\frac{GM_\star}{a^2R}\left(\frac{R}{r}-1\right)\right].
\end{align}
The initial profile of the outflowing disk angular momentum $J$ takes, in 2D models,
the explicit form
\begin{align}\label{numappr5}
J(R,z)=\rho(R,z)\sqrt{-g_RR^3}, 
\end{align}
where the radial component $g_R(R,z)$ of the external gravity is given by the radial analog of Eq.~\eqref{zgrav}.
The initial radial and vertical components of momentum are 
\begin{align}\label{numappr6}
\Pi_R=\rho(R, z)V_R=0,\quad
\Pi_z=\rho(R, z)V_z=0, 
\end{align}
in the whole computational domain.
We describe boundary conditions in Sect.~\ref{numapproach} 
while some additional or specific initial and boundary conditions may be also described in Sections that refer to specific models.

\section{Numerical methods}
\label{numapproach}
For the disk modeling we have
developed and used two types of hydrodynamic codes based on different principles: 1) the operator-split (ZEUS-like) finite 
volume Eulerian numeric 
schema on staggered mesh for the 2D (relatively) smooth hydrodynamic calculations \citep{1992ApJS...80..753S} and 2) our own version of a single-step (unsplit, ATHENA-like) 
finite volume Eulerian hydrodynamic algorithm based on the Roe$^\prime$s method 
\citep{1981JCoPh..43..357R,Toro}. We originally developed the latter code
for a different astrophysical problem with highly discontinuous flows, however, we employed the code also to 
test our calculations of the 2D disk structure
(see the description in Appendices~\ref{roeadiaflare}, \ref{roeisoflare}, and \ref{roetimeflare}). 
We have calculated our models several times with various parameters using both the code types, 
obtaining practically identical results.

For the time-dependent calculations, we write the left-hand sides
of hydrodynamic Eqs. \eqref{base2} -- \eqref{base4} 
in conservative form \citep[see, e.g.,][]{1986ASIC..188..187N,1989nyjw.book.....H,1992ApJS...80..753S,1995A&A...299..523F,LeVeque1998Comput}
\begin{align}\label{numappr1}
\frac{\partial{\vec{u}}}{\partial t}+\vec{\nabla}\cdot\vec{F}(\vec{u})=0, 
\end{align}
where $\vec{u}=\rho,\,\rho\vec{V},\,\vec{R}\times\rho\vec{V}$ and $\vec{F}(\vec{u})=\rho\vec{V},\,\rho\vec{V}
\,|\,
\vec{V},\,
\vec{R}\times\rho\vec{V}
\,|\,
\vec{V}$ for the mass, momentum, and angular momentum
conservation equations, respectively, where $\times$ denotes the vector product and 
$|$ denotes the dyadic product of two vectors of velocity.

The source steps, that is, the right-hand sides
of Eqs.~\eqref{base3} and \eqref{base4}, express the action of external forces
(gravity) and internal pressure forces on the gas \citep[see][]{1986ASIC..188..187N} in the standard cylindrical coordinates. 
We solve finite-difference approximations of the following differential equations in the 
approximative Lagrangian form \citep{1992ApJS...80..753S},
\begin{align}\label{numappr2}
\frac{\d\rho}{\d t}&=0,\\
\frac{\d\Pi}{\d t}&=\rho\frac{V_\phi^2}{R}-\frac{\partial(\rho a^2)}{\partial R}-\rho\frac{GM_\star R}{(R^2+z^2)^{3/2}}-\frac{\partial Q}{\partial R},\\
\frac{\d J}{\d t}&=f_{\text{visc}}^{(2)},
\end{align}
where $\Pi=\rho V_R$ is the radial momentum density, $J$ is
the angular momentum density, and $f_{\text{visc}}^{(2)}$ is the second-order viscosity term derived in Eq.~\eqref{base5}. 

We use the following boundary conditions for particular hydrodynamic quantities at the inner boundary (at stellar surface): here the density $\rho$ and the angular momentum $J$ are fixed
(assuming  constant mass flow and the permanent Keplerian azimuthal velocity) while the radial momentum flow $\Pi_R$ and the vertical momentum flow $\Pi_z$
are free,~that is,~the quantities are extrapolated there from the neighboring grid interior values as a $0$th-order extrapolation. 
We set the free (outflowing) outer boundary conditions for all the quantities.
We assume the lower and upper (vertical) boundary conditions as outflowing (or alternatively periodic, which does not significantly affect the calculation), see also  
\citet{2014A&A...569A..23K,kurfurst2015thesis}.

In order to smooth out the discontinuities which occur when shocks are present, 
the operator-split staggered mesh type of numeric schema involves the artificial viscosity \citep{1950JAP....21..232V}
which is in general described by a tensor \citep{zbMATH03332291}. 
We employ (in the ZEUS-like code) merely the 
scalar
artificial viscosity $Q$ 
(providing the same practical results as the tensor)
in the explicit form \citep{1986ASIC..188..187N,1998JCoPh.144...70C}, 
\begin{align}\label{numappr3}
Q_{i,k}=\rho_{i,k}\,\Delta V_{R,i,k}\,[-\text{C}_1a_{i,k}+\text{C}_2\text{min}(\Delta V_{R,i,k},0)], 
\end{align}
where $\Delta V_{R,i,k}=V_{R,i+1,k}-V_{R,i,k}$ is the forward difference of the radial velocity component, $a$ is the sound speed and
the lower indices $i,k$ denote the $i$-th and $k$-th radial and vertical spatial grid cells. The second term scaled by a constant $\text{C}_2=1.0$ is 
the quadratic artificial viscosity \citep[see][]{1998JCoPh.144...70C} used in compressive zones.
We use the linear viscosity term for damping
oscillations in stagnant regions of the flow \citep{1986ASIC..188..187N}. We use this term with the coefficient $\text{C}_1=0.5$ only
rarely, when some numeric oscillations occur, for example, in the inner disk region near the stellar surface. 

We use spherical coordinates in order to define the maximally uniform mesh covering the rotationally oblate stellar surface. 
The spherical coordinate 
$\vartheta$
(stellar colatitude - not to be confused with the $\theta$ coordinate in the flaring coordinate system, 
see Appendix~\ref{flarecoords}) 
ranges approximately from $0$ to $\pi/2$ assuming
that the irradiation comes to the ``upper'' disk ``surface'' only from the ``upper'' stellar hemisphere. 
Following the Roche model (for a rigidly rotating star with most of its mass concentrated to stellar center) 
we obtain the spherical radial distance $r$ of each grid point on the stellar
surface for each given colatitude 
$\vartheta$
by finding the root of the equation \citep[e.g.,][page 24]{2009pfer.book.....M}
\begin{align}\label{numappr7}
1-\frac{r}{R_\star}+\frac{4}{27}\left(\frac{\Omega}{\Omega_\text{crit}}\right)^2\left(\frac{r}{R_\star}\right)^3\sin^2
\vartheta
=0,
\end{align}
where $\Omega$ is the angular velocity of the stellar rotation, $R_\star$ is the stellar polar radius and $\Omega_\text{crit}$ is the angular velocity of the star
in case of critical rotation.
The stellar azimuthal 
$\varphi$ 
direction ranges from $-\pi$ to $\pi$ in order to take into account also the stellar
irradiation that emerges from the stellar surface behind the meridional circle impinging the disk regions with high vertical coordinate $z$. 
Both stellar 
$\vartheta$
and $\varphi$ domains are divided into 100 grid-cell intervals.

From the computational point of view, the fundamental problem of the 2D self-consistent disk modeling
in the $R$-$z$ plane is the disproportional increase of the disk vertical scale height $H$ within a
distance of a few stellar radii (from Eq.~\eqref{diskh} follows $H\propto R^{3/2}$ in the inner Keplerian region and $H\propto R^2$
in the outer angular momentum conserving region). To avoid this difficulty, we
have developed a specific type of non-orthogonal 
(we call it ``flaring'') grid, which is a kind of hybrid of cylindrical and spherical spatial mesh.
We convert Eqs.~\eqref{numappr2} - \eqref{numappr6} into the ``flaring'' system using transformation equations introduced 
in Appendix~\ref{flarecoords} where we 
analyze the grid geometry in detail. The aim of introducing such a grid
was to exclude the points with high $z$ and low $R$ 
(see Fig.~\ref{fig3}) from the calculation. The disk density is very low in
these regions, therefore they are irrelevant for the disk dynamics. On the other
hand, we have to involve the regions with high $z$ at greater distance from the star \citep[cf.][]{kurfurst2015thesis}. 
We also describe the explicit form of the Roe matrices used in the single-step code version, re-calculated for the 
``flaring'' coordinate system, in Sects.~\ref{roeadiaflare} and \ref{roeisoflare}.
The gas hydrodynamic computations on the flaring grid
successfully run up to (at least) $1000\,R_\star$.
On the other hand, we did not 
use the grid for magnetohydrodynamic
calculations for its great mathematical complexity. Another possibility may be the usage
of a kind of structured quadrilateral mesh, presented, for example, in \citet{2002cfd..book.....C}
with static mesh refinement multigrid hierarchy.
The viability of the flaring system for various purposes and conditions will be the subject of additional testing.

We have also made attempts to calculate some models with the vertical
hydrostatic equilibrium included merely as an initial state, while in the further evolution, the
hydrostatic equilibrium is numerically switched off. However, such calculations have not yet been
successful due to a violent computational instability after a few time steps.
The likely reason is too overly temporal and spatial variation of all quantities
(unlike,~e.g.,~in the isothermal and non-viscous model of \citet{2016MNRAS.458.2323K}), including the crossing of the transforming wave (see Sect.~\ref{chartime}). 
Further testing and fine-tuning of the numeric schema is need at this point.
For this reason the current numerical calculations are performed by updating the vertical hydrostatic balance within each time step.

In the operator-split smooth hydrodynamic code, we calculate the time step $\Delta t$ employing the predefined
Courant-Friedrichs-Lewy (CFL) number $C_0\le 1/2$ \citep{1992ApJS...80..753S}.
The directionally split algorithm (where for simplicity we write the standard cylindrical $z$ coordinate while the ``flaring'' form of the algorithm may be an analog 
of Eq.~\eqref{Roetimes}) is
\begin{align}\label{time1}
\Delta t_{R}=\frac{\Delta R_i}{|V_{R(i,k)}|},\,
\Delta t_{z}=\frac{\Delta z_k}{|V_{z(i,k)}|},\,\Delta t_{a}=\frac{\text{min}\left(\Delta R_i,\Delta z_k\right)}{a_{i,k}},
\end{align}
where $a$ is the isothermal speed of sound.
We  denote the contribution of the numerical viscosity to the time increment as $\Delta t_{\text{vis}}$ where \citep{1992ApJS...80..753S}
\begin{align}\label{time3}
\Delta t_{\text{vis}}=\frac{1}{4\text{C}_2}\text{min}\left(\frac{\Delta R_i}{|\Delta V_{R(i,k)}|},
\frac{\Delta z_k}{|\Delta V_{z(i,k)}|}\right).
\end{align}
The factor $4$ in the denominator of Eq.~\eqref{time3} results from the transformation of the momentum equation into the diffusion equation due to the inclusion of the
artificial viscosity. The time step is calculated using the relation \citep{1986ASIC..188..187N,1992ApJS...80..753S} 
\begin{align}\label{time4}
\Delta t=C_0\left[\left(\Delta t_R\right)^{-2}+
\left(\Delta t_z\right)^{-2}+\left(\Delta t_a\right)^{-2}+\left(\Delta t_\text{vis}\right)^{-2}\right]^{-1/2};
\end{align}
the algorithm thus checks the Courant stability theorem by controlling the time step according to all the relevant physical conditions.

We calculate the interface between the optically thin and optically thick regions in the disk along the line-of-sight from stellar pole
(see the basic description in Sect.~\ref{tempstruct} and details in Appendix~\ref{optdepth}). 
We use the method of short characteristics for tracing the path of each ray that passes through 
the nodes of the grid (see Fig.~\ref{shortrays}). For each node of the disk 
grid located in the optically thin domain we find the distance $d$ (see Eq.~\eqref{diskirrad5}) from each ``visible'' node of the stellar surface grid.
We calculate the impinging stellar irradiation $\mathcal{F}_\text{irr}$ according to Eq.~\eqref{centerflux} as a numerical quadrature
neglecting the absorption of the flux in the optically thin region. We assume the flux $\mathcal{F}_\text{irr}$ is fully 
thermalized in the optically thin region taking into account radiative cooling using the optically thin approximation (see Sect.~\ref{tempstruct}).
The temperature profile in the optically thick domain is calculated as the diffusion approximation using Eqs.~\eqref{temp1}-\eqref{temp11} in Sect.~\ref{tempstruct}.

We performed all the calculations, employing both the methods introduced in this Section, on a numerical grid logarithmically scaled 
in the radial direction, with the outer radius of the computational domain $500\,R_\text{eq}$. The purpose of using such a relatively large
domain is to achieve a region with smooth profiles in the characteristics of the outer boundary, which are necessary for the proper convergence of the models.
The results introduced in the following Sections are therefore merely the illustrative fragments of the total computational domain.
The number of spatial grid cells is 500 in the radial and 200 in the vertical direction
(where the $R$ distance is logarithmically scaled).
The computational time was approximately one week (the maximum number of parallel processes used was 100). The physical time of the convergence of the models 
is identical to dynamical times introduced in Sect.~\ref{chartime}.

\section{Results of numerical models}\label{numresults}
\subsection{Disk evolution time}
\label{chartime}
\begin{figure}[t]
\begin{center}
\centering\resizebox{0.95\hsize}{!}{\includegraphics{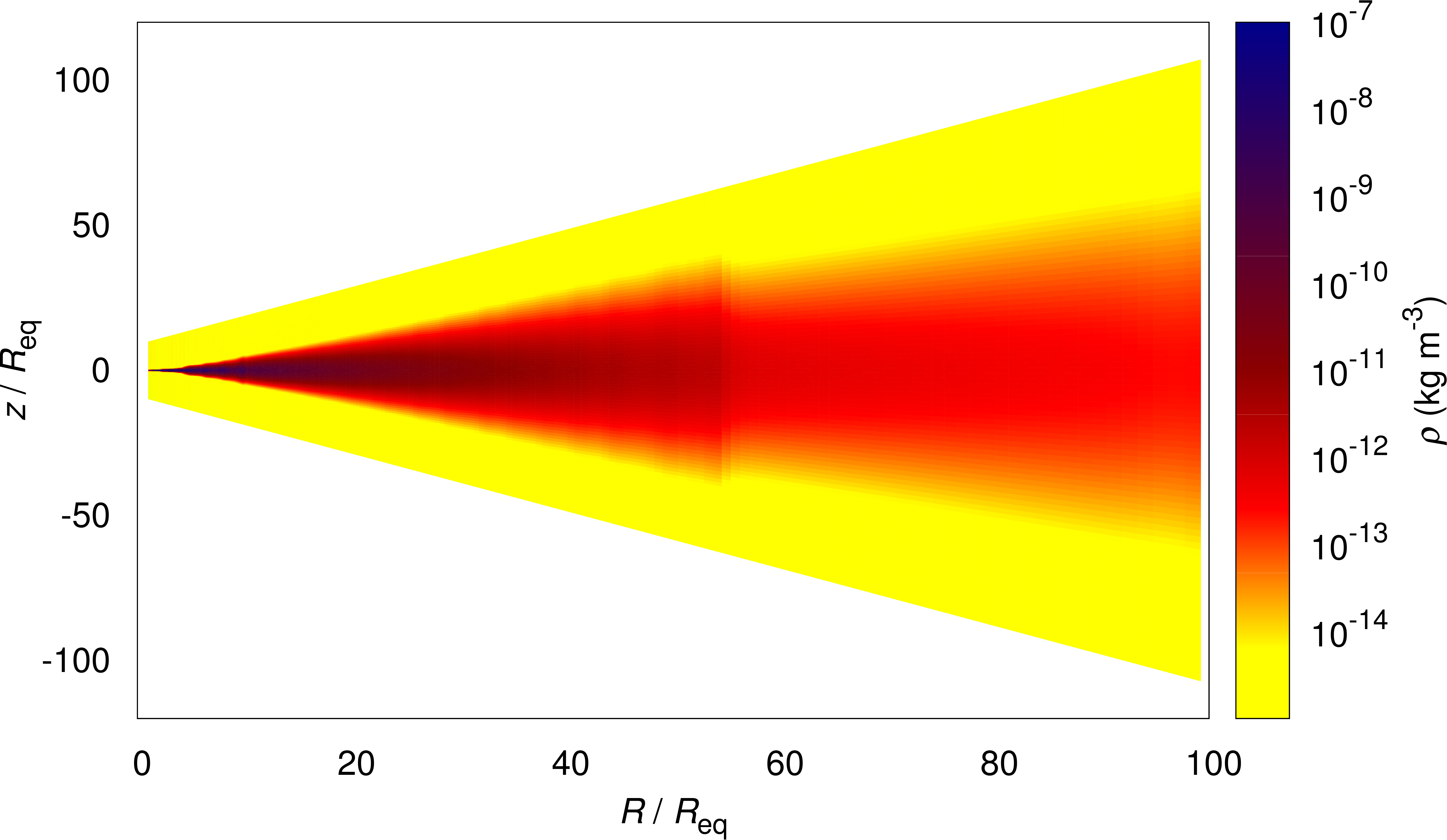}}
\caption{Snapshot of the density wave
propagating  in the distance $R\approx 55 R_{\text{eq}}$ in the converging 
2D model of the isothermal outflowing disk (cf.~the disk-forming density wave in 1D models in \citealt{2014A&A...569A..23K}),
calculated in the $R$-$z$ ($R$-$\theta$) plane. 
The elapsed time since the start of the simulation is approximately 80 days. The parameters 
of the critically rotating B0-type parent star are 
introduced in Table~\ref{table1}. 
The sonic point is beyond the displayed 
domain ($R_\text{s}\approx 5.5\times 10^2\,R_{\text{eq}}$). The model is calculated using the flaring 
grid whose prescription is given in Appendix~\ref{flarecoords}. 
The initial disk midplane density is determined by the stellar mass-loss rate $\dot{M}=10^{-9}\,M_{\odot}\,\text{yr}^{-1}$
and the viscosity is parameterized as $\alpha=\alpha_0=0.025$ \citep{2013MNRAS.428.2255P}.}
\label{denseadvect}
\end{center}
\end{figure}
\begin{table}[t]
\caption{Stellar parameters for B0-type star \citep[][see also \citealt{2014A&A...569A..23K}]{1988BAICz..39..329H}
and for typical Pop III star \citep[e.g.,][]{2001A&A...371..152M} used in the models. 
A selected metallicity $Z$ for B0-type star may roughly correspond to an average of Magellanic Clouds
\citep[e.g.,][]{2013A&ARv..21...69R} while we assume for simplicity $Z=0$ in the models of dense disks, corresponding to Pop~III stars.}
\label{table1}
\begin{center}
\begin{tabular}{ccccc}
\multicolumn{5}{l}{B0-type star:}\\
\hline
$M_\star(M_\odot)$  & $R_\star(R_\odot)$ & $R_\text{eq}(R_\odot)$ & $\langle T_\text{eff}\rangle(\text{K})$ & $Z$\\\hline
$14.5$ & $5.8$ & $8.7$ & $30\,000$ & $0.005$\\
\hline\\
\multicolumn{5}{l}{Pop III star:}\\
\hline
$M_\star(M_\odot)$  & $R_\star(R_\odot)$ & $R_\text{eq}(R_\odot)$ & $\langle T_\text{eff}\rangle(\text{K})$ & $Z$\\\hline
$40$ & $30$ & $45$ & $25\,000$ & $0$\\
\hline
\end{tabular}
\end{center}
\end{table} 
In the 2D hydrodynamic time-dependent models we recognize the same transforming wave as in our 1D models 
\citep[described in detail in][]{2014A&A...569A..23K} 
that converges the disk initial state to its final stationary state. The wave occurs during the disk-developing phase in all types of disks 
(Sects.~\ref{massiveresults} and \ref{lessmassiveresults}) and it may effectively determine 
the timescale of the circumstellar disk growth and dissipation periods \citep[e.g.,][]{1984ApJ...287L..39G,2003A&A...402..253S}. 
In the subsonic region the wave establishes nearly hydrostatic equilibrium in the radial direction
(see Eq.~\eqref{base3}), while its speed 
approximately equals the speed of sound. According to our 1D models \citep{2014A&A...569A..23K} the wave propagates
in the supersonic region as a shock wave
that is slowed down, compared to the subsonic region, to approx.~$0.5$-$0.3$~$a$. However, we cannot yet verify this prediction
in 2D self-consistent models 
since we have not yet been able to extend the 
calculations significantly above the sonic point distance due to the enormous computational cost.

In an isothermal disk (with $p=0$ in Eq.~\eqref{timeevol1})
the propagation time of the transforming wave in the subsonic region is the dynamical time 
(see the discussion in Sect.~5.2. of our foregoing paper \citet{2014A&A...569A..23K})
\begin{align}
\label{tdyn}
t_\text{dyn}\approx R/a.
\end{align}
We note that the dynamical time is almost independent of the viscosity while it significantly increases with decreasing temperature.
Figure \ref{denseadvect} demonstrates the snapshot of the time evolution of the density in a 2D
isothermal model calculated 
using the flaring grid (see Sect.~\ref{flarecoords}).
The Figure shows the transforming wave that is currently propagating roughly in the middle of the radial domain. 
The central star is in this case the B0-type star with the parameters 
given in Table~\ref{table1}, with the viscosity parameter 
$\alpha=\alpha_0=0.025$ \citep{2013MNRAS.428.2255P}.
The initial and boundary conditions are identical to those described in Sects.~\ref{basehydroeqs} and \ref{numapproach};
we however employ the initial (Keplerian) azimuthal velocity profile defined as $V_{\phi,\,\text{ini}}=\sqrt{-g_RR}$
where $g_R(R,z)$ is the radial component of gravitational acceleration.
The initial disk midplane density is determined by the mass-loss rate $\dot{M}=10^{-9}\,M_{\odot}\,\text{yr}^{-1}$ (selected as a free parameter).
For example for the two distances, $50\,R_{\text{eq}}$ and $100\,R_{\text{eq}}$, the 2D model gives 
from Eq.~\eqref{tdyn} $t_{\text{dyn}}\approx 0.6\,\text{yr}$ and $1.2\,\text{yr}$, respectively.

To derive an estimate of
the dynamical time in our nonisothermal models, we neglect the
vertical variations of temperature and fit the disk midplane temperature by a
power law Eq.~\eqref{timeevol1}. The dynamical time is
from Eqs.~\eqref{tdyn} and \eqref{timeevol1}
\begin{align}\label{timeevol2}
t_\text{dyn}=\frac{1}{a_0R_\text{eq}^{p/2}}\int_{R_\text{eq}}^{R}R^{p/2}\dd R,
\end{align}
where $a_0^2=kT_0/(\mu m_\text{u})$ is the speed of sound at the base of the disk, $k$ is the Boltzmann constant, $\mu$ is the mean molecular weight 
(for simplicity we hereafter use $\mu=0.62$ for the fully ionized hydrogen-helium plasma), and $m_\text{u}$ is 
the atomic mass unit. Using this relation we can compare the values of
$t_\text{dyn}$ in the non-isothermal models.
As an example, the analytical estimate of the dynamical time
in the model introduced in Sect.~\ref{massiveresults} with $\dot{M}=10^{-6}\,M_{\odot}\,\textup{yr}^{-1}$, given by Eq.~\eqref{timeevol2}
(with  $T_0\approx 80\,000\,\text{K}$ and $p\approx 0.7$ derived from the fit of
the numerical model),
is $t_{\text{dyn}}\approx 4.4\,\text{yr}$ for the distance $50\,R/R_\text{eq}$ 
while the numerical model gives the value $t_{\text{dyn}}\approx 
3.9\,\text{yr}$.

\begin{figure}[t]
\begin{center}
\centering\resizebox{1.\hsize}{!}{\includegraphics{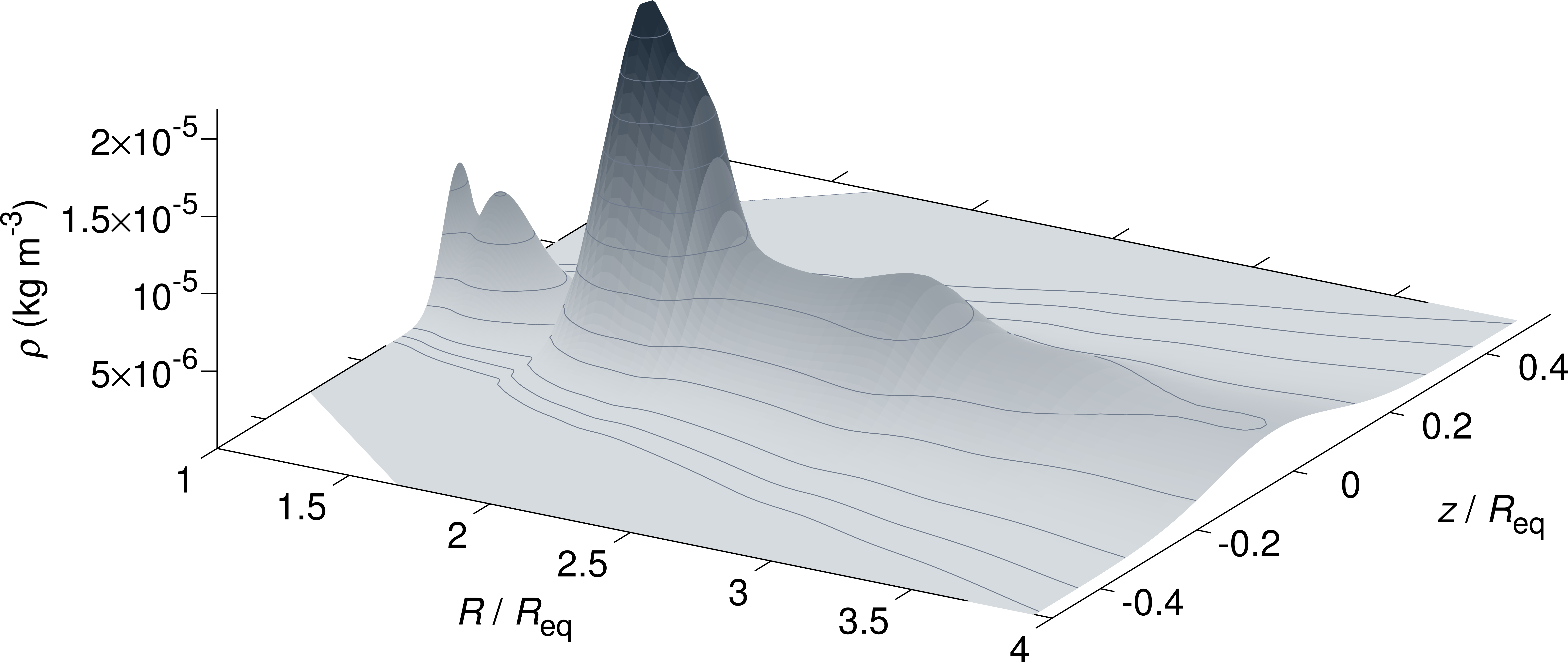}}
\caption{2D graph of the density of the converged model of a
circumstellar viscous outflowing disk of a (near) 
critically rotating star in the very inner region up to 5 stellar radii, corresponding to the disk mass-loss rate 
$\dot{M}=10^{-6}\,M_{\odot}\,\text{yr}^{-1}$, with the constant viscosity parameter $\alpha=\alpha_0=0.1$.
The parameters of the star correspond to average parameters of Pop III stars (see Table \ref{table1})
or sgB[e] stars \citep[e.g.,][]{2007A&A...463..627K}. The sonic point is in this case at a distance of approximately~$2\times 10^4\,R_\text{eq}$.
The density profile shows in this case irregular bumps in the very inner disk region. 
The contours mark the density levels (from lower to higher) $10^{-9}$, $10^{-8}$, $10^{-7}$, 
$10^{-6}$, $2.5\times 10^{-6}$, $5\times 10^{-6}$ $[\text{kg}\,\text{m}^{-3}]$, 
and so on, with a constant increment of $2.5\times 10^{-6}\,\text{kg}\,\text{m}^{-3}$.}
\label{densedisk-6-01-0_very_short}
\end{center}
\end{figure}
\begin{figure}[t]
\begin{center}
\centering\resizebox{1.\hsize}{!}{\includegraphics{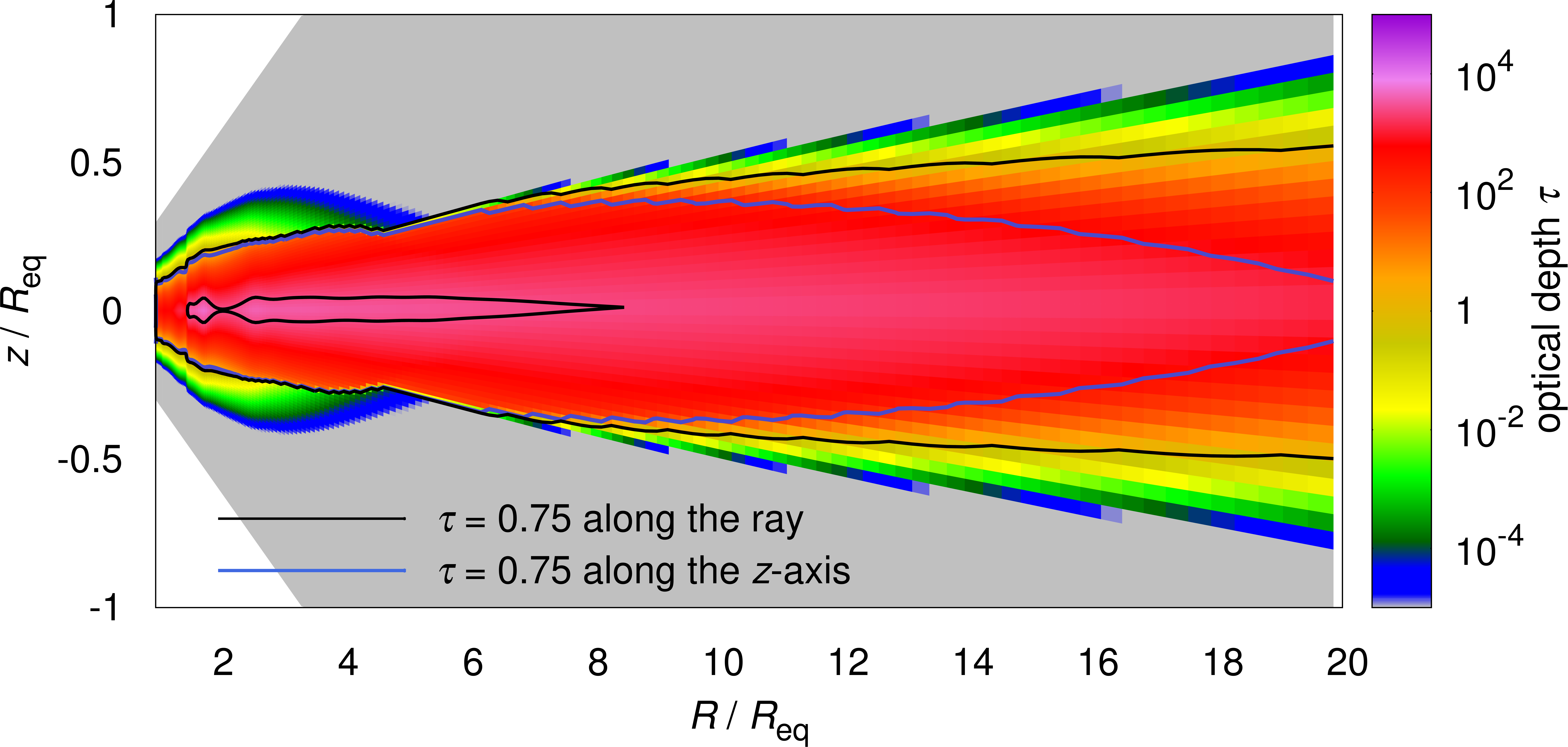}}
\centering\resizebox{1.\hsize}{!}{\includegraphics{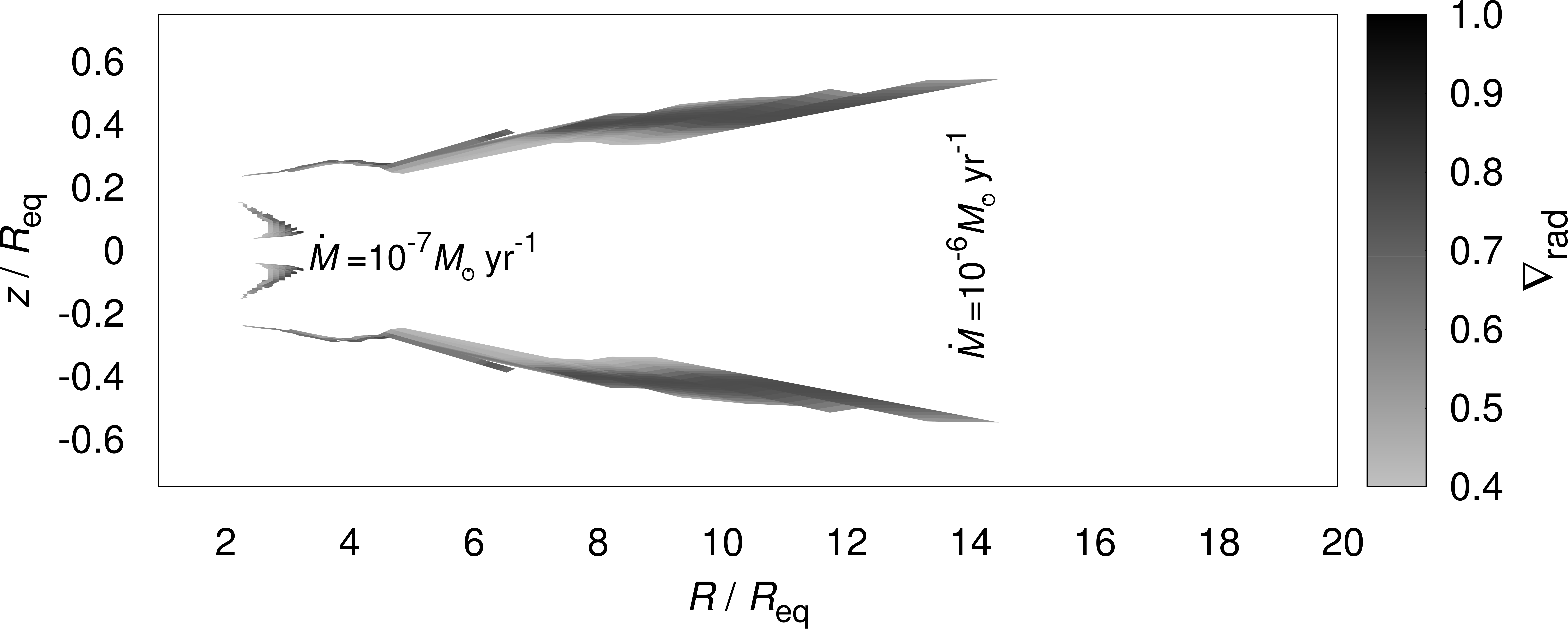}}
\caption{
\textit{Upper panel}: 
Color map of the profile of the optical depth in the inner part (up to 20 stellar equatorial radii) of the dense circumstellar outflowing disk of a (near) 
critically rotating star with the same parameters as in Fig.~\ref{densedisk-6-01-0_very_short}. The optical depth is calculated using the method described 
in Appendix~\ref{optdepth}. 
The outer black contour traces the optical depth level $\tau=0.75$ that is calculated along the line-of-sight from the stellar pole 
(described as ``along the ray'' in the Figure legend), 
while the blue contour traces the same optical depth calculated along 
the vertical ($z$ axis) direction. The inner black contour traces the optical depth level $\tau=2\,500$ that is calculated along the line-of-sight from the stellar pole. 
\textit{Lower panel}: Map of the convective zones with $\nabla_\text{rad}>\nabla_\text{ad}$ on the same scale. The large zones refer to the 
disk mass-loss rate $\dot{M}=10^{-6}\,M_{\odot}\,\text{yr}^{-1}$ while the small wings near the disk base refer to $\dot{M}=10^{-7}\,M_{\odot}\,\text{yr}^{-1}$.}
\label{optdepth_short}
\end{center}
\end{figure}

\subsection{Models of massive dense disks with disk mass-loss rate $\dot{M}> 10^{-8}\,M_{\odot}\,\textup{yr}^{-1}$}\label{massiveresults}
\begin{figure}[t]
\begin{center}
\centering\resizebox{1.\hsize}{!}{\includegraphics{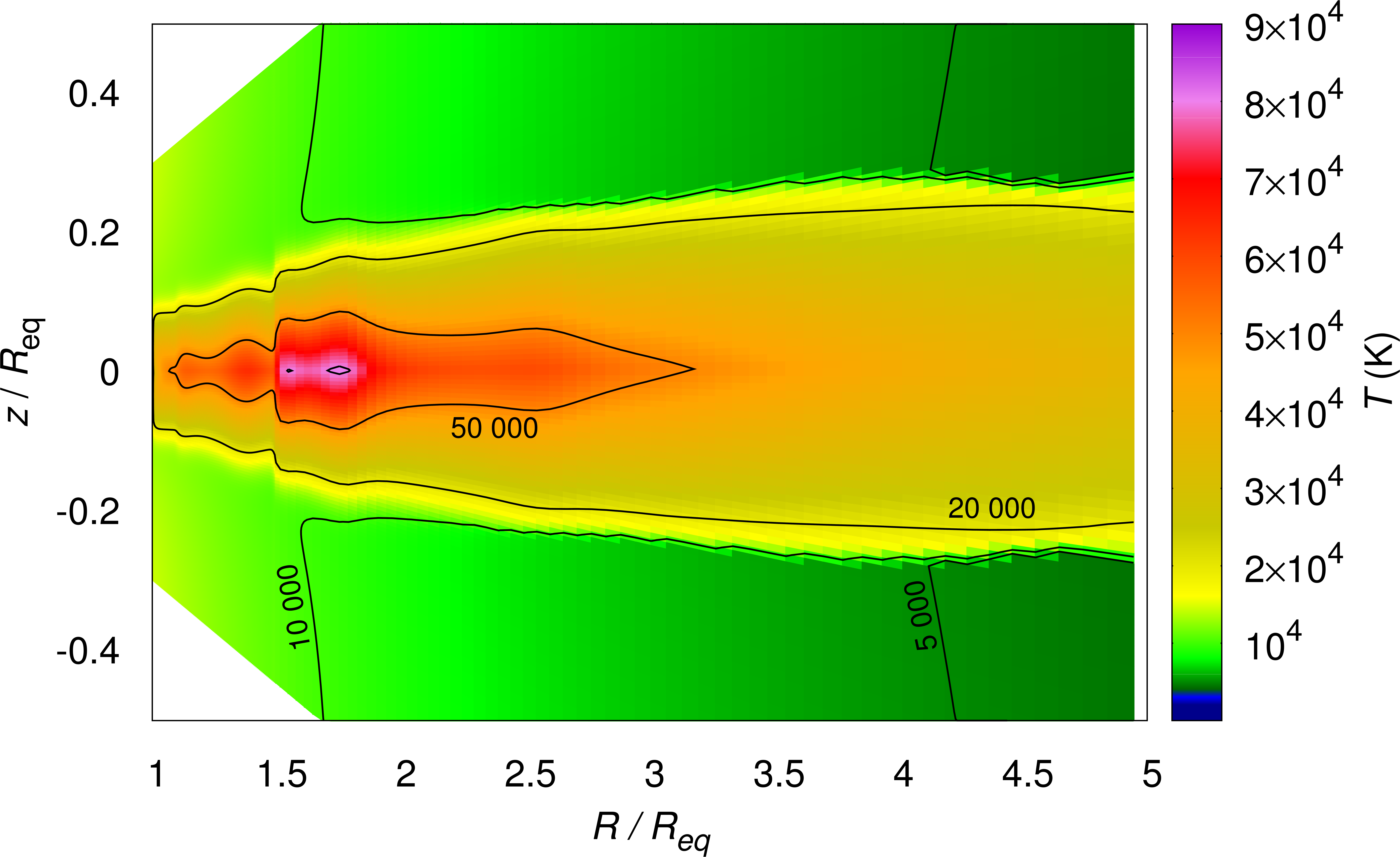}}
\centering\resizebox{1.\hsize}{!}{\includegraphics{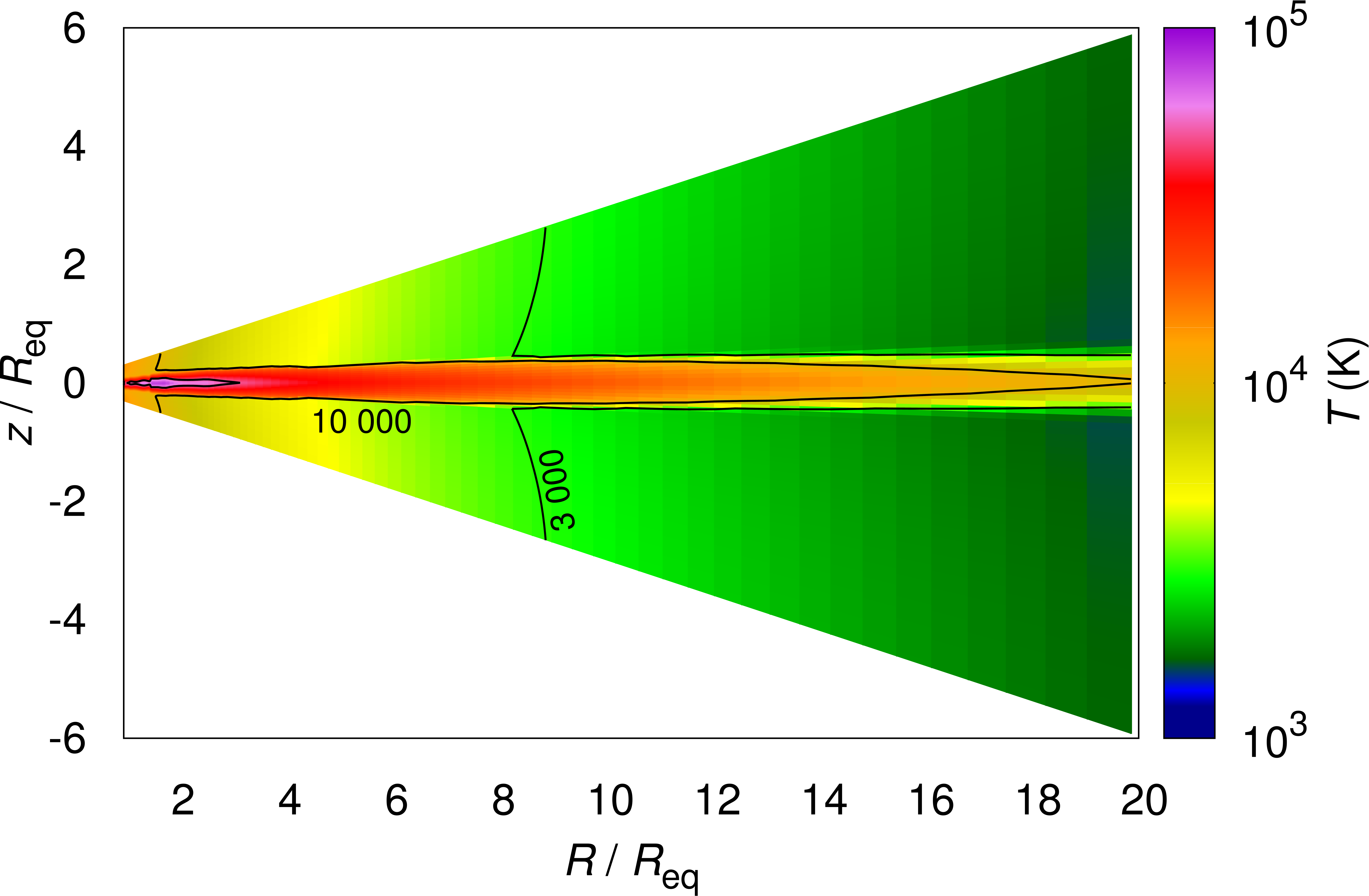}}
\caption{\textit{Upper panel}: The color map of the temperature distribution in the very inner part (up to 5 stellar equatorial radii) of the dense circumstellar outflowing disk of a (near) 
critically rotating star with the same parameters as in Fig.~\ref{densedisk-6-01-0_very_short}.
The region of significantly increased temperature near the disk midplane is generated by the viscous heating. 
The contours mark the temperature levels 5\,000\,K, 10\,000\,K, 20\,000\,K, 50\,000\,K and 80\,000\,K.
\textit{Lower panel}: As in the upper panel, in the inner part up to 20 stellar equatorial radii. 
The strip of highly increased temperature near the disk midplane is generated by the viscous heating. 
The contours mark the temperature levels 3\,000\,K, 10\,000\,K and 50\,000\,K.}
\label{tempdisk-6-01-0_very_short}
\end{center}
\end{figure}
\begin{figure}[t]
\begin{center}
\centering\resizebox{1.\hsize}{!}{\includegraphics{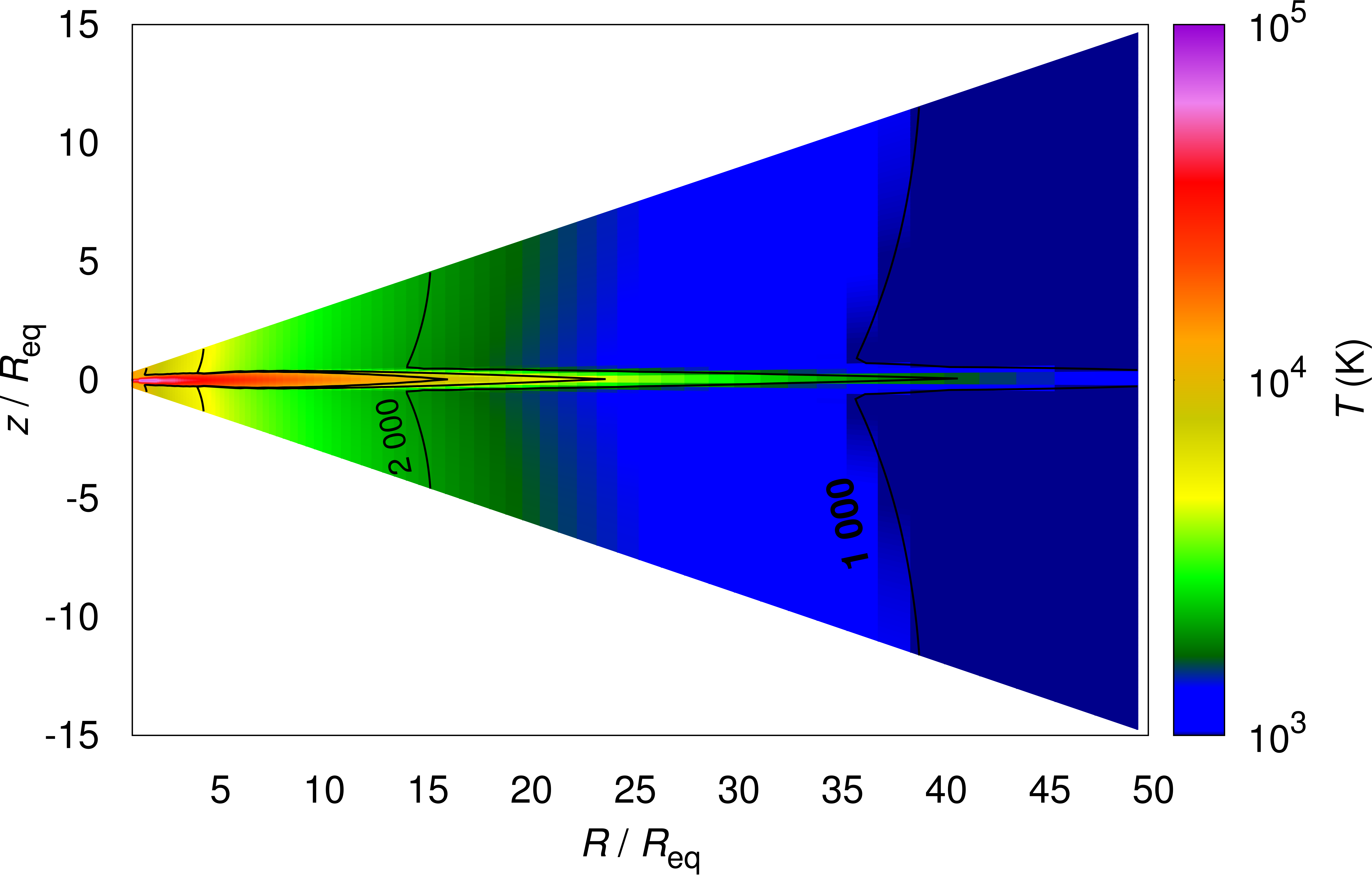}}\\
\vspace{0.2cm}
\centering\resizebox{1.\hsize}{!}{\includegraphics{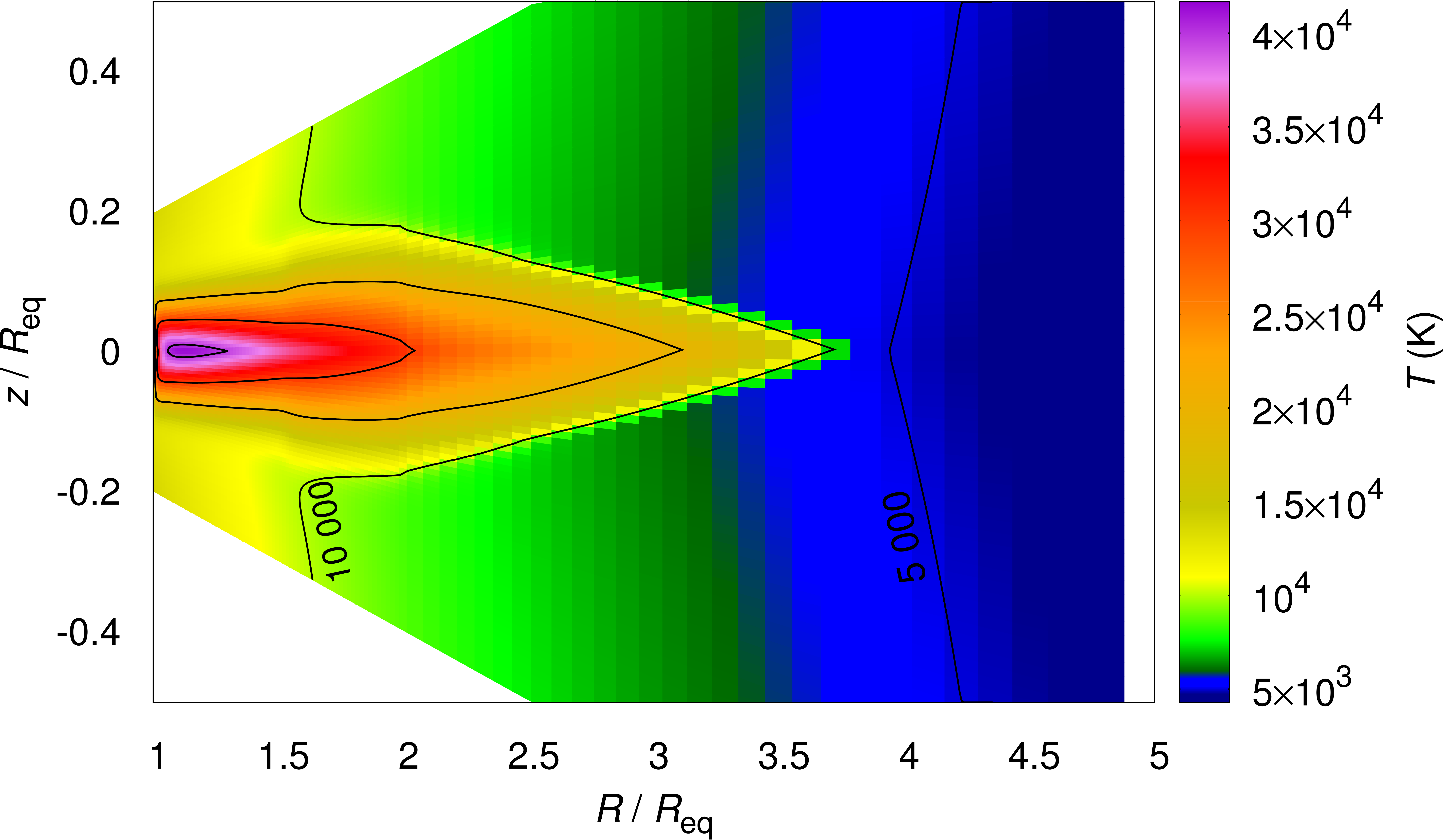}}
\caption{\textit{Upper panel}: As in Fig.~\ref{tempdisk-6-01-0_very_short}, up to the intermediate distance of 50 stellar equatorial radii. 
The near-equatorial strip of enhanced temperature
exceeds in this case the distance $50\,R_\text{eq}$.
The contours mark the temperature levels 1\,000\,K, 2\,000\,K, 5\,000\,K and 10\,000\,K.
\textit{Lower panel}: The color map of the temperature distribution in the very inner part (up to 5 stellar equatorial radii) of the dense circumstellar outflowing disk of a (near) 
critically rotating star with the same parameters as in Fig.~\ref{densedisk-6-01-0_very_short}, however, 
the parameterized disk mass-loss rate is in this case $\dot{M}=10^{-7}\,M_{\odot}\,\text{yr}^{-1}$. 
The region of increased temperature generated by the viscous heating near the disk midplane is in this case significantly reduced and extends up to only
approximatively $4\,R_\text{eq}$.
The contours mark the temperature levels 5\,000\,K, 10\,000\,K, 20\,000\,K, 30\,000\,K and 40\,000\,K.}
\label{tempdisk-6-01-0_mid}
\end{center}
\end{figure}

We have examined a great variety of models with different parameters, from the highest disk mass-loss rate $\dot{M}=10^{-5}\,M_{\odot}\,\text{yr}^{-1}$ that may correspond to 
the most massive outflowing disks of Pop~III~stars \citep{2001A&A...371..152M,2008A&A...478..467E} or sgB[e]s
\citep{2007A&A...463..627K,2010A&A...517A..30K}
to the lowest ones, $\dot{M}=10^{-11}\,M_{\odot}\,\text{yr}^{-1}$, corresponding to classical Be star disks \citep[e.g.,][]{2008ApJ...684.1374C}.
The reason why we introduce disk models with the mass-loss rate higher or lower than approximately 
$\dot{M}\approx 10^{-8}\,M_{\odot}\,\text{yr}^{-1}$ separately in different Sections is 
that the higher 
ones may correspond to massive disks of giant stars like sgB[e]s or Pop III stars,
while the lower ones dominantly represent the disks of classical Be stars
(although the transition may not be accurate nor sharp).

Our calculation showed that the results of the models with disk mass-loss rates
$\dot{M}=10^{-5}\,M_{\odot}\,\text{yr}^{-1}$ and
$\dot{M}=10^{-6}\,M_{\odot}\,\text{yr}^{-1}$ are qualitatively similar and
differ
only by the density and temperature.
Figure~\ref{densedisk-6-01-0_very_short}
shows the distribution of the self-consistently calculated density in the very inner region of the disk (up to $5\,R_\text{eq}$)
in the model
with
disk mass-loss rate
$\dot{M}=10^{-6}\,M_{\odot}\,\text{yr}^{-1}$ and the constant viscosity $\alpha=\alpha_0=0.1$.
The parameters of the central 
Pop III (or sgB[e]) star are introduced in Table~1 (we assume for simplicity
zero metallicity
in this type of disks).

Irregular bumps of the density occur in these high-density models that may be connected with the (more or less regular) 
drops in radial velocity $V_R$
to negative values, 
which indicate the material infall that periodically increases the angular momentum of the inner disk, creating
the density waves (cf.~the density waves in the model with lower density, but with the high-viscosity parameter, in the 
lower panel in Fig.~\ref{densedisk-8-01-0_very_short}, 
vs. the smooth density profile in the model with relatively low density and viscosity in the upper panel of Fig.~\ref{densedisk-8-01-0_very_short}). 
The 
time scale
of this process is however quite fast; of the order of hours or days.
For a better estimation, we need to simulate this particular 
disk behavior on much shorter time-scales, probably using the local computational schema focused on a small region of a disk.

\begin{figure}[t]
\begin{center}
\centering\resizebox{0.925\hsize}{!}{\includegraphics{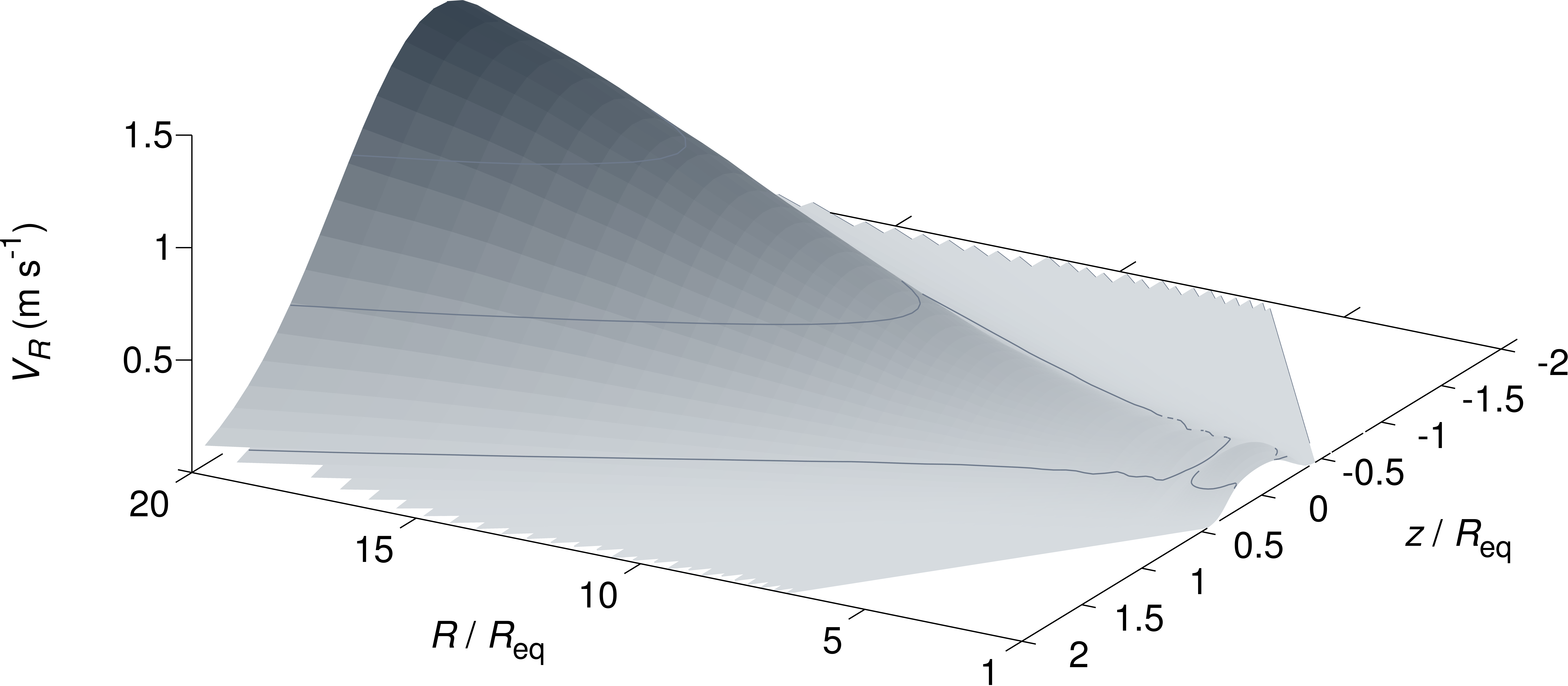}}\\
\centering\resizebox{0.95\hsize}{!}{\includegraphics{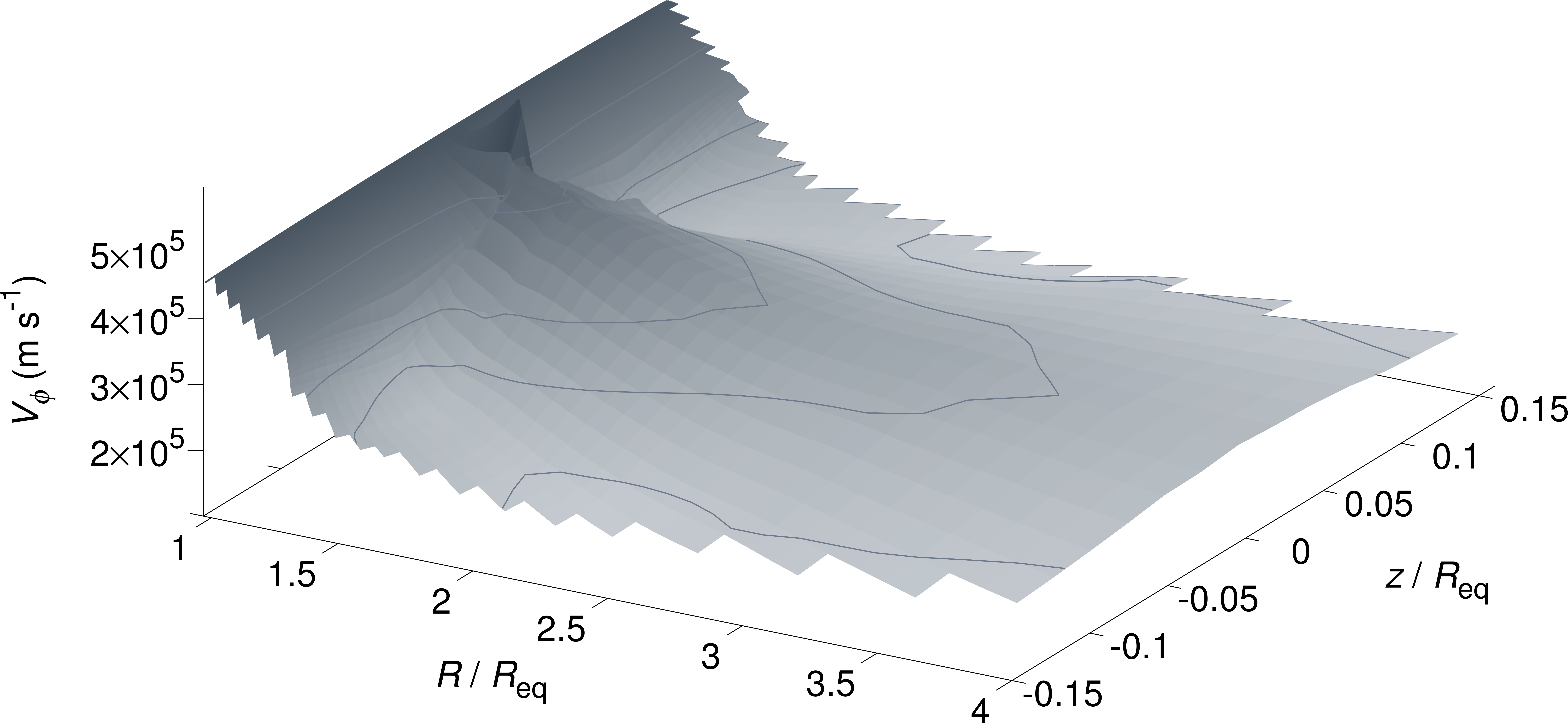}}
\caption{\textit{Upper panel}: 2D profile of disk radial velocity $V_R$ calculated up to the distance $R=20\,R_\text{eq}$ 
for the Pop III star's disk with parameters and rates given in Table~\ref{table1} and Fig.~\ref{tempdisk-6-01-0_very_short}.
Contours mark the radial velocity $0.1$ and $0.5\,\text{m}\,\text{s}^{-1}$.
\textit{Lower panel}: 2D profile of disk azimuthal velocity $V_\phi$ calculated up to the same distance for the same model as in the upper panel.
Contours mark the azimuthal velocity $2\times 10^5$, $2.5\times 10^5$, and $3\times 10^5\,\text{m}\,\text{s}^{-1}$.}
\label{velos}
\end{center}
\end{figure}

We attach in Fig.~\ref{velos} the graphs of radial velocity $V_R$ and azimuthal velocity $V_\phi$ in the inner disk region 
(up to 20 stellar radii) of the converged disk, 
calculated within the 2D model of the disk of the Pop III star with parameters given in Table~\ref{table1} with $\dot{M}=10^{-6}\,M_\odot\,\text{yr}^{-1}$.
As we may expect kinematically, both the velocities show maximum vertical values in the disk midplane while with increasing $|z|$ the velocities are lower.
However, the calculations do not include the effects of ablation of the disk's surface layers
due to strong radiation \citep[cf.][]{2016MNRAS.458.2323K}.
Equation~\eqref{base2} (with use of Eq.~\eqref{sigi}) implies that $\dot{M}=\text{const.}$ throughout the stationary disk \citep[see also][]{2014A&A...569A..23K}.
We do not show the profile of the angular-momentum-loss rate $\dot{J}(R)$ since it basically corresponds to the 
profiles introduced in \citet{2014A&A...569A..23K}, that is, the rate increases up to approximately the sonic 
point radius.

All the models with $\dot{M}>10^{-9}\,M_{\odot}\,\text{yr}^{-1}$ (see 
Figs.~\ref{densedisk-6-01-0_very_short} and \ref{densedisk-8-01-0_very_short}
as well as the corresponding chart of the temperature in Fig.~\ref{tempdisk-6-01-0_very_short})
show the permanent bump of the density (and of
pressure and temperature) at the radius
$1.5\text{ - }3.5\,R_\text{eq}$ produced by large
viscous friction together with the increase of the rate of external heating at the particular distance (due to stellar oblateness). 
This causes the rapid increase of disk vertical scale height 
(see Eq.~\eqref{diskh}, which becomes significant in the models
with high disk density.
This, in turn, leads to formation of the ``inner torus'' in the models 
(cf. the accretion disk morphology in \citet{2003ARA&A..41..555B}, however, the formation mechanism is in this case likely quite different).

We also examined the zones of possible
influence of convection 
(see the lower panel in Fig.~\ref{optdepth_short} which maps the convective zones with $\nabla_\text{rad}>\nabla_\text{ad}$;
see also the description in Sect.~\ref{tempstruct}), however we
do not model the convective process itself. 
In agreement with \citet{1991MNRAS.250..432L}, we found the formation of convective zones 
only in the models with $\dot{M}\ge 10^{-7}\,M_{\odot}\,\text{yr}^{-1}$. The largest disk convective zone (as expected)
develops in the model with $\dot{M}\ge 10^{-6}\,M_{\odot}\,\text{yr}^{-1}$; its
inner boundary is at the radius
$R\approx 2\,R_\text{eq}$, and it is developed in the 
relatively narrow strip near the line of optical depth level
$\tau\approx 1$ (see Fig.~\ref{optdepth_short}) up to the radius
$R\approx 14\,R_\text{eq}$. 
The convective zone is drastically reduced to the area between $R\approx 2\text{ - }3\,R_\text{eq}$ and $|z|\approx 0.05\text{ - }0.15\,R_\text{eq}$
in the model with 
$\dot{M}=10^{-7}\,M_{\odot}\,\text{yr}^{-1}$.
The convective zones, 
which may also be connected with and enhanced by the effects of hydrogen ionization, 
contribute to the increase of the
disk optical depth (cf. \citealt{1991MNRAS.250..432L}). However, the permanent bump (although more subtle and more distant) occurs also in the models 
without the convection (see, e.g., Fig.~\ref{densedisk-8-01-0_very_short}).

The upper panel in 
Fig. \ref{optdepth_short}
shows the distribution of the optical depth in the very massive disk model with $\dot{M}=10^{-6}\,M_{\odot}\,\text{yr}^{-1}$.
The highest optical depths in the disk midplane in particular models in this Section are $\tau\approx 4000$ at the radius
$R\approx 1.5\,R_\text{eq}$ in the model with $\dot{M}=10^{-6}\,M_{\odot}\,\text{yr}^{-1}$, while this is reduced to 
$\tau\approx 200$ at the radius
$R\approx 1.1\,R_\text{eq}$ in the model with $\dot{M}=10^{-7}\,M_{\odot}\,\text{yr}^{-1}$. 

The resulting radial-vertical temperature structure of the dense disk models up to 5, 20 and 50 stellar radii is plotted in 
Figs.~\ref{tempdisk-6-01-0_very_short} - \ref{tempdisk-6-01-0_mid}.
The maximum temperatures are $T\approx 80\,000\,\text{K}$ and $T\approx 40\,000\,\text{K}$ in the disks with $\dot{M}=10^{-6}\,M_{\odot}\,\text{yr}^{-1}$ and
$\dot{M}=10^{-7}\,M_{\odot}\,\text{yr}^{-1}$, respectively, and
the regions with maximum temperature roughly correspond to the regions with highest optical depth.
The disks with $\dot{M}\ge 10^{-6}\,M_{\odot}\,\text{yr}^{-1}$ show (with particular stellar parameters) 
significant strip of increased temperature up to the radius
$R\approx 50\,R_\text{eq}$.
In this point we note that the sonic point radius
for the disks with given parameters is 
approximatively at $2\times 10^4\,R_\text{eq}$. We also tested the models with higher 
disk base viscosity, however, the resulting temperature profiles differ only by a few percents. 
The models with radially decreasing viscosity (where $n>0$ from Eq.~\eqref{base1})
give the same results as the introduced models with constant viscosity, the differences in 
the models with decreasing viscosity are evident only at much larger distances from the central star
\citep[see our 1D models in][]{2014A&A...569A..23K}.

The detection of strong [OI] line emission in sgB[e]
disks \citep{2007A&A...463..627K} implies a presence of
high density region that is neutral in hydrogen ($6000\,\text{K}\lesssim T\lesssim 8000\,\text{K}$).
Assuming a viscous disk, the temperatures near the disk core are too high due to the viscous heat generation.
However, since the temperature contour $10\,000\,\text{K}$ is inside the optically thick 
region ($\tau\gtrsim 0.75$), the neutral hydrogen emitting layer could
possibly be located
near the transition from an optically thin to thick region. 

\subsection{Models of less massive disks with disk mass-loss rate $\dot{M}\le 10^{-8}\,M_{\odot}\,\textup{yr}^{-1}$ 
that may correspond to classical decretion disks of Be stars}\label{lessmassiveresults}
\begin{figure}[t]
\begin{center}
\centering\resizebox{0.95\hsize}{!}{\includegraphics{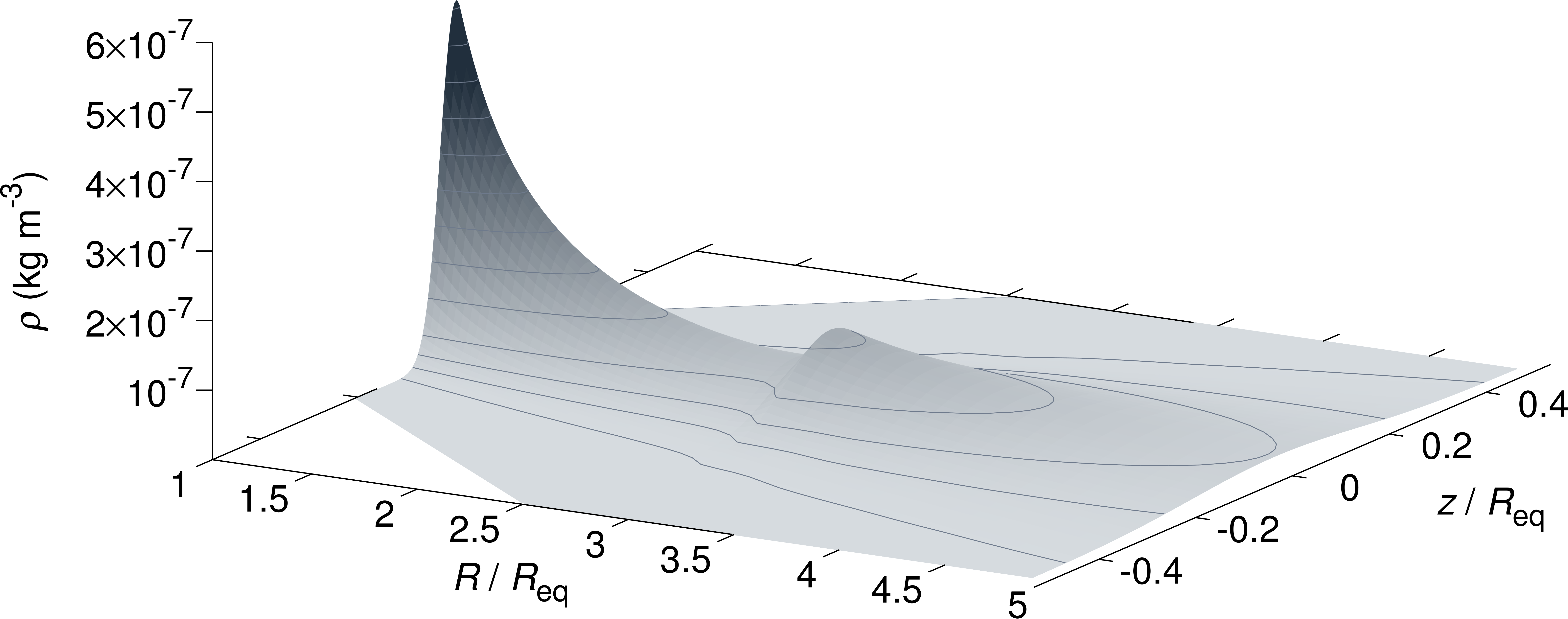}}\\
\vspace{0.2cm}
\centering\resizebox{0.95\hsize}{!}{\includegraphics{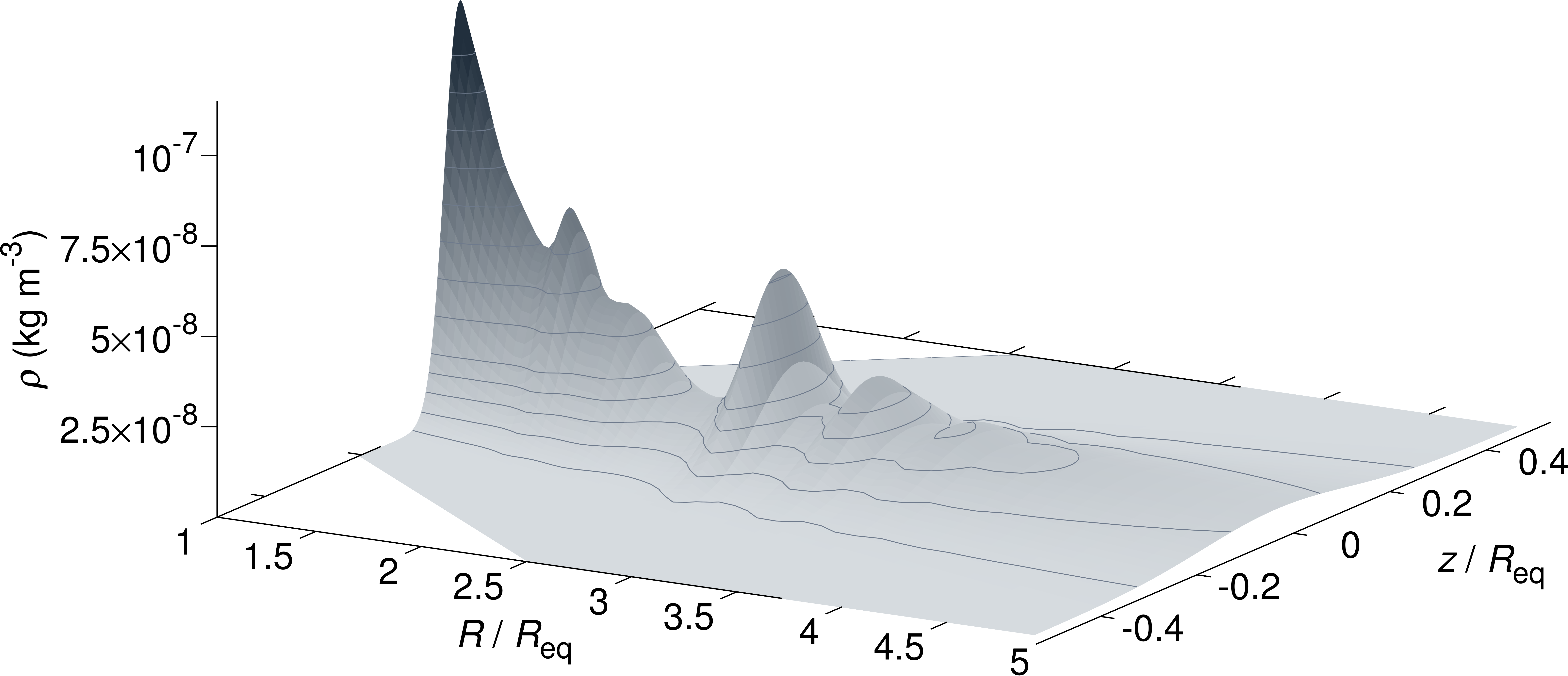}}
\caption{\textit{Upper panel}: 2D graph of the density of a converged model of a
circumstellar viscous outflowing disk of a critically rotating B0-type
star in the very inner region up to 5 stellar radii, corresponding to the disk mass-loss rate 
$\dot{M}=10^{-8}\,M_{\odot}\,\text{yr}^{-1}$, with the constant viscosity parameter $\alpha=\alpha_0=0.1$.
The sonic point radius is in this case approximately
$2.5\times 10^4\,R_\text{eq}$ and
the density profile is relatively smooth. The contours mark the density values (from lower to higher) $2.5\times 10^{-8}$, 
$5\times 10^{-8}$, $10^{-7}$, $1.5\times 10^{-7}$, $2\times 10^{-7}$ $[\text{kg}\,\text{m}^{-3}]$, and so on, with 
a constant increment of $0.5\times 10^{-7}\,\text{kg}\,\text{m}^{-3}$.
\textit{Lower panel}: Same graph, but with the viscosity coefficient $\alpha=\alpha_0=1$.
The density profile shows significant roughly periodic waves that occur in the case of a viscosity parameter $\alpha\gtrsim 0.5$. 
The contours mark the density levels (from lower to higher) $2.5\times 10^{-8}$, 
$5\times 10^{-8}$, $7.5\times 10^{-8}$ and $10^{-7}$ $[\text{kg}\,\text{m}^{-3}]$.}
\label{densedisk-8-01-0_very_short}
\end{center}
\end{figure}
\begin{figure}[t]
\begin{center}
\centering\resizebox{1.\hsize}{!}{\includegraphics{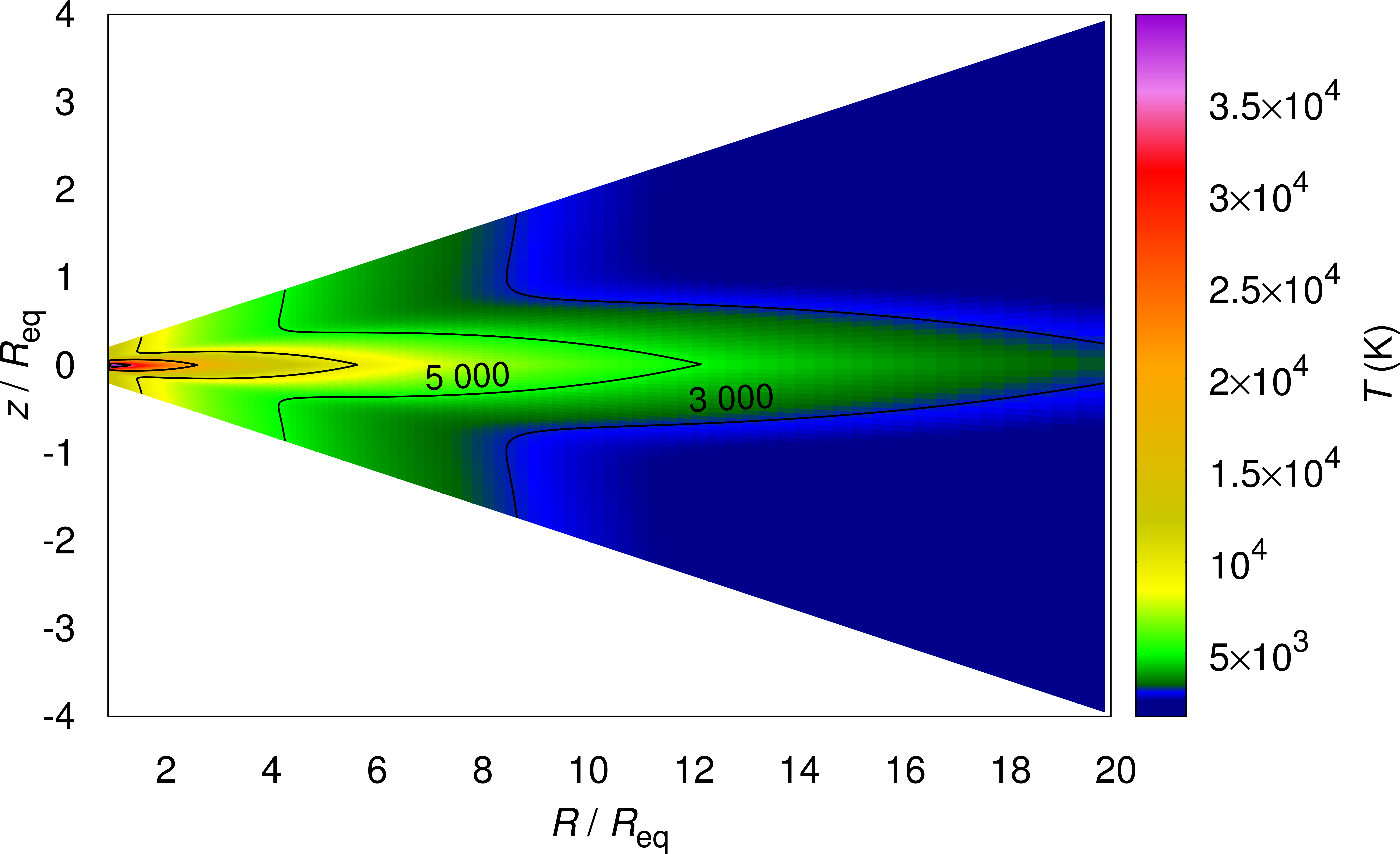}}\\
\vspace{0.1cm}
\centering\resizebox{1.\hsize}{!}{\includegraphics{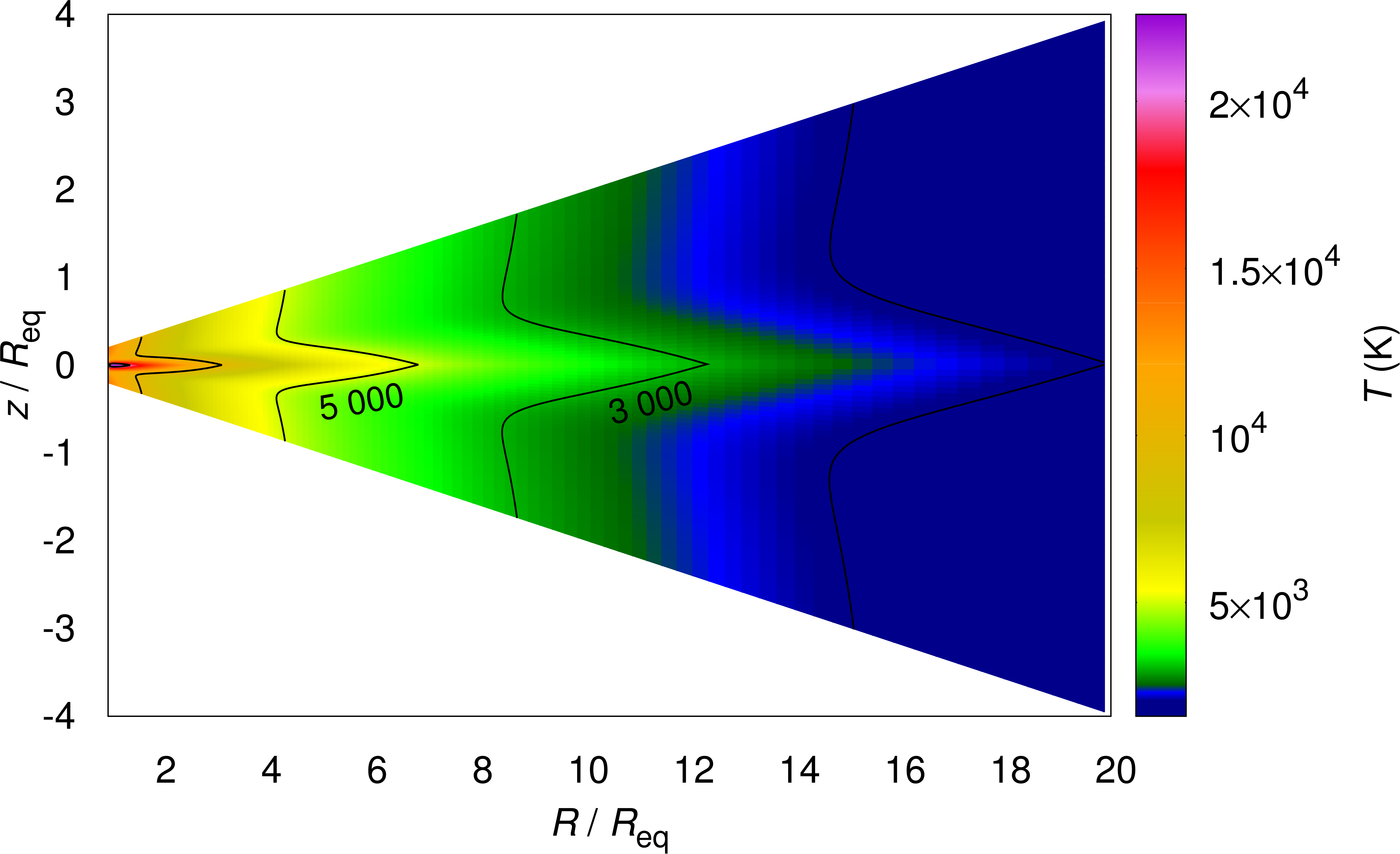}}
\caption{\textit{Upper panel}: The color map of the temperature distribution in the inner part (up to 20 stellar equatorial radii) 
of the circumstellar outflowing disk of a 
critically rotating star with the same parameters as in Fig.~\ref{densedisk-8-01-0_very_short} ($\dot{M}=10^{-8}\,M_{\odot}\,\text{yr}^{-1}$, $\alpha=\alpha_0=0.1$).
The region of increased temperature generated by the viscous heating near the disk midplane extends in this case to a distance of 
approximately $20\,R_\text{eq}$.
The contours mark the temperature levels 3\,000\,K, 5\,000\,K, 10\,000\,K, 20\,000\,K (and 35\,000\,K at the base of the disk).
\textit{Lower panel}: Color map of the temperature distribution in the inner part (up to $20\,R_\text{eq}$)
of the circumstellar outflowing disk of a 
critically rotating star with the same parameters as in Fig.~\ref{densedisk-8-01-0_very_short}, with the disk mass-loss rate 
$\dot{M}=10^{-9}\,M_{\odot}\,\text{yr}^{-1}$, with the constant viscosity parameter $\alpha=\alpha_0=0.1$.
The region of increased temperature generated by the viscous heating near the disk midplane is significant to a distance of 
approximately $15\,R_\text{eq}$.
The contours mark the temperature levels 2\,000\,K, 3\,000\,K, 5\,000\,K, 10\,000\,K (and 20\,000\,K at the base of the disk).}
\label{densedisk-8-01-0_temp_short}
\end{center}
\end{figure}
\begin{figure}[t]
\begin{center}
\centering\resizebox{1.\hsize}{!}{\includegraphics{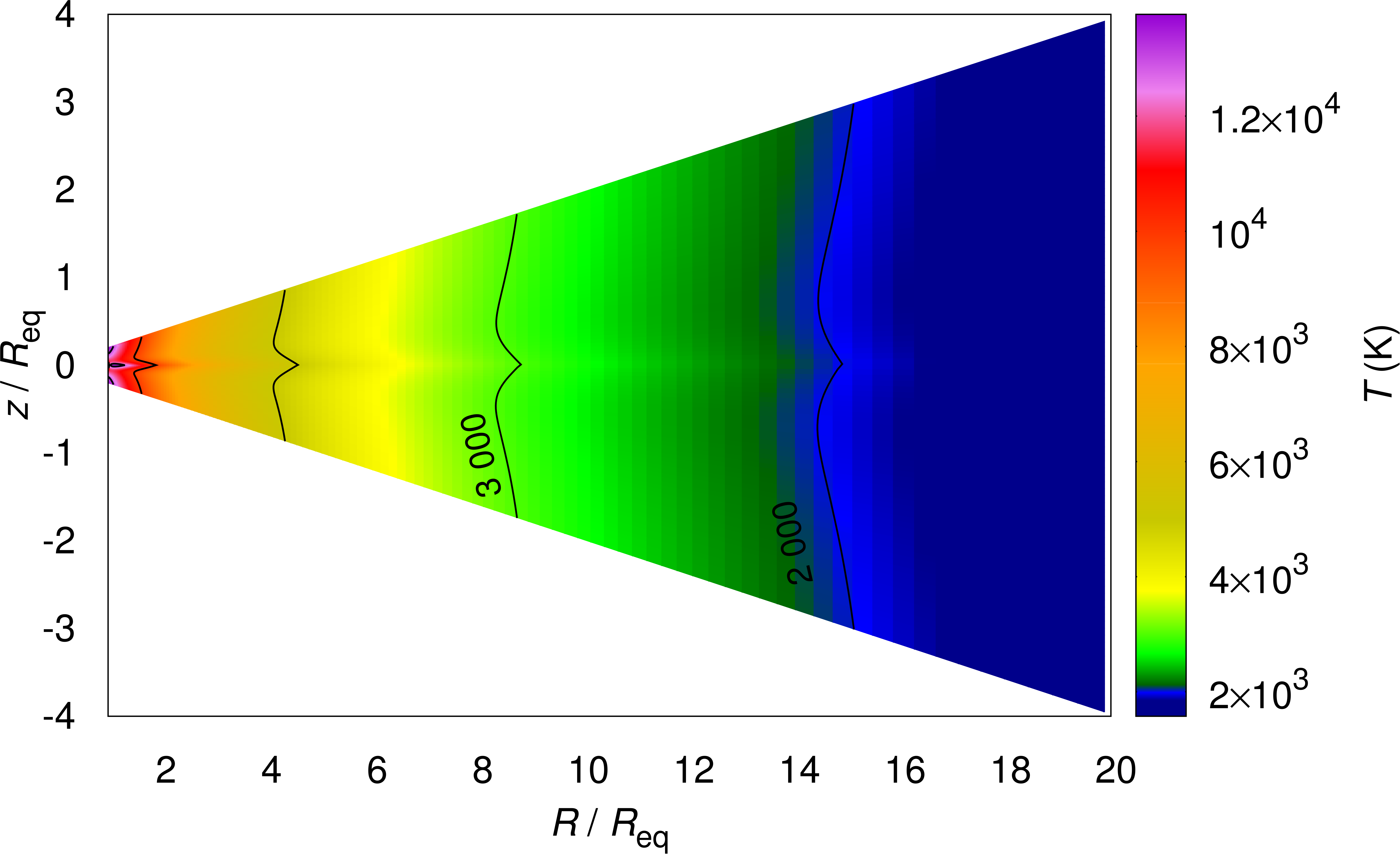}}\\
\vspace{0.1cm}
\centering\resizebox{0.95\hsize}{!}{\includegraphics{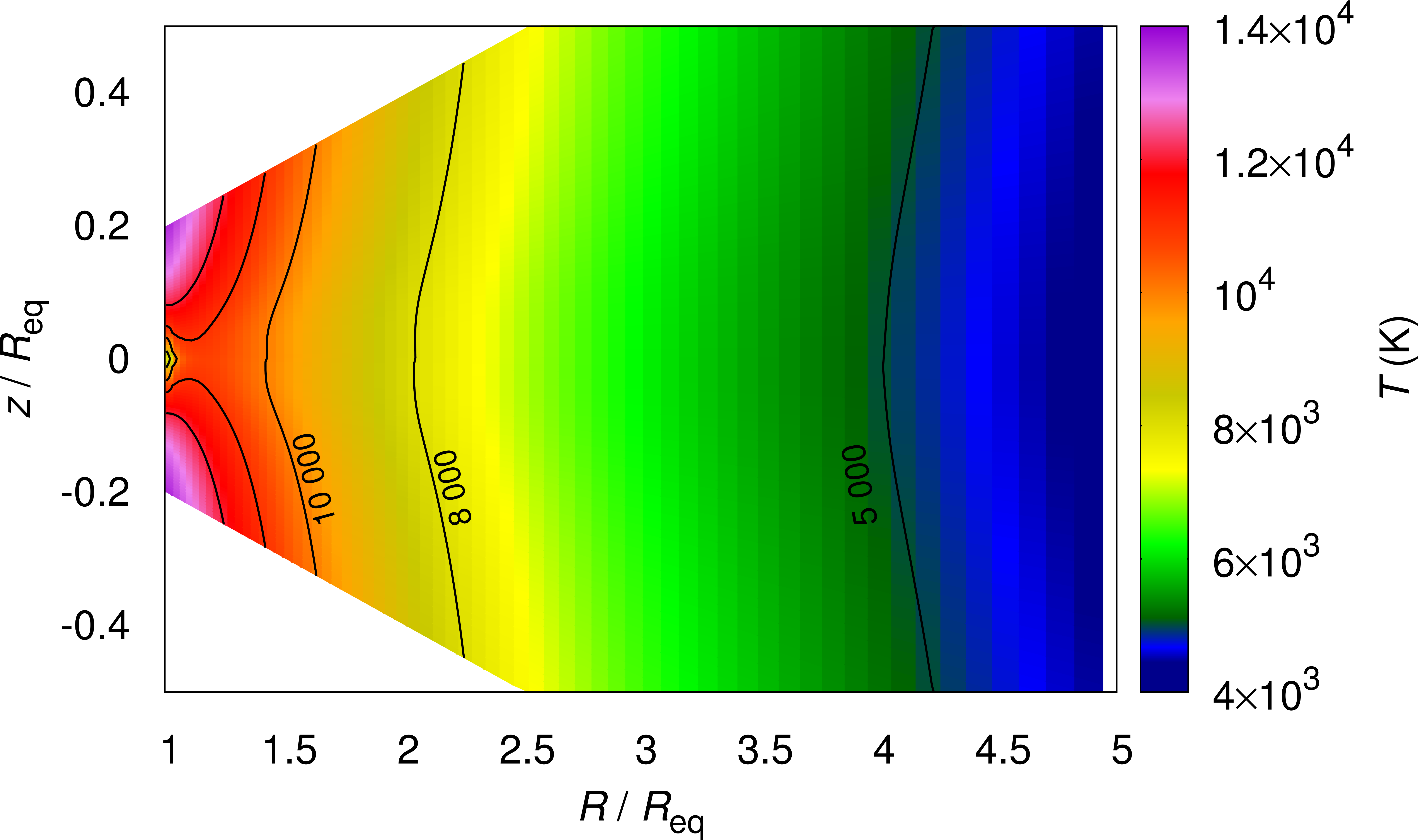}}
\caption{\textit{Upper panel}: Color map of the temperature distribution in the inner part (up to $20\,R_\text{eq}$)
of the circumstellar outflowing disk of a 
critically rotating star with the same stellar parameters as in Fig.~\ref{densedisk-8-01-0_very_short}, with the disk mass-loss rate 
$\dot{M}=10^{-10}\,M_{\odot}\,\text{yr}^{-1}$, with the constant viscosity parameter $\alpha=\alpha_0=0.1$.
The strip of increased temperature generated by the viscous heating begins to be visible near the disk midplane. 
The contours mark the temperature levels 2\,000\,K, 3\,000\,K, 5\,000\,K, 10\,000\,K (and 12\,000\,K at the base of the disk).
\textit{Lower panel}: The color map of the temperature distribution in the very inner part (up to 5 stellar equatorial radii) of the circumstellar outflowing disk of a 
critically rotating star with the same stellar parameters as in Fig.~\ref{densedisk-8-01-0_very_short}, corresponding however to a disk mass-loss rate 
$\dot{M}=10^{-11}\,M_{\odot}\,\text{yr}^{-1}$, with a constant viscosity parameter $\alpha=\alpha_0=0.1$.
The strip of increased temperature near the disk midplane generated by the viscous heating for this $\dot{M}$ disappears. 
The contours mark the temperature levels 5\,000\,K, 8\,000\,K, 10\,000\,K, 11\,000\,K, and 12\,000\,K.}
\label{densedisk-10-01-0_temp_short}
\end{center}
\end{figure}

The disks with a disk mass-loss rate equal to or lower than $\dot{M}=10^{-8}\,M_{\odot}\,\text{yr}^{-1}$ 
may correspond to the classical Be stars$^\prime$ disks \citep[e.g.,][]{2008ApJ...684.1374C,2011A&A...527A..84K,2013A&A...553A..25G}. The
assumed parameters of the central B0-type star in this Section are introduced in Table \ref{table1}.
Figure \ref{densedisk-8-01-0_very_short}
shows the difference in the profiles of density in the very inner region in the disk with $\dot{M}=10^{-8}\,M_{\odot}\,\text{yr}^{-1}$,
for the constant viscosities $\alpha=0.1$ and $\alpha=1$, respectively. While the density profile for $\alpha=0.1$ is relatively smooth, 
the same profile for $\alpha=1$ shows the 
bumps (waves) that are characteristic for the models with higher density 
(see Sect.~\ref{massiveresults}). Consistently, 
the disk base density for $\alpha=1$ is about
4-5 times lower than for $\alpha=0.1$ which corresponds to a higher radial velocity of the material outflow in the inner disk regions 
for higher values of $\alpha$ \citep[see the models with 
different viscosity parameters $\alpha$ in][]{2011A&A...527A..84K}.

The temperature structures of the disk models with
$\dot{M}=10^{-8}\,M_{\odot}\,\text{yr}^{-1}$,
$\dot{M}=10^{-9}\,M_{\odot}\,\text{yr}^{-1}$, and 
$\dot{M}=10^{-10}\,M_{\odot}\,\text{yr}^{-1}$ up to $20\,R_\text{eq}$
are plotted in Figs.~\ref{densedisk-8-01-0_temp_short} and \ref{densedisk-10-01-0_temp_short}, respectively.
The maximum temperatures in the disk cores that may be generated by the viscous heating
decrease with decreasing disk mass-loss rate from
$T\approx 38\,000\,\text{K}$ in the disks with $\dot{M}=10^{-8}\,M_{\odot}\,\text{yr}^{-1}$
and $T\approx 22\,000\,\text{K}$ in the disks with $\dot{M}=10^{-9}\,M_{\odot}\,\text{yr}^{-1}$ to
$T\approx 13\,500\,\text{K}$ in the disks with $\dot{M}=10^{-10}\,M_{\odot}\,\text{yr}^{-1}$. 
The maximum optical depths within the same regions are $\tau\approx 165$, $\tau\approx 20$, and $\tau\approx 2.2$,  respectively. However, for 
$\dot{M}\le 10^{-11}\,M_{\odot}\,\text{yr}^{-1}$ , the disk optical depth is $\tau<0.5$ throughout the disk and 
the contribution of the viscosity to the temperature structure becomes negligible due to low disk density
(lower panel of Fig.~\ref{densedisk-10-01-0_temp_short}).

\section{Discussion}\label{discuss}

To test the assumption of heat equilibrium (neglecting the advection term; see Sect.~\ref{basehydroeqs}), 
we compared the energy efficiency of external and internal sources for the disk models with 
$\dot{M}\ge 10^{-6}\,M_{\odot}\,\text{yr}^{-1}$ (the highest examined mass loss rates) in the following way:
we quantified the radiative energy density of the stellar irradiation that impinges the disk 
on one side and the excess of energy density generated in the disk 
(kinetic energy due to the radial outflow 
plus the energy produced by pressure and viscosity) on the other side. The ratio of energy densities of the external irradiation to the internal sources
in the model is approximately $2$ orders of magnitude at the base of the disk while it approaches $3$ orders of magnitude at the 
radius $10\,R_\text{eq}$. Obviously, in the models with lower $\dot{M}$ this ratio is higher. 
This comparison shows that it is not necessary to solve full energy equation to
obtain the disk temperature.

Our 2D calculations of the disk-core temperature structure are based on a diffusion approximation within the optically thick domain of the disk, 
where the optical depth $\tau\geq 2/3$, and include
the contribution of heating generated by viscous friction. 
In the surrounding optically thin environment we calculate the temperature assuming a local thermodynamic equilibrium of the gas with 
the impinging stellar irradiation, including the effects of radiative loss in the optically thin approximation
accounting for heavier elements (see Sect.~\ref{tempstruct}), 
while neglecting the contribution of the viscous heating. We calculate self-consistently the evolution of the gas-dynamic and thermal structure 
of the disk within each time-step of the convergence of our time-dependent 2D models. 

Similar disk models provided by, for example, \citet{2008ApJ...684.1374C} or \citet{2009ApJ...699.1973S} 
employ
NLTE radiative codes for the calculations of the optically thin disk temperature structure. 
They however use fixed hydrodynamic distribution of the disk gas, calculated 
prior to their NLTE models, and do not take into account the viscous friction in the core of the disk. On the other hand
we take into account only the free electrons produced by ionization of hydrogen, 
and do not include bound-bound opacities.

Comparing the basic results of our calculations with published models, we see that
the viscous heat effect begins to occur near the disk equatorial plane in the models with $\dot{M}=10^{-10}\,M_\odot\,\text{yr}^{-1}$. 
The total effects of calculated viscous heat are in accordance with \citet{1991MNRAS.250..432L} 
who found the domination of viscous heat generation rate over the irradiative
heating rate only when $\dot{M}\gtrsim 10^{-5}\,M_\odot\,\text{yr}^{-1}$, regarding the vertically integrated viscous heat.
For example, in the model of the Pop III star (see Table~\ref{table1}) with $\dot{M}=10^{-6}\,M_\odot\,\text{yr}^{-1}$, where in the distance $1.5\,R_\text{eq}$
the parameters $a^2\approx 10^9 \text{m}^2\text{s}^{-2}$, $\Omega=7.2\times 10^{-6}\,\text{s}^{-1}$, and $\Sigma\approx 2.5\times 10^5\,\text{kg}\,\text{m}^{-2}$,
we obtain $\mathcal{F}_\text{irr}(R)\approx 1.5\times 10^9\,\text{W}\,\text{m}^{-2}$ while the vertically integrated 
$F_\text{vis}(R)\approx 2\times 10^8\,\text{W}\,\text{m}^{-2}$ (the latter value 
is in fact
even lower because we assumed here for simplicity
the maximum disk core temperature $80\,000\,\text{K}$ within the whole vertical disk profile, while in 
reality
the temperature vertically decreases).
The disk base temperatures 
$T_0 \approx 17\,500\,\text{K}$ and $T_0 \approx  21\,000\,\text{K}$
in the models for $\dot{M}\geq 10^{-9}\,M_\odot\,\text{yr}^{-1}$ in Sects.~\ref{massiveresults} and \ref{lessmassiveresults}, respectively, 
correspond to approximately 70\% of the stellar effective temperature $\langle T_{\text{eff}}\rangle = 25\,000\,\text{K}$
and $\langle T_{\text{eff}}\rangle = 30\,000\,\text{K}$. This
agrees with the calculations of 
\citet{2008ApJ...684.1374C}, for example, who found approximately $T_0/\langle T_{\text{eff}}\rangle \approx 0.7$ within the 
nearly isothermal solution of Keplerian viscous disk with the disk mass-loss rate $\dot{M}=5\times 10^{-11}\,M_\odot\,\text{yr}^{-1}$.
The reason why the disk base temperature is lower than the mean effective stellar
temperature is the self-shadowing of the star itself due to the stellar oblateness and equatorial darkening together with 
the self-shadowing of the optically thick gas in the inner disk region \citep{kurfurst2015thesis}. Unlike \citet{2008ApJ...684.1374C}, 
we have not obtained a similar radial temperature profile, that is, 
a rapid temperature decrease to approx.~$0.4\,T_{\text{eff}}$ at the radius $R\approx (3\text{\,-\,}4)\,R_{\text{eq}}$ 
followed by its increase to approx.~$0.6\,T_{\text{eff}}$ at the radius $R\approx 5\,R_{\text{eq}}$
in the disk midplane,
which results from inclusion of the optically thin radiative equilibrium in a significantly rarefied gas at larger distance.
\citet{2009ApJ...699.1973S} derive
the disk base temperature $T_0=13\,500\,\text{K}$ in a circumstellar viscous disk of Be star 
$\gamma$ Cas with $\langle T_{\text{eff}}\rangle\approx 25\,000\,\text{K}$.
Several different density models of \citet{2009ApJ...699.1973S} with $\rho_0=10^{-9}\,\text{kg}\,\text{m}^{-3}$, 
$\rho_0=5\times 10^{-9}\,\text{kg}\,\text{m}^{-3}$ and $\rho_0=5\times 10^{-8}\,\text{kg}\,\text{m}^{-3}$ 
(noting that the disk mass-loss rate $\dot{M}=10^{-11}\,M_\odot\,\text{yr}^{-1}$ 
roughly corresponds to $\rho_0=10^{-8}\,\text{kg}\,\text{m}^{-3}$), respectively, show, in the 2D temperature distributions of the $\gamma$ Cas disk, 
the development of a cool region near the equatorial plane with steep vertical temperature gradients. The vertical temperature profiles in those models range
approximately from 10\,000\,K in the disk midplane to the maximum 15\,000\,K in the 
model with highest density and approximately from 8\,000\,K in the disk midplane to a maximum of 15\,000\,K in the model with lowest density, 
considering the radial domain up to 5 stellar radii. 
In our model with similar parameters (lower panel of Fig.~\ref{densedisk-10-01-0_temp_short}), the disk vertical temperature profile varies within the range
(10\,000\,K-12\,000\,K), but only in the very short radial region up to 1-2 stellar radii. However, for $\dot{M}=10^{-11}\,M_\odot\,\text{yr}^{-1}$ our calculation is 
less relevant, since we do not take into account the optically thin radiative equilibrium in the relatively very low-density gas. Another question refers to 
the behavior of the models of B0 star disks 
with $\dot{M}=10^{-10}\,M_\odot\,\text{yr}^{-1}$ and $\dot{M}=10^{-9}\,M_\odot\,\text{yr}^{-1}$ when taking into account the NLTE calculations.

We have also fitted the models to examine the radial profiles of the disk midplane density $\rho_\text{eq}\sim R^n$ 
in the final stationary state. 
Avoiding the regions with irregularities, we found the slopes of $\rho_\text{eq}$ in the models of very dense disks with $\dot{M}=10^{-6}\,M_\odot\,\text{yr}^{-1}$
corresponding to the slope parameter $n$ between $2.85$ and $3.0$ (where the usual analytical solution gives the value $3.5$ - see Sect.~\ref{instate}) in the region 
between approximately $R=6\,R_\text{eq}$ and $R=20\,R_\text{eq}$. The slope parameter steepens to $4.0$ in the region between approximately 
$R=30\,R_\text{eq}$ and $R=50\,R_\text{eq}$, and the slope parameter stabilizes at approximately $3.2$ 
in the region above the distance $R\geq 100\,R_\text{eq}$. For the less massive disks with 
$\dot{M}\leq 10^{-8}\,M_\odot\,\text{yr}^{-1}$ the value of the same slope parameter within the same region is approximately $3.125$. This
agrees with observational results of \citet{2017A&A...601A..74K}, who determine
the values of the disk midplane density slope parameter between approximately $3.0$ and $3.5$ for the Be stars' disks that are in a steady state.
They also predict that disks with this parameter higher than $3.5$ should correspond to disks that are in the process of formation, and that disks with
this parameter lower than $3$ are in the process of dissipation; the variations of the values of the slope parameter in the case of very massive disks indicate 
that these processes occur up to a certain distance (see Sect.~\ref{massiveresults}).

The occurrence of the more or less irregular instabilities (density waves) in the models with 
$\dot{M}\geq 10^{-7}\,M_\odot\,\text{yr}^{-1}$ (or with very high $\alpha$ parameter) roughly corresponds to cases of the convection development 
due to the steep radiative gradient 
(see Fig.~\ref{optdepth_short}). 
In regards to this point, our current calculations do not confirm either the violation of the Richardson criterion due to the rotation-driven instabilities 
\citep[see, e.g.,][page 294]{2009pfer.book.....M}, 
or that these perturbations seem to be associated with turbulence generated by hydrodynamic (or WKB) 
density waves \citep{2003ARA&A..41..555B,2011Natur.470..475B}.
We regard these instabilities as ``viscous instabilities'' caused by development of pressure bumps in the region near the base of the disk
\citep{2012A&A...538A.114P,2017arXiv170900246D}. The idea of a viscous disk 
that may become dynamically unstable and even lead to the formation of concentric rings was also suggested by \citet{2005MNRAS.362..361W}. However,
although the driving mechanisms of the instabilities may be, in the referred papers, different (nor are they are mathematically analyzed in detail), 
all the authors employ the initial infinitesimally small bump or wiggle in pressure (viscosity or diffusion) to 
describe the process analytically within the perturbation theory.
From disk theory \citep[e.g.,][]{1981ARA&A..19..137P,2002apa..book.....F}, as well as from Eqs.~\eqref{base2} - \eqref{base5} with use of Eq.~\eqref{sigi}, 
it follows that $\nu(R)\Sigma(R)=\text{const.}$ During the numerical convergence, the ``initial perturbation'' 
(given by the difference between the approximate initial state and the final converged state) relaxes in cases where 
the $\nu(R)\Sigma(R)$ product is relatively small while in the case of high values of the product the perturbations permanently 
propagate as a referred ``viscous instability''.
The detailed analysis of the effects of time-dependent perturbations and their distribution and evolution within the disk will be the subject of
future work.

This process may also be provoked by mutual dependence of density, viscosity, and temperature 
(see, e.g., Figs.~\ref{densedisk-6-01-0_very_short}, \ref{optdepth_short}, and \ref{tempdisk-6-01-0_very_short}) when 
the decrease in temperature causes the decrease in density and viscous friction (and vice versa) in the inner part of the disk, which allows the material to fall back.
However, the determination of the mechanism as well as the frequency of these perturbations 
will need to be the subject of further detailed studies (we cannot even actually rule out the effects of numerical artefacts 
caused by a number of mutually entangled calculations with extreme values of variables).

The nature of sgB[e] stars' disks is
an open question. There remains even a possibility of the existence of
a system of rings of high-density material in the surroundings of
these objects,
containing gas and dust \citep[see, e.g.,][]{2013A&A...549A..28K}. The
formation mechanism of such a disk (or ring structure) is also unclear. We know
about at least two sgB[e] candidates for
rapid rotators with the rotation velocity reaching a substantial part of the
critical velocity: the stars LHA 115-S 23
\citep{2000ASPC..214...26Z,2008A&A...487..697K} and LHA 115-S
65, both in the Small Magellanic Cloud \citep{2010A&A...517A..30K}. As another
example of this
type of objects we may consider the Galactic eccentric binary system GG Car
with a circumbinary ring, where one of the possible scenarios refers to the 
primary component as a classical Be star during the previous evolution of the
system \citep{2013A&A...549A..28K}.

Our models may provide significant corroboration of some aspects of a viscous decretion disk scenario. 
The performed results may, for example, confirm an agreement of the behavior of the instabilities in the
theoretical models with the observed dissipation curve of time-variable dissipating disks. 
This provides a tool to improve an estimate of the disk
viscosity parameter $\alpha$ and of the disk mass-loss rate $\dot{M}$ \citep[cf.,][who conclude that 
higher values of $\alpha$ parameter result from turbulent viscosity induced by disk instability]{2012ApJ...744L..15C}.
We may expect also a kind of observational consequence in case of very high temperatures in the 
cores of very dense disks (see Sect.~\ref{massiveresults}). Even if the cores are shielded by colder upper layers, the total radiative emission  
on lower frequencies should be significant. The turbulent instabilities may also
lead to the enhancement of the magnetic field by a dynamo effect.

\section{Summary and conclusions}\label{sumconcl}
We calculated axisymmetric, 2D, time-dependent models of circumstellar outflowing disks of
critically or near-critically rotating stars in the radial-vertical plane. 
We developed for this purpose two types of our own numerical Eulerian hydrodynamic code that employ full Navier-Stokes viscosity: the operator-split finite volume algorithm for (relatively) smooth hydrodynamic calculations,
and the unsplit finite volume algorithm (based on the Roe's method)
for the calculations of problems with sharp discontinuities and/or high Mach numbers. Furthermore, we calculated most of the performed models 
using both methods to compare the obtained numeric results.
Our 2D models of the self-consistently calculated
time-dependent density-velocity structure 
performed in the disk $R$-$z$ ($R$-$\theta$) plane employ the effects of
impinging stellar irradiation of the rotationally oblate star and the internal
viscous heating of the disk matter.

The models of very dense viscous disks with
$\dot{M}>10^{-8}\,M_\odot\,\text{yr}^{-1}$, which
may correspond to disk or disk-like environments of Pop III stars or sgB[e]
stars, show 
large strips of high density, optical depth and temperature that are generated by the viscosity, extending to a significant distance. Consequently, the 
calculations point to the existence of the convective zones in the models with $\dot{M}\ge 10^{-7}\,M_\odot\,\text{yr}^{-1}$ that may
affect the disk behavior \citep[see][]{1991MNRAS.250..432L};
namely they may alter the temperature and therefore the density profile and may provoke the enhancement of instabilities.

The models of the disks with the lower disk mass-loss rates, which
may correspond to classical Be star disks, show
less pronounced viscous-heated disk midplane strips which disappear for
$\dot{M}< 10^{-10}\,M_\odot\,\text{yr}^{-1}$. The higher values of the $\alpha$ parameter of viscosity and/or high mass-loss rates lead to unstable disk behavior,
producing waves or bumps in the inner disk region that consistently occur in the
profiles of density, pressure, radial velocity, optical depth, and temperature. 

The fundamental computational problem is the enormous
ratio of the disk vertical scale height to the radial extent of the disk which represents several orders of magnitude within a 
distance of a few stellar radii from the central star.
To overcome this problem, we developed a specific non-orthogonal ``flaring disk''
grid (see Sect.~\ref{numapproach} and Appendix~\ref{flarecoords} for the detailed mathematical description). 
However, we are still not able to model
the self-consistent 2D global disk density and thermal structure 
significantly above the sonic point radius.

The models of disks of Pop III stars where we assume the enormous disk mass-loss rates \citep[e.g.,][]{2001A&A...371..152M} 
may contribute to a better understanding of the rate 
of supply of interstellar space with gas as well as with radiation generated in the disks due to viscosity heating.
The nature and structure of disks around sgB[e] stars (at least some of them) might be 
explained by the viscous disk model. The spectroscopic analyses reveal the Keplerian rotation
velocity of the observed disk or rings. Many computations are still needed, however, to
explain the origin and evolution of the suggested ring structure where no radial flow has been
detected so far \citep{2013A&A...549A..28K}. Is a structure of the concentric rings that may extend up to
hundreds of stellar radii a result of stellar pulsations or sub-critical rotation with high $\alpha$ viscosity parameter?
Taking into account the high disk mass-loss rates and therefore the high densities and optical depths
in the inner disks, there appears to be (in the case of viscous disk models) a substantial contribution of viscous heating in these regions. However, 
the high densities produce an intensive 
radiative shielding in the inner disk, providing thus the possibility for a neutral hydrogen occurrence
in a narrow strip near the transition from an optically thin to optically thick zone.
We prove the existence of convective zones in the disks with
$\dot{M}\geq10^{-7}\,M_\odot\,\text{yr}^{-1}$. Such zones can be relatively
extended; for example, they appear in the radial range
($2$\,-\,$14$)\,$R_\text{eq}$ in the disk with
$\dot{M}=10^{-6}\,M_\odot\,\text{yr}^{-1}$.

\begin{acknowledgements}
Access to computing and storage facilities owned by parties and projects contributing to the 
National Grid Infrastructure MetaCentrum, provided under the programme
"Projects of Large Infrastructure for Research, Development, and Innovations" (LM2010005) is appreciated.
This work was supported by the grant GA \v{C}R 16-01116S.
\end{acknowledgements}

\bibliographystyle{aa} 
\bibliography{bibliography} 

\begin{thebibliography}{78}
\expandafter\ifx\csname natexlab\endcsname\relax\def\natexlab#1{#1}\fi

\bibitem[{{Arfken} \& {Weber}(2005)}]{2005mmp..book.....A}
{Arfken}, G.~B. \& {Weber}, H.~J. 2005, {Mathematical methods for physicists
  6th ed.}

\bibitem[{{Balbus}(2003)}]{2003ARA&A..41..555B}
{Balbus}, S.~A. 2003, \araa, 41, 555

\bibitem[{{Balbus}(2011)}]{2011Natur.470..475B}
{Balbus}, S.~A. 2011, \nat, 470, 475

\bibitem[{{Caramana} {et~al.}(1998){Caramana}, {Shashkov}, \&
  {Whalen}}]{1998JCoPh.144...70C}
{Caramana}, E.~J., {Shashkov}, M.~J., \& {Whalen}, P.~P. 1998, Journal of
  Computational Physics, 144, 70

\bibitem[{{Carciofi}(2011)}]{2011IAUS..272..325C}
{Carciofi}, A.~C. 2011, in IAU Symposium, Vol. 272, Active OB Stars: Structure,
  Evolution, Mass Loss, and Critical Limits, ed. C.~{Neiner}, G.~{Wade},
  G.~{Meynet}, \& G.~{Peters}, 325--336

\bibitem[{{Carciofi} \& {Bjorkman}(2008)}]{2008ApJ...684.1374C}
{Carciofi}, A.~C. \& {Bjorkman}, J.~E. 2008, \apj, 684, 1374

\bibitem[{{Carciofi} {et~al.}(2012){Carciofi}, {Bjorkman}, {Otero}, {Okazaki},
  {{\v S}tefl}, {Rivinius}, {Baade}, \& {Haubois}}]{2012ApJ...744L..15C}
{Carciofi}, A.~C., {Bjorkman}, J.~E., {Otero}, S.~A., {et~al.} 2012, \apjl,
  744, L15

\bibitem[{{Carlsson} \& {Leenaarts}(2012)}]{2012A&A...539A..39C}
{Carlsson}, M. \& {Leenaarts}, J. 2012, \aap, 539, A39

\bibitem[{{Chung}(2002)}]{2002cfd..book.....C}
{Chung}, T.~J. 2002, {Computational Fluid Dynamics}, 1036

\bibitem[{{Davies} {et~al.}(2005){Davies}, {Oudmaijer}, \&
  {Vink}}]{2005A&A...439.1107D}
{Davies}, B., {Oudmaijer}, R.~D., \& {Vink}, J.~S. 2005, \aap, 439, 1107

\bibitem[{{Dullemond} \& {Penzlin}(2017)}]{2017arXiv170900246D}
{Dullemond}, C.~P. \& {Penzlin}, A.~B.~T. 2017, ArXiv e-prints
  [\eprint[arXiv]{1709.00246}]

\bibitem[{{Ekstr{\"o}m} {et~al.}(2008){Ekstr{\"o}m}, {Meynet}, {Maeder}, \&
  {Barblan}}]{2008A&A...478..467E}
{Ekstr{\"o}m}, S., {Meynet}, G., {Maeder}, A., \& {Barblan}, F. 2008, \aap,
  478, 467

\bibitem[{{Feldmeier}(1995)}]{1995A&A...299..523F}
{Feldmeier}, A. 1995, \aap, 299, 523

\bibitem[{{Frank} {et~al.}(2002){Frank}, {King}, \&
  {Raine}}]{2002apa..book.....F}
{Frank}, J., {King}, A., \& {Raine}, D.~J. 2002, {Accretion Power in
  Astrophysics: Third Edition}, 398

\bibitem[{{Granada} {et~al.}(2013){Granada}, {Ekstr{\"o}m}, {Georgy}, {Krti{\v
  c}ka}, {Owocki}, {Meynet}, \& {Maeder}}]{2013A&A...553A..25G}
{Granada}, A., {Ekstr{\"o}m}, S., {Georgy}, C., {et~al.} 2013, \aap, 553, A25

\bibitem[{{Guinan} \& {Hayes}(1984)}]{1984ApJ...287L..39G}
{Guinan}, E.~F. \& {Hayes}, D.~P. 1984, \apjl, 287, L39

\bibitem[{{Harmanec}(1988)}]{1988BAICz..39..329H}
{Harmanec}, P. 1988, Bulletin of the Astronomical Institutes of Czechoslovakia,
  39, 329

\bibitem[{{Heger} \& {Langer}(1998)}]{1998A&A...334..210H}
{Heger}, A. \& {Langer}, N. 1998, \aap, 334, 210

\bibitem[{{Heyrovsk{\'y}}(2007)}]{2007ApJ...656..483H}
{Heyrovsk{\'y}}, D. 2007, \apj, 656, 483

\bibitem[{{Hillier}(2006)}]{2006ASPC..355...39H}
{Hillier}, D.~J. 2006, in Astronomical Society of the Pacific Conference
  Series, Vol. 355, Stars with the B[e] Phenomenon, ed. M.~{Kraus} \& A.~S.
  {Miroshnichenko}, 39

\bibitem[{{Hirsch}(1989)}]{1989nyjw.book.....H}
{Hirsch}, C. 1989, {Numerical computation of internal and external flows. Vol.
  1 - Fundamentals of Numerical Discretization}

\bibitem[{{Kee} {et~al.}(2016){Kee}, {Owocki}, \&
  {Sundqvist}}]{2016MNRAS.458.2323K}
{Kee}, N.~D., {Owocki}, S., \& {Sundqvist}, J.~O. 2016, \mnras, 458, 2323

\bibitem[{{Klement} {et~al.}(2017){Klement}, {Carciofi}, {Rivinius},
  {Matthews}, {Vieira}, {Ignace}, {Bjorkman}, {Mota}, {Faes}, {Bratcher},
  {Cur{\'e}}, \& {{\v S}tefl}}]{2017A&A...601A..74K}
{Klement}, R., {Carciofi}, A.~C., {Rivinius}, T., {et~al.} 2017, \aap, 601, A74

\bibitem[{{Kraus} {et~al.}(2007){Kraus}, {Borges Fernandes}, \& {de
  Ara{\'u}jo}}]{2007A&A...463..627K}
{Kraus}, M., {Borges Fernandes}, M., \& {de Ara{\'u}jo}, F.~X. 2007, \aap, 463,
  627

\bibitem[{{Kraus} {et~al.}(2010){Kraus}, {Borges Fernandes}, \& {de
  Ara{\'u}jo}}]{2010A&A...517A..30K}
{Kraus}, M., {Borges Fernandes}, M., \& {de Ara{\'u}jo}, F.~X. 2010, \aap, 517,
  A30

\bibitem[{{Kraus} {et~al.}(2008){Kraus}, {Borges Fernandes}, {Kub{\'a}t}, \&
  {de Ara{\'u}jo}}]{2008A&A...487..697K}
{Kraus}, M., {Borges Fernandes}, M., {Kub{\'a}t}, J., \& {de Ara{\'u}jo}, F.~X.
  2008, \aap, 487, 697

\bibitem[{{Kraus} {et~al.}(2013){Kraus}, {Oksala}, {Nickeler}, {Muratore},
  {Borges Fernandes}, {Aret}, {Cidale}, \& {de Wit}}]{2013A&A...549A..28K}
{Kraus}, M., {Oksala}, M.~E., {Nickeler}, D.~H., {et~al.} 2013, \aap, 549, A28

\bibitem[{{Krti\v cka} {et~al.}(2015){Krti\v cka}, {Kurf\" urst}, \& {Krti\v
  ckov\' a}}]{2015A&A...573A..20K}
{Krti\v cka}, J., {Kurf\" urst}, P., \& {Krti\v ckov\' a}, I. 2015, \aap, 573,
  A20

\bibitem[{{Krti{\v c}ka} {et~al.}(2011){Krti{\v c}ka}, {Owocki}, \&
  {Meynet}}]{2011A&A...527A..84K}
{Krti{\v c}ka}, J., {Owocki}, S.~P., \& {Meynet}, G. 2011, \aap, 527, A84

\bibitem[{{Kurf\" urst}(2015)}]{kurfurst2015thesis}
{Kurf\" urst}, P. 2015, PhD thesis, Masaryk University, Brno, Czech Republic

\bibitem[{{Kurf{\"u}rst}(2012)}]{2012IAUS..282..257K}
{Kurf{\"u}rst}, P. 2012, in IAU Symposium, Vol. 282, From Interacting Binaries
  to Exoplanets: Essential Modeling Tools, ed. M.~T. {Richards} \& I.~{Hubeny},
  257--258

\bibitem[{{Kurf{\"u}rst} {et~al.}(2014){Kurf{\"u}rst}, {Feldmeier}, \& {Krti{\v
  c}ka}}]{2014A&A...569A..23K}
{Kurf{\"u}rst}, P., {Feldmeier}, A., \& {Krti{\v c}ka}, J. 2014, \aap, 569, A23

\bibitem[{{Kurf{\"u}rst} {et~al.}(2017){Kurf{\"u}rst}, {Feldmeier}, \& {Krti{\v
  c}ka}}]{2017ASPC..508...17K}
{Kurf{\"u}rst}, P., {Feldmeier}, A., \& {Krti{\v c}ka}, J. 2017, in
  Astronomical Society of the Pacific Conference Series, Vol. 508, The B[e]
  Phenomenon: Forty Years of Studies, ed. A.~{Miroshnichenko}, S.~{Zharikov},
  D.~{Kor{\v c}{\'a}kov{\'a}}, \& M.~{Wolf}, 17

\bibitem[{{Kurf\"urst} \& {Krti\v cka}(2012)}]{2012ASPC..464..223K}
{Kurf\"urst}, P. \& {Krti\v cka}, J. 2012, in Astronomical Society of the
  Pacific Conference Series, Vol. 464, Circumstellar Dynamics at High
  Resolution, ed. A.~C. {Carciofi} \& T.~{Rivinius}, 223

\bibitem[{{Landau} \& {Lifshitz}(1982)}]{9780750628969}
{Landau}, L.~D. \& {Lifshitz}, E.~M. 1982, Mechanics: Volume 1 (Course of
  Theoretical Physics S) (Butterworth-Heinemann)

\bibitem[{{Lee} {et~al.}(1991){Lee}, {Osaki}, \& {Saio}}]{1991MNRAS.250..432L}
{Lee}, U., {Osaki}, Y., \& {Saio}, H. 1991, \mnras, 250, 432

\bibitem[{{Leenaarts} {et~al.}(2011){Leenaarts}, {Carlsson}, {Hansteen}, \&
  {Gudiksen}}]{2011A&A...530A.124L}
{Leenaarts}, J., {Carlsson}, M., {Hansteen}, V., \& {Gudiksen}, B.~V. 2011,
  \aap, 530, A124

\bibitem[{{LeVeque} {et~al.}(1998){LeVeque}, {Mihalas}, {Dorfi}, \&
  {M{\"u}ller}}]{LeVeque1998Comput}
{LeVeque}, R.~J., {Mihalas}, D., {Dorfi}, E.~A., \& {M{\"u}ller}, E. 1998,
  {Computational Methods for Astrophysical Fluid Flow}

\bibitem[{{Maeder}(1999)}]{1999A&A...347..185M}
{Maeder}, A. 1999, \aap, 347, 185

\bibitem[{{Maeder}(2009)}]{2009pfer.book.....M}
{Maeder}, A. 2009, {Physics, Formation and Evolution of Rotating Stars}

\bibitem[{{Marigo} {et~al.}(2001){Marigo}, {Girardi}, {Chiosi}, \&
  {Wood}}]{2001A&A...371..152M}
{Marigo}, P., {Girardi}, L., {Chiosi}, C., \& {Wood}, P.~R. 2001, \aap, 371,
  152

\bibitem[{{McGill} {et~al.}(2011){McGill}, {Sigut}, \&
  {Jones}}]{2011ApJ...743..111M}
{McGill}, M.~A., {Sigut}, T.~A.~A., \& {Jones}, C.~E. 2011, \apj, 743, 111

\bibitem[{{Meynet} {et~al.}(2006){Meynet}, {Ekstr{\"o}m}, \&
  {Maeder}}]{2006A&A...447..623M}
{Meynet}, G., {Ekstr{\"o}m}, S., \& {Maeder}, A. 2006, \aap, 447, 623

\bibitem[{{Mihalas}(1978)}]{1978stat.book.....M}
{Mihalas}, D. 1978, {Stellar atmospheres /2nd edition/}

\bibitem[{{Mihalas} \& {Mihalas}(1984)}]{1984oup..book.....M}
{Mihalas}, D. \& {Mihalas}, B.~W. 1984, {Foundations of radiation
  hydrodynamics}

\bibitem[{{Norman} \& {Winkler}(1986)}]{1986ASIC..188..187N}
{Norman}, M.~L. \& {Winkler}, K.-H.~A. 1986, in NATO Advanced Science
  Institutes (ASI) Series C, Vol. 188, NATO Advanced Science Institutes (ASI)
  Series C, ed. K.-H.~A. {Winkler} \& M.~L. {Norman}, 187

\bibitem[{{Okazaki}(1991)}]{1991PASJ...43...75O}
{Okazaki}, A.~T. 1991, \pasj, 43, 75

\bibitem[{{Okazaki}(2001)}]{2001PASJ...53..119O}
{Okazaki}, A.~T. 2001, \pasj, 53, 119

\bibitem[{{Okazaki}(2007)}]{2007ASPC..361..230O}
{Okazaki}, A.~T. 2007, in Astronomical Society of the Pacific Conference
  Series, Vol. 361, Active OB-Stars: Laboratories for Stellare and
  Circumstellar Physics, ed. A.~T. {Okazaki}, S.~P. {Owocki}, \& S.~{Stefl},
  230

\bibitem[{{Penna} {et~al.}(2013){Penna}, {S{\c a}dowski}, {Kulkarni}, \&
  {Narayan}}]{2013MNRAS.428.2255P}
{Penna}, R.~F., {S{\c a}dowski}, A., {Kulkarni}, A.~K., \& {Narayan}, R. 2013,
  \mnras, 428, 2255

\bibitem[{{Pinilla} {et~al.}(2012){Pinilla}, {Birnstiel}, {Ricci}, {Dullemond},
  {Uribe}, {Testi}, \& {Natta}}]{2012A&A...538A.114P}
{Pinilla}, P., {Birnstiel}, T., {Ricci}, L., {et~al.} 2012, \aap, 538, A114

\bibitem[{{Pringle}(1981)}]{1981ARA&A..19..137P}
{Pringle}, J.~E. 1981, \araa, 19, 137

\bibitem[{{Rivinius} {et~al.}(2013){Rivinius}, {Carciofi}, \&
  {Martayan}}]{2013A&ARv..21...69R}
{Rivinius}, T., {Carciofi}, A.~C., \& {Martayan}, C. 2013, \aapr, 21, 69

\bibitem[{{Roe}(1981)}]{1981JCoPh..43..357R}
{Roe}, P.~L. 1981, Journal of Computational Physics, 43, 357

\bibitem[{{Rosner} {et~al.}(1978){Rosner}, {Tucker}, \&
  {Vaiana}}]{1978ApJ...220..643R}
{Rosner}, R., {Tucker}, W.~H., \& {Vaiana}, G.~S. 1978, \apj, 220, 643

\bibitem[{{Schulte-Ladbeck} {et~al.}(1994){Schulte-Ladbeck}, {Clayton},
  {Hillier}, {Harries}, \& {Howarth}}]{1994ApJ...429..846S}
{Schulte-Ladbeck}, R.~E., {Clayton}, G.~C., {Hillier}, D.~J., {Harries}, T.~J.,
  \& {Howarth}, I.~D. 1994, \apj, 429, 846

\bibitem[{{Schulz}(1964)}]{zbMATH03332291}
{Schulz}, W.~D. 1964, J. Math. Phys., 5, 133

\bibitem[{{Schwarzschild}(1958)}]{1958ses..book.....S}
{Schwarzschild}, M. 1958, {Structure and evolution of the stars.}

\bibitem[{{Shakura} \& {Sunyaev}(1973)}]{1973A&A....24..337S}
{Shakura}, N.~I. \& {Sunyaev}, R.~A. 1973, \aap, 24, 337

\bibitem[{{Shu}(1992)}]{1992pavi.book.....S}
{Shu}, F.~H. 1992, {The physics of astrophysics. Volume II: Gas dynamics.}

\bibitem[{{Sigut} \& {Jones}(2007)}]{2007ApJ...668..481S}
{Sigut}, T.~A.~A. \& {Jones}, C.~E. 2007, \apj, 668, 481

\bibitem[{{Sigut} {et~al.}(2009){Sigut}, {McGill}, \&
  {Jones}}]{2009ApJ...699.1973S}
{Sigut}, T.~A.~A., {McGill}, M.~A., \& {Jones}, C.~E. 2009, \apj, 699, 1973

\bibitem[{{Skinner} \& {Ostriker}(2010)}]{2010ApJS..188..290S}
{Skinner}, M.~A. \& {Ostriker}, E.~C. 2010, \apjs, 188, 290

\bibitem[{{Smak}(1989)}]{1989AcA....39..201S}
{Smak}, J. 1989, \actaa, 39, 201

\bibitem[{{\v Stefl} {et~al.}(2003){\v Stefl}, {Baade}, {Rivinius}, {Otero},
  {Stahl}, {Budovi{\v c}ov{\'a}}, {Kaufer}, \& {Maintz}}]{2003A&A...402..253S}
{\v Stefl}, S., {Baade}, D., {Rivinius}, T., {et~al.} 2003, \aap, 402, 253

\bibitem[{{Stone} {et~al.}(2008){Stone}, {Gardiner}, {Teuben}, {Hawley}, \&
  {Simon}}]{2008ApJS..178..137S}
{Stone}, J.~M., {Gardiner}, T.~A., {Teuben}, P., {Hawley}, J.~F., \& {Simon},
  J.~B. 2008, \apjs, 178, 137

\bibitem[{{Stone} \& {Norman}(1992)}]{1992ApJS...80..753S}
{Stone}, J.~M. \& {Norman}, M.~L. 1992, \apjs, 80, 753

\bibitem[{{Th{\' e}} {et~al.}(1994){Th{\' e}}, {de Winter}, \& {P{\'
  e}rez}}]{1994A&AS..104..315T}
{Th{\' e}}, P.~S., {de Winter}, D., \& {P{\' e}rez}, M.~R. 1994, \aaps, 104

\bibitem[{{Toro}(1999)}]{Toro}
{Toro}, E.~F. 1999, {Riemann Solvers and Numerical Methods for Fluid Dynamics:
  A Practical Introduction}, 2nd edn.

\bibitem[{{Townsend} {et~al.}(2004){Townsend}, {Owocki}, \&
  {Howarth}}]{2004MNRAS.350..189T}
{Townsend}, R.~H.~D., {Owocki}, S.~P., \& {Howarth}, I.~D. 2004, \mnras, 350,
  189

\bibitem[{{van Hamme}(1993)}]{1993AJ....106.2096V}
{van Hamme}, W. 1993, \aj, 106, 2096

\bibitem[{{von Neumann} \& {Richtmyer}(1950)}]{1950JAP....21..232V}
{von Neumann}, J. \& {Richtmyer}, R.~D. 1950, Journal of Applied Physics, 21,
  232

\bibitem[{{von Zeipel}(1924)}]{1924MNRAS..84..665V}
{von Zeipel}, H. 1924, \mnras, 84, 665

\bibitem[{{Wade} \& {Rucinski}(1985)}]{1985A&AS...60..471W}
{Wade}, R.~A. \& {Rucinski}, S.~M. 1985, \aaps, 60, 471

\bibitem[{{W{\"u}nsch} {et~al.}(2005){W{\"u}nsch}, {Klahr}, \&
  {R{\'o}{\.z}yczka}}]{2005MNRAS.362..361W}
{W{\"u}nsch}, R., {Klahr}, H., \& {R{\'o}{\.z}yczka}, M. 2005, \mnras, 362, 361

\bibitem[{{Zel'dovich} \& {Raizer}(1967)}]{1967pswh.book.....Z}
{Zel'dovich}, {\relax Ya}.~B. \& {Raizer}, {\relax Yu}.~P. 1967, {Physics of
  shock waves and high-temperature hydrodynamic phenomena}

\bibitem[{{Zickgraf}(2000)}]{2000ASPC..214...26Z}
{Zickgraf}, F. 2000, in Astronomical Society of the Pacific Conference Series,
  Vol. 214, IAU Colloq. 175: The Be Phenomenon in Early-Type Stars, ed. M.~A.
  {Smith}, H.~F. {Henrichs}, \& J.~{Fabregat}, 26

\bibitem[{{Zickgraf}(1998)}]{1998ASSL..233....1Z}
{Zickgraf}, F.-J. 1998, in Astrophysics and Space Science Library, Vol. 233,
  B[e] stars, ed. A.~M. {Hubert} \& C.~{Jaschek}, 1

\end{thebibliography}

\onecolumn
\begin{appendix}
\section{Flaring disk coordinate system}\label{flarecoords}

\subsection{Basic transformation equations}\label{baseflareeqs}

\noindent We introduce the unique coordinate system that fits the geometry of circumstellar disks, taking into account the axisymmetricity and the vertical 
hydrostatic equilibrium as well as the ``\textit{flaring disk}'' geometry. The inclusion of the flaring angle causes major 
difficulties of numerical schemes based on standard cylindrical coordinates.
To avoid the geometrical discrepancy, we developed a new computational grid, and we call it the flaring disk coordinate system. 

\begin{figure} [t]
\centering\resizebox{0.54\hsize}{!}{\includegraphics{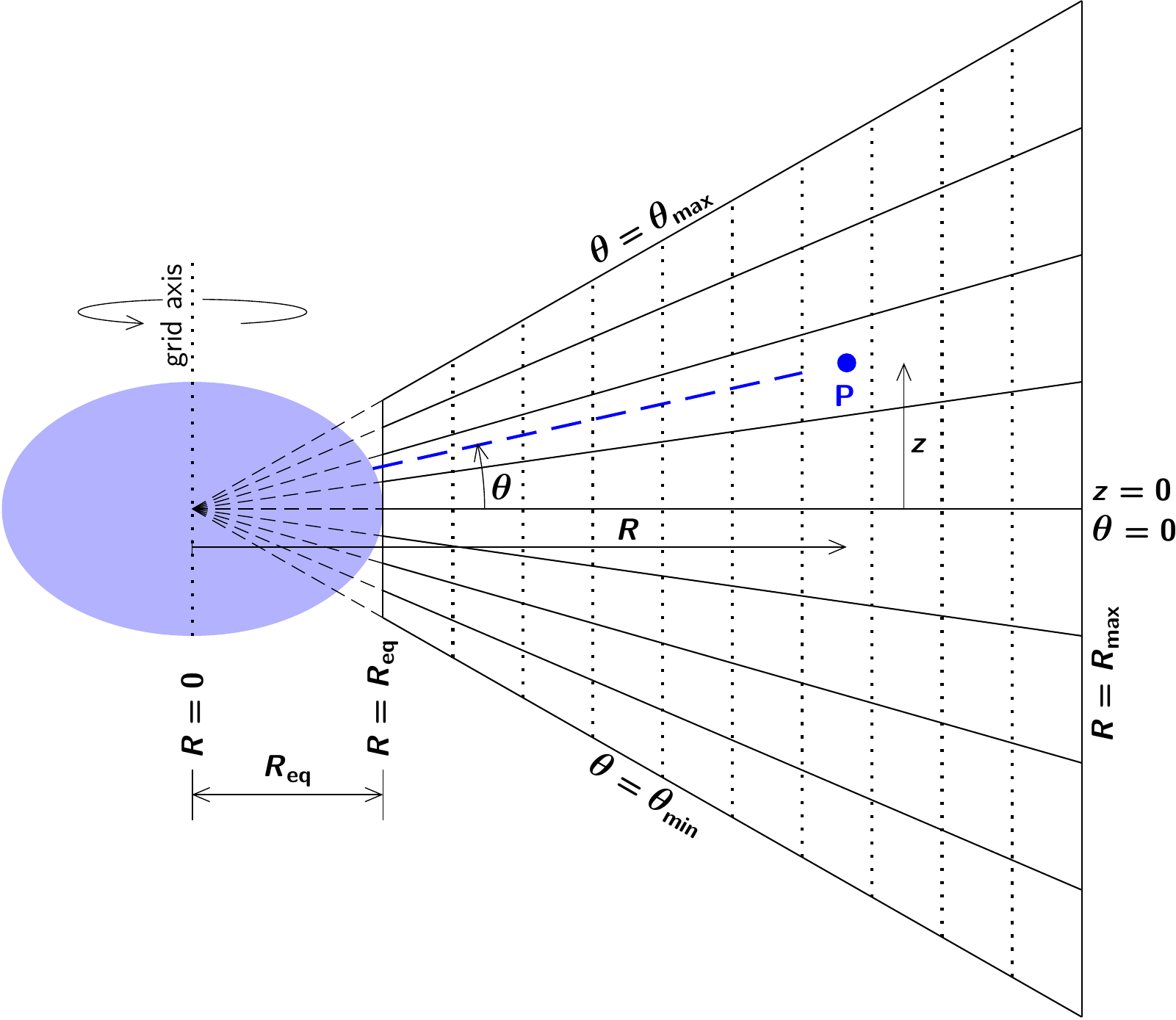}}
\caption{Schema of the flaring disk (wedge-cylindrical) coordinate system in radial-vertical ($R$-$\theta$) plane. Coordinates 
of arbitrary point $P$ are $R,\phi,\theta$. Intersection of coordinate surface $\theta=\arctan\left(z/R\right)$ with the $R$-$\theta$ plane
is highlighted by the blue dashed line. The input parameters of the numerical coordinate system (see the description in Appendix~\ref{flarecoords}) are given by the coordinates
$R_\text{eq},\,R_{\max},\,\theta_{\min}\text{ and }\theta_{\max}$.}
\label{fig3}
\end{figure}

Figure \ref{fig3} illustrates the system in radial-vertical ($R$-$\theta$) plane. The radial and azimuthal coordinates are identical with
standard cylindrical coordinate system, and
we thus denote them $R$ and $\phi$. The third coordinate $\theta$ is defined as the spherical angle
calculated in positive and negative direction from the equatorial plane.
Transformation equations from the flaring into Cartesian coordinates are 
\begin{align}\label{flare1}
x=R\cos\phi,\quad y=R\sin\phi,\quad z=R\tan\theta.
\end{align} 
Transformation equations of the unit basis vectors 
(cf.~the notation and basic principles introduced in \citet{2005mmp..book.....A}) 
from Cartesian to the flaring disk coordinates are 
\begin{align}\label{flare3}
\hat{\vec{R}}=\hat{\vec{x}}\cos\phi+\hat{\vec{y}}\sin\phi,\quad
\hat{\vec{\phi}}=-\hat{\vec{x}}\sin\phi+\hat{\vec{y}}\cos\phi,\quad
\hat{\vec{\theta}}=-(\hat{\vec{x}}\cos\phi+\hat{\vec{y}}\sin\phi)\sin\theta+\hat{\vec{z}}\cos\theta.
\end{align}
The inverse transformation of the unit basis vectors is
\begin{align}\label{flare4} 
\hat{\vec{x}}=\hat{\vec{R}}\cos\phi-\hat{\vec{\phi}}\sin\phi,\quad
\hat{\vec{y}}=\hat{\vec{R}}\sin\phi+\hat{\vec{\phi}}\cos\phi,\quad
\hat{\vec{z}}=\frac{\hat{\vec{R}}\sin\theta+\hat{\vec{\theta}}}{\cos\theta}.
\end{align}
In the flaring disk coordinate system all the basis vectors are non-constant, and their directions vary in dependence on azimuthal and flaring angle. 
Angular and time derivatives of the unit basis vectors, covariant metric tensor $g_{ij}$ for coordinates $R,\phi,\theta$, respectively, 
and the transformation Jacobian $J$ from the Cartesian to the flaring disk coordinate system, are
\begin{equation}
\begin{aligned}
\frac{\partial\hat{\vec{R}}}{\partial{\phi}} & = \hat{\vec{\phi}},\\
\frac{\partial\hat{\vec{R}}}{\partial{\theta}} & =0,\\
\frac{\partial\hat{\vec{R}}}{\partial t} & =\hat{\vec{\phi}}\dot{\phi}, 
\end{aligned}\quad
\begin{aligned}
\frac{\partial\hat{\vec\phi}}{\partial\phi} & =-\hat{\vec{R}},\\
\frac{\partial\hat{\vec{\phi}}}{\partial{\theta}} & =0,\\
\frac{\partial\hat{\vec{\phi}}}{\partial t} & =-\hat{\vec{R}}\dot{\phi},        
\end{aligned}\quad
\begin{aligned}
\frac{\partial\hat{\vec{\theta}}}{\partial\phi} & =-\hat{\vec{\phi}}\sin\theta,\\
\frac{\partial\hat{\vec{\theta}}}{\partial\theta} & =-\frac{\hat{\vec{R}}+\hat{\vec{\theta}}\sin\theta}{\cos\theta},\\
\frac{\partial\hat{\vec{\theta}}}{\partial t} & =-\frac{\hat{\vec{R}}\dot{\theta}}{\cos\theta}
-\hat{\vec{\phi}}\dot{\phi}\sin\theta-\hat{\vec{\theta}}\dot{\theta}\tan\theta,
\end{aligned}\quad
\begin{aligned}
g_{ij}=
\begin{pmatrix}\dfrac{1}{\cos^2\theta} & 0 & \dfrac{R\sin\theta}{\cos^3\theta}\\[8pt] 0 & R^2 & 0 \\[2pt]
\dfrac{R\sin\theta}{\cos^3\theta} & 0 & \dfrac{R^2}{\cos^4\theta}
\end{pmatrix},\quad
J=\frac{R^2}{\cos^2\theta}.
\end{aligned}
\end{equation}

\subsection{Basic differential operators}\label{flarediiffs}
Gradient of a scalar function $f=f(R,\phi,\theta)$ is
\begin{align}\label{flare12}
\vec{\nabla}f=\hat{\vec{R}}\frac{\partial f}{\partial R}+
\hat{\vec{\phi}}\frac{1}{R}\frac{\partial f}{\partial\phi}+
\hat{\vec{\theta}}\frac{\cos\theta}{R}\frac{\partial f}{\partial\theta}.
\end{align}
The gradient of an arbitrary vector field $\vec{A}(R,\phi,\theta)$ for coordinates $R,\phi,\theta$, respectively, is the matrix $\vec{\nabla}\vec{A}$ whose elements are
\begin{gather}\label{flare13}
(\vec{\nabla}\vec{A})_{11}=\frac{\partial A_R}{\partial R}\hat{\vec{R}}\hat{\vec{R}},\quad(\vec{\nabla}\vec{A})_{12}=
\frac{\partial A_\phi}{\partial R}\hat{\vec{R}}\hat{\vec{\phi}},\quad
(\vec{\nabla}\vec{A})_{13}=\frac{\partial A_\theta}{\partial R}\hat{\vec{R}}\hat{\vec{\theta}},\nonumber\\
(\vec{\nabla}\vec{A})_{21}=\left(\frac{1}{R}\frac{\partial A_R}{\partial\phi}-\frac{A_{\phi}}{R}\right)\hat{\vec{\phi}}\hat{\vec{R}},\quad
(\vec{\nabla}\vec{A})_{22}=\left(\frac{1}{R}\frac{\partial A_\phi}{\partial\phi}+\frac{A_R}{R}-\frac{A_\theta\sin\theta}{R}\right)\hat{\vec{\phi}}\hat{\vec{\phi}},\quad
(\vec{\nabla}\vec{A})_{23}=\frac{1}{R}\frac{\partial A_\theta}{\partial\phi}\hat{\vec{\phi}}\hat{\vec{\theta}},\\
(\vec{\nabla}\vec{A})_{31}=\left(\frac{\cos\theta}{R}\frac{\partial A_R}{\partial\theta}-\frac{A_\theta}{R}\right)\hat{\vec{\theta}}\hat{\vec{R}},\quad
(\vec{\nabla}\vec{A})_{32}=\frac{\cos\theta}{R}\frac{\partial A_\phi}{\partial\theta}\hat{\vec{\theta}}\hat{\vec{\phi}},\quad
(\vec{\nabla}\vec{A})_{33}=\left(\frac{\cos\theta}{R}\frac{\partial A_\theta}{\partial\theta}-\frac{A_\theta\sin\theta}{R}\right)\hat{\vec{\theta}}\hat{\vec{\theta}}.\nonumber
\end{gather}
Divergence of an arbitrary vector field $\vec{A}(R,\phi,\theta)$, defined as a dot product 
of the gradient vector (Eq.~\eqref{flare12}) and an arbitrary vector 
$A_R\hat{\vec{R}}+A_{\phi}\hat{\vec{\phi}}+A_\theta\hat{\vec{\theta}}$ (with nonzero dot product of different basis vectors
$\hat{\vec{R}}\cdot\hat{\vec{\theta}}=-\sin\theta$), is
\begin{align}\label{flare14b}
\vec{\nabla}\cdot\vec{A}=
\frac{1}{R}\frac{\partial}{\partial R}\left(R A_R\right)+
\frac{1}{R}\frac{\partial A_{\phi}}{\partial\phi}+
\frac{\cos\theta}{R}\frac{\partial A_\theta}{\partial\theta}-
\frac{\sin\theta}{R}\left[
\frac{\partial}{\partial R}\left(R A_\theta\right)+\cos\theta\frac{\partial A_R}{\partial\theta}\right].
\end{align}
Curl of a vector $\vec{A}(R,\phi,\theta)$ is defined as a cross product of 
gradient vector (Eq.~\eqref{flare12}) and an arbitrary vector
$A_R\hat{\vec{R}}+A_{\phi}\hat{\vec{\phi}}+A_\theta\hat{\vec{\theta}}$ 
where the cross products of different basis vectors are
\begin{align}\label{flare17} 
\hat{\vec{R}}\times\hat{\vec{\phi}}=
\frac{\hat{\vec{R}}\sin\theta+\hat{\vec{\theta}}}{\cos\theta},\quad
\hat{\vec{\phi}}\times\hat{\vec{\theta}}=
\frac{\hat{\vec{R}}+\hat{\vec{\theta}}\sin\theta}{\cos\theta},\quad
\hat{\vec{\theta}}\times\hat{\vec{R}}=
\hat{\vec{\phi}}\cos\theta.
\end{align}
Curl of a vector in the flaring disk system takes the form 
\begin{align}\label{flare18}
\vec\nabla\times\vec{A}=\hat{\vec{R}}
\left\{\frac{\tan\theta}{R}\left[\frac{\partial}{\partial R}\left(R A_\phi\right)-\frac{\partial A_R}{\partial\phi}\right]+
\frac{1}{R}\left(\frac{1}{\cos\theta}\frac{\partial A_\theta}{\partial\phi}-\frac{\partial A_\phi}{\partial\theta}\right)\right\}+
\hat{\vec{\phi}}
\left\{\frac{\cos\theta}{R}\left[\cos\theta\frac{\partial A_R}{\partial\theta}-
\frac{\partial}{\partial R}\left(R A_\theta\right)\right]\right\}+\nonumber\\
+\,\hat{\vec{\theta}}
\left\{\frac{1}{R\cos\theta}\left[\frac{\partial}{\partial R}\left(R A_\phi\right)-\frac{\partial A_R}{\partial\phi}\right]+
\frac{\sin\theta}{R}\left(\frac{1}{\cos\theta}\frac{\partial A_\theta}{\partial\phi}-\frac{\partial A_\phi}{\partial\theta}\right)\right\}.
\end{align}
The Laplacian operator in the flaring disk system becomes
\begin{align}\label{flare19} 
\Delta=\frac{1}{R}\frac{\partial}{\partial R}\left(R\frac{\partial}{\partial R}\right)+
\frac{1}{R^2}\frac{\partial^2}{\partial\phi^2}+
\frac{\cos\theta}{R^2}\frac{\partial}{\partial\theta}\left(\cos\theta\frac{\partial}{\partial\theta}\right)-
\frac{2\sin\theta\cos\theta}{R}\frac{\partial^2}{\partial R\,\partial\theta}.
\end{align}

\subsection{Volumes and surfaces}\label{flarevolsurfs}

\noindent Volumes and surfaces of computational grid cells are used in the method of finite volume (see Sect.~\ref{numapproach}). 
The volume bounded by coordinate surfaces (the surfaces with constant coordinates $R_2,R_1,\phi_1,\phi_2,\theta_1,\theta_2$) is 
\begin{align}\label{volsurf1}
V=\frac{R_2^3-R_1^3}{3}\Big(\phi_2-\phi_1\Big)
\Big(|\tan\theta_2|-|\tan\theta_1|\Big).
\end{align} 
The areas of corresponding grid cell surfaces (where the subscripts refer to the constant surface coordinate) are 
\begin{gather}\label{volsurf2}
S_{\!R}=
R^2\,\left(\phi_2-\phi_1\right)\left(|\tan\theta_2|-|\tan\theta_1|\right),\quad
S_{\!\phi}=
\frac{R_2^2-R_1^2}{2}\left(|\tan\theta_2|-|\tan\theta_1|\right),\quad
S_{\!\theta}=
\frac{R_2^2-R_1^2}{2}
\frac{(\phi_2-\phi_1)}{\cos\theta}.
\end{gather}
\subsection{Velocity and acceleration}\label{flareaccel}
The vectors of position $\vec{r}$ and velocity $\vec{V}=\d\vec{r}/\d t=V_R\hat{\vec{R}}+V_\phi\hat{\vec{\phi}}+V_\theta\hat{\vec{\theta}}$ 
in the flaring disk coordinates take the form
\begin{align}\label{velaccel1}
\vec{r}=x\hat{\vec{x}}+y\hat{\vec{y}}+z\hat{\vec{z}}=
\frac{\hat{\vec{R}}R+\hat{\vec{\theta}}R\sin\theta}{\cos^2\theta},\quad
\vec{V}=
\hat{\vec{R}}\left(\frac{\dot{R}+R\dot{\theta}\,\tan\theta}{\cos^2\theta}\right)+
\hat{\vec{\phi}}R\dot{\phi}+
\hat{\vec{\theta}}\left(\frac{\dot{R}\tan\theta}{\cos\theta}+\frac{R\dot{\theta}}{\cos^3\theta}\right).
\end{align}
We obtain
the components of the vector of acceleration $\vec{a}=\d\vec{V}/\d t=a_R\hat{\vec{R}}+a_\phi\hat{\vec{\phi}}+
a_\theta\hat{\vec{\theta}}$
by differentiation of the velocity in Eq.~\eqref{velaccel1},
\begin{align}\label{velaccel2}
a_R=\frac{\ddot{R}+R\ddot{\theta}\tan\theta+2\dot{\theta}\,(\dot{R}+R\dot{\theta}\tan\theta)\tan\theta}{\cos^2\theta}-R\dot{\phi}^2,\quad
a_\phi=R\ddot{\phi}+2\dot{R}\dot{\phi},\quad
a_\theta=\frac{1}{\cos\theta}\left[\ddot{R}\tan\theta+\frac{R\ddot{\theta}+2\dot{\theta}\,(\dot{R}+R\dot{\theta}\,\tan\theta)}{\cos^2\theta}\right].
\end{align}
From Eqs.~\eqref{velaccel1} and \eqref{velaccel2} we express the components of the velocity vector as
\begin{align}\label{velaccel3}
\dot{R}=V_R-V_\theta\sin\theta,\,\dot{\phi}=\frac{V_\phi}{R},\,\,\dot{\theta}=\frac{(V_\theta-V_R\sin\theta)\cos\theta}{R}. 
\end{align}
Following the identity $\d\vec{V}/\d t=\partial\vec{V}/\partial t+\vec{V}\cdot\vec{\nabla}\vec{V}$, 
we write the components of the vector of acceleration in terms of the velocity vector components,
\begin{gather}\label{velaccel4}
a_R=\frac{\partial V_R}{\partial t}+\underbrace{V_R\frac{\partial V_R}{\partial R}+
\frac{V_\phi}{R}\frac{\partial V_R}{\partial\phi}
+V_\theta\frac{\cos\theta}{R}\frac{\partial V_R}{\partial\theta}}_{\left(\vec{V}\cdot\vec{\nabla}\right)V_R}-
\frac{V_\phi^2+V_\theta^2}{R}+\frac{V_R V_\theta\sin\theta}{R},\\
\label{velaccel5}
a_\phi=\frac{\partial V_\phi}{\partial t}+\underbrace{V_R\frac{\partial V_\phi}{\partial R}+
\frac{V_\phi}{R}\frac{\partial V_\phi}{\partial\phi}
+V_\theta\frac{\cos\theta}{R}\frac{\partial V_\phi}{\partial\theta}}_{\left(\vec{V}\cdot\vec{\nabla}\right)V_\phi}+
\frac{V_R V_\phi}{R}-\frac{V_\phi V_\theta\sin\theta}{R},\\
\label{velaccel6}
a_\theta=\frac{\partial V_\theta}{\partial t}+\underbrace{V_R\frac{\partial V_\theta}{\partial R}+
\frac{V_\phi}{R}\frac{\partial V_\theta}{\partial\phi}
+V_\theta\frac{\cos\theta}{R}\frac{\partial V_\theta}{\partial\theta}}_{\left(\vec{V}\cdot\vec{\nabla}\right)V_\theta}-
\frac{V_\theta^2\sin\theta}{R}+\frac{V_RV_\theta\sin^2\theta}{R}.
\end{gather}
The underbraced terms on the right-hand sides of Eqs.~\eqref{velaccel4}-\eqref{velaccel6} express the nonlinear advection while the other terms 
represent 
the apparent centrifugal and Coriolis accelerations (``fictitious forces''). This perfectly corresponds to a
general theorem that in the classical rotating frame 
with constant angular velocity $\Omega$ 
(see, e.g.,~\citet{9780750628969}, p.~128),
the only apparent accelerations
are
the centrifugal and Coriolis 
(new apparent accelerations may occur,~e.g.,~in general relativity). 

From Eqs.~\eqref{flare1} and \eqref{velaccel3} we express the velocity vector components $V_R,V_\phi,V_\theta$ in terms of the velocity vector 
components in standard cylindrical coordinate system, $V_{R,\,\text{cyl}},V_{\phi,\,\text{cyl}},V_z$, obtaining
\begin{align}
V_R=V_{R,\,\text{cyl}}+V_z\,\tan\theta,\,\,V_\phi=V_{\phi,\,\text{cyl}},\,\,V_\theta=\frac{V_z}{\cos\theta}.
\end{align}
Taking into account the vertical hydrostatic balance, $V_z=0$, the hydrodynamic acceleration terms, Eqs.~\eqref{velaccel4} - \eqref{velaccel6}, 
become however identical to corresponding terms in standard cylindrical coordinates. Noting here that the flaring geometry of the disks together with
the vertical hydrostatic equilibrium assumption are the main reasons for introducing this coordinate system, this is an important 
simplification of its mathematical complexity. Some small difficulties connected with the nonorthogonal flaring disk coordinate system thus mainly remain
at the technical level (in the correct implementation of the grid).

\subsection{Calculation of the disk optical depth}
\label{optdepth}
\begin{figure} [t]
\centering\resizebox{0.33\hsize}{!}{\includegraphics{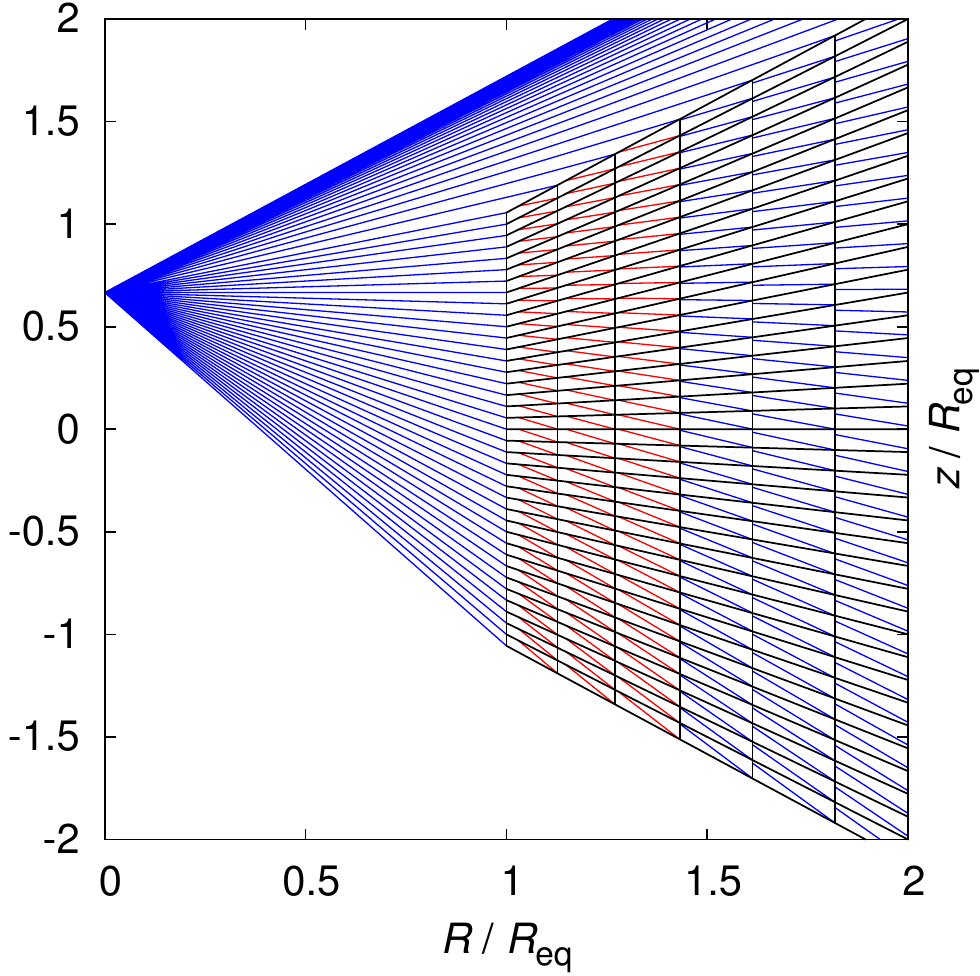}}
\caption{Schematic graph of the entirely inner part of the flaring disk grid in radial-vertical $(R$-$\theta)$ plane with marked ray-tracing for calculation of the 
disk optical depth. All the rays come from one point (from the ``north'' pole of a critically rotating star where $R_\text{eq}=1.5\,R_\star$).
The grid cells are depicted as black lines, the blue rays intersect the vertical cell interfaces while the red lines enter the grid cells through their upper flaring 
interface (or the lower flaring interface for the not plotted rays that emerge from the stellar ``south'' pole). See Sect.~\ref{optdepth} for the description.}
\label{shortrays}
\end{figure}
We trace the optical depth along the rays showed schematically in Fig.~\ref{shortrays} using the short characteristics method (we
plan a more advanced calculation with large number of radiation sources on the stellar
surface within the future work).
We denote $\delta$ the angle of a diagonal of a computational grid cell,
\begin{align}\label{short1}
(\tan\delta)_{i,k}=\frac{z_{i,k}-z_{i-1,k+1}}{R_i-R_{i-1}}.
\end{align}
Denoting the radial and vertical coordinate values of intersection of the ray with vertical or ``flaring disk'' grid cell interface, $R^\prime$ and $z^\prime$ , respectively, and denoting the vertical coordinate value of intersection of the ray with ``flaring disk'' grid cell interface,  
$z^{\prime\prime}$ , we
may obtain the following relations,
\begin{gather}\label{short2}
R^\prime_{i,k}=\frac{R_\star R_i}{R_\star+z_{i,k+1}-z_{i,k}},\quad z^\prime_{i,k}=z_{i,k}+\frac{(R_\star-z_{i,k})(R_{i}-R_{i-1})}{R_{i}},\quad
z^{\prime\prime}_{i,k}=\frac{R^\prime_{i,k}z_{i,k+1}}{R_{i}},
\end{gather}
where $R_\star$ is the stellar polar radius.
For the calculation of the optical depth (cf. Eq.~\eqref{temp8}) we have to distinguish two cases of the $\delta$ slope in respect to the source of rays,
\begin{align}\label{short3}
\delta_{i,k}\geq \frac{(z_{i,k}-R_\star)}{R_i}\quad\text{or}\quad\delta_{i,k}<\frac{(z_{i,k}-R_\star)}{R_i}. 
\end{align}
We use the piecewise linear interpolation of the quantity $\chi=\kappa\rho$. 
The first and second case in Eq.~\eqref{short3} respectively imply,  
\begin{align}\label{short4}
\Delta\tau_{i,k} = \left[\chi_{i,k}(R^\prime,z^{\prime\prime})+\chi_{i,k}(R,z)\right]\,\Delta s_{i,k}\,/2,\quad
\Delta\tau_{i,k} = \left[\chi_{i,k}(R_{i-1},z^{\prime})+\chi_{i,k}(R,z)\right]\,\Delta s_{i,k}\,/2, 
\end{align}
where $\Delta\tau$ is the increase of $\tau$ and $\Delta s$ is the length of the path of the ray within one computational cell.
The optical depth for each grid point $i,k$ is then calculated as $\tau_{i,k}=\tau^\prime_{i,k}+\Delta\tau_{i,k}$ where
$\tau^\prime_{i,k}$ is the linearly interpolated optical depth in points ($R^\prime_{i,k},z^{\prime\prime}_{i,k}$) or ($R_{i-1},z^\prime_{i,k}$), according to Eq.~\eqref{short3}.
The described principle takes into account only the optical depth tracing on one side of the disk 
(for the rays that emerge, e.g., from the ``north'' stellar pole according to Fig.~\ref{shortrays}), and
for the other side of the disk the principle applies analogously.

\subsection{Eigensystems of the flaring disk coordinates using the Roe's method: adiabatic hydrodynamics}
\label{roeadiaflare}
For adiabatic hydrodynamics in the \textit{flaring disk coordinate frame} with the conservative variables $\vec{U}=(\rho,\Pi_R,\Pi_\phi,\Pi_\theta,E) = (u_1,\ldots,u_5)$ 
we write Eq.~\eqref{numappr1} in 3D component form
\citep[see the formalism and terminology given in, e.g.,][]{Toro,2008ApJS..178..137S},
\begin{align}\label{adiaroe0}
\frac{\partial\vec{U}}{\partial t}+\xi,\xi^\prime\frac{\partial\vec{F}}{\partial R}+\frac{1}{R}\frac{\partial\vec{G}}{\partial\phi}
+\frac{\cos\theta}{R}\eta,\eta^\prime\frac{\partial\vec{H}}{\partial\theta}=\zeta,\zeta^\prime\vec{S},
\end{align}
where $\vec{F},\,\vec{G},\,\vec{H}$ are the vectors of fluxes in the $R,\,\phi,\,\theta$ directions, respectively, and 
$\vec{S}$ is the vector of geometric source terms. The factors $\xi=1,\xi^\prime=\sin\theta$ stay at particular $R,\theta$ components of the flux $\vec{F}$,
factors $\eta=1,\eta^\prime=\sin\theta$ stay at particular $\theta,R$ components of the flux $\vec{H}$,
factors $\zeta=1,\zeta^\prime=\sin\theta$ stay at particular $R,\theta$ components of the geometric source term $\vec{S}$. 
The explicit conservative form of the basic adiabatic hydrodynamic equations (equation of continuity, three components of momentum equation and energy equation, omitting the source terms) 
in the \textit{flaring disk coordinate frame} is 
\begin{gather}
\frac{\partial\rho}{\partial t}+\frac{\partial}{\partial R}\left(\rho V^\prime\right)+\frac{1}{R}\frac{\partial}{\partial\phi}\left(\rho V_\phi\right)
+\frac{\cos\theta}{R}\left[\frac{\partial}{\partial\theta}\left(\rho V_\theta\right)-\sin\theta\frac{\partial}{\partial\theta}\left(\rho V_R\right)\right]=
-\frac{\rho V^\prime}{R},
\label{adiaroe1}\\[1pt]
\frac{\partial\left(\rho V_R\right)}{\partial t}+\frac{\partial}{\partial R}\left[\rho\left(V_RV^\prime+H\right)-E\right]+
\frac{1}{R}\frac{\partial}{\partial\phi}\left(\rho V_RV_\phi\right)
+\frac{\cos\theta}{R}\left[\frac{\partial}{\partial\theta}\left(\rho V_RV_\theta\right)-\sin\theta\frac{\partial}{\partial\theta}\left(\rho V_R^2\right)\right]=
-\frac{\rho V_RV^\prime}{R},
\label{adiaroe2}\\[1pt]
\frac{\partial\left(\rho V_\phi\right)}{\partial t}+\frac{\partial}{\partial R}\left(\rho V_\phi V^\prime\right)+
\frac{1}{R}\frac{\partial}{\partial\phi}\left[\rho\left(V_\phi^2+H\right)-E\right]
+\frac{\cos\theta}{R}\left[\frac{\partial}{\partial\theta}\left(\rho V_\phi V_\theta\right)-\sin\theta\frac{\partial}{\partial\theta}\left(\rho V_RV_\phi\right)\right]=
-\frac{\rho V_\phi V^\prime}{R},
\label{adiaroe3}\\[1pt]
\frac{\partial\left(\rho V_\theta\right)}{\partial t}+\frac{\partial}{\partial R}\left(\rho V_\theta V^\prime\right)+
\frac{1}{R}\frac{\partial}{\partial\phi}\left(\rho V_\phi V_\theta\right)
+\frac{\cos\theta}{R}\left\{\frac{\partial}{\partial\theta}\left[\rho\left(V_\theta^2+H\right)-E\right]-
\sin\theta\frac{\partial}{\partial\theta}\left(\rho V_R V_\theta\right)\right\}
=-\frac{\rho V_\theta V^\prime}{R},
\label{adiaroe4}\\[1pt]
\frac{\partial E}{\partial t}+\frac{\partial}{\partial R}\left(\rho V^\prime H\right)+\frac{1}{R}\frac{\partial}{\partial\phi}\left(\rho V_\phi H\right)
+\frac{\cos\theta}{R}\left[\frac{\partial}{\partial\theta}\left(\rho V_\theta H\right)-\sin\theta\frac{\partial}{\partial\theta}\left(\rho V_R H\right)\right]=
-\frac{\rho V^\prime H}{R}.
\label{adiaroe5}
\end{gather}
We hereafter substitute $V_{R-\theta}=V_R-V_\theta$, $V_{\theta-R}=V_\theta-V_R$, $V^\prime=V_{R-\theta}\sin\theta$, $H$ is the enthalpy, $H=\left(E+P\right)/\rho$,
and the velocity vector $\vec{V}=(V_R,V_\phi,V_\theta)$.
The explicit components of $\vec{F}$, $\vec{G}$, $\vec{H}$, $\vec{S}$, are
\begin{gather}\label{adiaroe6}
\vec{F}= 
\begin{bmatrix}
\rho V_{R-\theta}\\[2pt]\rho V_RV_{R-\theta}+P\\[2pt]\rho V_\phi V_{R-\theta}\\[2pt]\rho V_\theta V_{R-\theta}\\[2pt](E+P)V_{R-\theta}
\end{bmatrix},\quad
\vec{G}=
\begin{bmatrix}
\rho V_\phi\\[2pt]\rho V_RV_\phi\\[2pt]\rho V_\phi^2+P\\[2pt]\rho V_\phi V_\theta\\[2pt](E+P)V_\phi
\end{bmatrix},\quad
\vec{H}= 
\begin{bmatrix}
\rho V_{\theta-R}\\[2pt]\rho V_RV_{\theta-R}\\[2pt]\rho V_\phi V_{\theta-R}\\[2pt]\rho V_\theta V_{\theta-R}+P\\[2pt](E+P)V_{\theta-R}
\end{bmatrix},\quad
\vec{S}=-\frac{1}{R}
\begin{bmatrix}
\rho V_{R-\theta}\\[2pt]\rho\vec{V}V_{R-\theta}\\[2pt](E+P)V_{R-\theta}
\end{bmatrix}.
\end{gather}
The vectors $\vec{F},\vec{G},\vec{H}$ written in terms of the five conservative variables $\vec{U} = (u_1,\ldots,u_5)$, 
corresponding to Eq.~\eqref{numappr1} in the flaring disk coordinates, are
\begin{align}\label{adiaroe7}
\vec{F}(\vec{u}),\vec{G}(\vec{u}),\vec{H}(\vec{u}) = 
\begin{pmatrix}
u_2-u_4^*,u_3,u_4-u_2^*\\[4pt]\dfrac{(u_2-u_4^*)u_2}{u_1}+(\gamma-1)\left[u_5-\dfrac{u_2^2+u_3^2+u_4^2}{2u_1}\right],\dfrac{u_3u_2}{u_1},\dfrac{(u_4-u_2^*)u_2}{u_1}\\[2pt]
\dfrac{(u_2-u_4^*)u_3}{u_1},\dfrac{u_3^2}{u_1}+(\gamma-1)\left[u_5-\dfrac{u_2^2+u_3^2+u_4^2}{2u_1}\right],\dfrac{(u_4-u_2^*)u_3}{u_1}\\[2pt]
\dfrac{(u_2-u_4^*)u_4}{u_1},\dfrac{u_3u_4}{u_1},\dfrac{(u_4-u_2^*)u_4}{u_1}+(\gamma-1)\left[u_5-\dfrac{u_2^2+u_3^2+u_4^2}{2u_1}\right]\\[2pt]
\left[\gamma u_5-(\gamma-1)\dfrac{u_2^2+u_3^2+u_4^2}{2u_1}\right]\left(\dfrac{u_2-u_4^*}{u_1},\dfrac{u_3}{u_1},\dfrac{u_4-u_2^*}{u_1}\right)
\end{pmatrix},
\end{align}
where the asterisk denotes the factor $\sin\theta$ staying in front of the derivative of the particular term.
Equation~\eqref{adiaroe7} demonstrates the proper swapping of the corresponding components of momentum \citep[cf., e.g.,][]{2008ApJS..178..137S,2010ApJS..188..290S}. 

The adiabatic conservative Jacobian matrices $\mathbf{A_1^\mathbf{C}}=\dfrac{\partial\vec{F}(\vec{u})}{\partial\vec{U}}$, 
$\mathbf{A_2^\mathbf{C}}=\dfrac{1}{R}\dfrac{\partial\vec{G}(\vec{u})}{\partial\vec{U}}$, and
$\mathbf{A_3^\mathbf{C}}=\dfrac{\mu}{R}\dfrac{\partial\vec{H}(\vec{u})}{\partial\vec{U}}$, 
where we hereafter denote $\mu=\cos\theta$, written explicitly in the directionally swapped form, are
\begin{align}\label{adiaroe8}
\mathbf{A_1^\mathbf{C}}= 
\begin{bmatrix}
0 & 1 & 0 &-\sin\theta & 0\\
-V_RV^\prime+(\gamma-1)\dfrac{V^2}{2} &-(\gamma-2)V_R+V^\prime &-(\gamma-1)V_\phi & V^{\prime\prime}-\gamma V_\theta & \gamma-1\\
-V_\phi V^\prime & V_\phi & V^\prime &-V_\phi\sin\theta & 0\\[2pt]
-V_\theta V^\prime & V_\theta & 0 & V^\prime-V_\theta\sin\theta & 0\\
-V^\prime H+(\gamma-1)\dfrac{V^\prime V^2}{2} &-(\gamma-1)V_RV^\prime+H &-(\gamma-1)V_\phi V^\prime &-(\gamma-1)V_\theta V^\prime-H\sin\theta & \gamma V^\prime
\end{bmatrix},
\end{align}
\begin{align}\label{adiaroe9}
\mathbf{A_2^\mathbf{C}}=\frac{1}{R}
\begin{bmatrix}
0 & 0 & 1 & 0 & 0\\[2pt]
-V_\phi V_R & V_\phi & V_R & 0 & 0\\
-V_\phi^2+(\gamma-1)\dfrac{V^2}{2} &-(\gamma-1)V_R &-(\gamma-3)V_\phi &-(\gamma-1)V_\theta & \gamma-1\\
-V_\phi V_\theta & 0& V_\theta & V_\phi & 0\\
-V_\phi H+(\gamma-1)\dfrac{V_\phi V^2}{2} &-(\gamma-1)V_RV_\phi &-(\gamma-1)V_\phi^2+H &-(\gamma-1)V_\phi V_\theta & \gamma V_\phi
\end{bmatrix},
\end{align}
\begin{align}\label{adiaroe10}
\mathbf{A_3^\mathbf{C}}=\frac{\mu}{R}
\begin{bmatrix}
0 &-\sin\theta & 0 & 1 & 0\\[2pt]
-V_RV^{\prime\prime} & V^{\prime\prime}-V_R\sin\theta & 0 & V_R & 0\\[2pt]
-V_\phi V^{\prime\prime} &-V_\phi\sin\theta & V^{\prime\prime} & V_\phi & 0\\
-V_\theta V^{\prime\prime}+(\gamma-1)\dfrac{V^2}{2} & V^\prime-\gamma V_R &-(\gamma-1)V_\phi &-(\gamma-2)V_\theta+V^{\prime\prime} & \gamma-1\\
-V^{\prime\prime}H+(\gamma-1)\dfrac{V^{\prime\prime}V^2}{2} &-(\gamma-1)V_RV^{\prime\prime}-H\sin\theta &-(\gamma-1)V_\phi V^{\prime\prime} &
-(\gamma-1)V_\theta V^{\prime\prime}+H & \gamma V^{\prime\prime}
\end{bmatrix},
\end{align}
where we hereafter substitute $V^{\prime\prime}=V_{\theta-R}\sin\theta$.
The corresponding eigenvalues $\lambda_1^\text{C},\lambda_2^\text{C},\lambda_3^\text{C}$, of the matrices $\mathbf{A_1^\mathbf{C}}$, $\mathbf{A_2^\mathbf{C}}$, $\mathbf{A_3^\mathbf{C}}$, are 
\begin{align}\label{adiaroe11}
\lambda_1^\text{C} = \left(V^\prime-a,V^\prime,V^\prime,V^\prime,V^\prime+a\right),\quad\lambda_2^\text{C} = \frac{1}{R}\left(V_\phi-a,V_\phi,V_\phi,V_\phi,V_\phi+a\right),\quad
\lambda_3^\text{C} = \frac{\mu}{R}\left(V^{\prime\prime}-a,V^{\prime\prime},V^{\prime\prime},V^{\prime\prime},V^{\prime\prime}+a\right),
\end{align}
where $a$ is the adiabatic speed of sound, $a^2=\gamma P/\rho$.
Corresponding right eigenvectors (following the directional swapping in Eqs.~\eqref{adiaroe8}-\eqref{adiaroe10}) are the columns of the matrices
\begin{align}\label{adiaroe12}
\mathbf{R_1^\mathbf{C}} = 
\begin{pmatrix}
1 & 0 & 0 & 1 & 1\\[2pt]
V_R-a & 0 & \sin\theta & V_R & V_R+a \\[2pt]
V_\phi & 1 & 0 & V_\phi & V_\phi\\[2pt]
V_\theta & 0 & 1 & V_\theta & V_\theta\\
H^\prime_{-} & V_\phi & \mu^2V_\theta & \dfrac{V^2}{2}&H^\prime_{+}
\end{pmatrix},\,\,
\mathbf{R_2^\mathbf{C}} = 
\begin{pmatrix}
1 & 0 & 1 & 0 & 1 \\[2pt]
V_R & 1 & V_R & 0 & V_R \\[2pt]
V_\phi-a & 0 & V_\phi & 0& V_\phi+a \\[2pt]
V_\theta & 0 & V_\theta & 1 & V_\theta \\
H-V_\phi a & V^\prime & \dfrac{V^2}{2} & V^{\prime\prime} & H+V_\phi a
\end{pmatrix},\,\,
\mathbf{R_3^\mathbf{C}} = 
\begin{pmatrix}
1 & 1 & 0 & 0 & 1 \\[2pt]
V_R & V_R & 1 & 0 & V_R \\[2pt]
V_\phi & V_\phi & 0 & 1 & V_\phi \\[2pt]
V_\theta-a & V_\theta & \sin\theta & 0 & V_\theta+a \\
H^{\prime\prime}_{-} & \dfrac{V^2}{2} & \mu^2V_R & V_\phi & H^{\prime\prime}_{+}
\end{pmatrix},
\end{align}
where $H^\prime_{-}=H-V^\prime a$, $H^\prime_{+}=H+V^\prime a$, $H^{\prime\prime}_{-}=H-V^{\prime\prime} a$ and $H^{\prime\prime}_{+}=H+V^{\prime\prime} a$.
The corresponding left eigenvectors are the rows of the matrices
\begin{align}\label{adiaroe13}
\mathbf{L_1^\mathbf{C}} = 
\begin{bmatrix}
\dfrac{(\gamma-1)V^2/2+V^\prime a}{2a^2} &-\dfrac{(\gamma-1)V^\prime+a}{2a^2} &-\dfrac{(\gamma-1)V_\phi}{2a^2} &-\dfrac{(\gamma-1)V^{\prime\prime}-a\sin\theta}{2a^2}&\dfrac{\gamma-1}{2a^2}\\[7pt]
-V_\phi & 0 & 1 & 0 & 0 \\[1pt]
-V_\theta & 0 & 0 & 1 & 0 \\
1-\dfrac{(\gamma-1)V^2}{2a^2} & \dfrac{(\gamma-1)V^\prime}{a^2} & \dfrac{(\gamma-1)V_\phi}{a^2} & \dfrac{(\gamma-1)V^{\prime\prime}}{a^2} & -\dfrac{\gamma-1}{a^2} \\
\dfrac{(\gamma-1)V^2/2-V^\prime a}{2a^2} &-\dfrac{(\gamma-1)V^\prime-a}{2a^2} &-\dfrac{(\gamma-1)V_\phi}{2a^2} &-\dfrac{(\gamma-1)V^{\prime\prime}+a\sin\theta}{2a^2} & \dfrac{\gamma-1}{2a^2}
\end{bmatrix},
\end{align}
\begin{align}\label{adiaroe14}
\mathbf{L_2^\mathbf{C}} = 
\begin{bmatrix}
\dfrac{(\gamma-1)V^2/2+V_\phi a}{2a^2} &-\dfrac{(\gamma-1)V^\prime}{2a^2} &-\dfrac{(\gamma-1)V_\phi+a}{2a^2} &-\dfrac{(\gamma-1)V^{\prime\prime}}{2a^2} & \dfrac{\gamma-1}{2a^2} \\[7pt]
-V_R & 1 & 0 & 0 & 0\\
1-\dfrac{(\gamma-1)V^2}{2a^2} & \dfrac{(\gamma-1)V^\prime}{a^2} & \dfrac{(\gamma-1)V_\phi}{a^2} & \dfrac{(\gamma-1)V^{\prime\prime}}{a^2}& -\dfrac{\gamma-1}{a^2} \\[7pt]
-V_\theta & 0 & 0 & 1 & 0\\
\dfrac{(\gamma-1)V^2/2-V_\phi a}{2a^2}&-\dfrac{(\gamma-1)V^\prime}{2a^2}&-\dfrac{(\gamma-1)V_\phi-a}{2a^2} &-\dfrac{(\gamma-1)V^{\prime\prime}}{2a^2} & \dfrac{\gamma-1}{2a^2}
\end{bmatrix},
\end{align}
\begin{align}\label{adiaroe15}
\mathbf{L_3^\mathbf{C}} = 
\begin{bmatrix}
\dfrac{(\gamma-1)V^2/2+V^{\prime\prime}a}{2a^2} &-\dfrac{(\gamma-1)V^{\prime}-a\sin\theta}{2a^2} &-\dfrac{(\gamma-1)V_\phi}{2a^2} &-\dfrac{(\gamma-1)V^{\prime\prime}+a}{2a^2} & \dfrac{\gamma-1}{2a^2}\\
1-\dfrac{(\gamma-1)V^2}{2a^2} & \dfrac{(\gamma-1)V^\prime}{a^2} & \dfrac{(\gamma-1)V_\phi}{a^2} & \dfrac{(\gamma-1)V^{\prime\prime}}{a^2} &-\dfrac{\gamma-1}{a^2}\\[7pt]
-V_R & 1 & 0 & 0 & 0 \\[1pt]
-V_\phi & 0 & 1 & 0 & 0 \\
\dfrac{(\gamma-1)V^2/2-V^{\prime\prime}a}{2a^2} &-\dfrac{(\gamma-1)V^{\prime}+a\sin\theta}{2a^2} & -\dfrac{(\gamma-1)V_\phi}{2a^2} &-\dfrac{(\gamma-1)V^{\prime\prime}-a}{2a^2} & \dfrac{\gamma-1}{2a^2}
\end{bmatrix}.
\end{align}

The same set of adiabatic equations \eqref{adiaroe1}-\eqref{adiaroe5} expressed in terms of primitive variables $\vec{W}=(\rho,V_R,V_\phi,V_\theta,P)=(\varw_1,\ldots,\varw_5)$ takes the form
\begin{gather}
\frac{\partial\rho}{\partial t}+\rho\left[\frac{\partial V^\prime}{\partial R}+\frac{1}{R}\frac{\partial V_\phi}{\partial\phi}+
\frac{\mu}{R}\left(\frac{\partial V_\theta}{\partial\theta}-\sin\theta\frac{\partial V_R}{\partial\theta}\right)\right]
+V^\prime\frac{\partial\rho}{\partial R}+\frac{V_\phi}{R}\frac{\partial\rho}{\partial\phi}+
\frac{\mu}{R}V^{\prime\prime}\frac{\partial\rho}{\partial\theta}=-\frac{\rho V^\prime}{R},
\label{adiaroe16}\\[1pt]
\frac{\partial V_R}{\partial t}+V^\prime\frac{\partial V_R}{\partial R}+\frac{V_\phi}{R}\frac{\partial V_R}{\partial\phi}+
\frac{\mu}{R}V^{\prime\prime}\frac{\partial V_R}{\partial\theta}+\frac{1}{\rho}\frac{\partial P}{\partial R}=0,
\label{adiaroe17}\\[1pt]
\frac{\partial V_\phi}{\partial t}+V^\prime\frac{\partial V_\phi}{\partial R}+\frac{V_\phi}{R}\frac{\partial V_\phi}{\partial \phi}+
\frac{\mu}{R}V^{\prime\prime}\frac{\partial V_\phi}{\partial \theta}+\frac{1}{\rho R}\frac{\partial P}{\partial\phi}=0,
\label{adiaroe18}\\[1pt]
\frac{\partial V_\theta}{\partial t}+V^\prime\frac{\partial V_\theta}{\partial R}+\frac{V_\phi}{R}\frac{\partial V_\theta}{\partial \phi}+
\frac{\mu}{R}\left(V^{\prime\prime}\frac{\partial V_\theta}{\partial\theta}+\frac{1}{\rho}\frac{\partial P}{\partial\theta}\right)=0,
\label{adiaroe19}\\[1pt]
\frac{\partial P}{\partial t}+\gamma P\left[\frac{\partial V^\prime}{\partial R}+\frac{1}{R}\frac{\partial V_\phi}{\partial\phi}+
\frac{\mu}{R}\left(\frac{\partial V_\theta}{\partial\theta}-\sin\theta\frac{\partial V_R}{\partial\theta}\right)\right]
+(\gamma V)^\prime\frac{\partial P}{\partial R}+\frac{V_\phi}{R}\frac{\partial P}{\partial\phi}+
\frac{\mu}{R}(\gamma V)^{\prime\prime}\frac{\partial P}{\partial\theta}=-\frac{\gamma PV^\prime}{R},
\end{gather}
where we substitute the expressions $(\gamma V)^\prime=V_R-\gamma V_\theta\sin\theta$,
$(\gamma V)^{\prime\prime}=V_\theta-\gamma V_R\sin\theta$.
The adiabatic primitive Roe matrices $\mathbf{A_1^\mathbf{P}}=\dfrac{\partial\vec{F}(\vec{\varw})}{\partial\vec{W}}$, $\mathbf{A_2^\mathbf{P}}=\dfrac{\partial\vec{G}(\vec{\varw})}{\partial\vec{W}}$, 
and $\mathbf{A_3^\mathbf{P}}=\dfrac{\partial\vec{H}(\vec{\varw})}{\partial\vec{W}}$, become
\begin{align}\label{adiaroe20}
\mathbf{A_1^\mathbf{P}} = 
\begin{bmatrix}
V^\prime & \rho & 0 &-\rho\sin\theta & 0 \\ 0 & V^\prime & 0 & 0 & \dfrac{1}{\rho} \\ 0 & 0 & V^\prime & 0 & 0 \\[1pt]
0 & 0 & 0 & V^\prime & 0 \\[1pt] 0 & \rho a^2 & 0 &-\rho a^2\sin\theta & \left(\gamma V\right)^\prime
\end{bmatrix},\quad
\mathbf{A_2^\mathbf{P}} = \frac{1}{R}
\begin{pmatrix}
V_\phi & 0 & \rho & 0 & 0 \\[1pt] 0 & V_\phi & 0 & 0 & 0 \\0 & 0 & V_\phi & 0 & \dfrac{1}{\rho}\\0 & 0 & 0 & V_\phi & 0 \\[1pt] 0 & 0 & \rho a^2 & 0 & V_\phi
\end{pmatrix},\quad
\mathbf{A_3^\mathbf{P}} = \frac{\mu}{R}
\begin{bmatrix}
V^{\prime\prime} &-\rho\sin\theta & 0 & \rho & 0 \\[1pt] 0 & V^{\prime\prime} & 0 & 0 & 0 \\[1pt] 0 & 0 & V^{\prime\prime} & 0 & 0 \\
0 & 0 & 0 & V^{\prime\prime} & \dfrac{1}{\rho} \\ 0 &-\rho a^2\sin\theta & 0 & \rho a^2 & \left(\gamma V\right)^{\prime\prime}
\end{bmatrix}.
\end{align}
The corresponding eigenvalues $\lambda_1^\text{P},\lambda_2^\text{P},\lambda_3^\text{P}$, of the matrices $\mathbf{A_1^\mathbf{P}}$, $\mathbf{A_2^\mathbf{P}}$, $\mathbf{A_3^\mathbf{P}}$, are 
\begin{align}\label{adiaroe21}
\lambda_1^\text{P} = \left(V^\prime-a,V^\prime,V^\prime,V^\prime,V^\prime+a\right),\quad
\lambda_2^\text{P} = \frac{1}{R}\left(V_\phi-a,V_\phi,V_\phi,V_\phi,V_\phi+a\right),\quad
\lambda_3^\text{P} = \frac{\mu}{R}\left(V^{\prime\prime}-a,V^{\prime\prime},V^{\prime\prime},V^{\prime\prime},V^{\prime\prime}+a\right).
\end{align}
The corresponding right eigenvectors (following the directional swapping in Eq.~\eqref{adiaroe20}) are the columns of the matrices
\begin{align}\label{adiaroe22}
\mathbf{R_1^\mathbf{P}} = 
\begin{pmatrix}
1 & 1 & 0 & 0 & 1 \\-\dfrac{a}{\rho} & 0 & 0 & \sin\theta & \dfrac{a}{\rho} \\ 0 & 0 & 1 & 0 & 0 \\[1pt] 0 & 0 & 0 & 1 & 0 \\[1pt] a^2 & 0 & 0 & 0 & a^2
\end{pmatrix},\quad
\mathbf{R_2^\mathbf{P}} = 
\begin{pmatrix}
1 & 1 & 0 & 0 & 1 \\[1pt] 0 & 0 & 1 & 0 & 0 \\-\dfrac{a}{\rho} & 0 & 0 & 0 & \dfrac{a}{\rho} \\ 0 & 0 & 0 &1 & 0 \\[1pt] a^2 & 0 & 0 & 0 & a^2
\end{pmatrix},\quad 
\mathbf{R_3^\mathbf{P}} = 
\begin{pmatrix}
1 & 1 & 0 & 0 & 1 \\[1pt] 0 & 0 & 0 & 1 & 0 \\[1pt] 0 & 0 & 1 & 0 & 0 \\-\dfrac{a}{\rho} & 0 & 0 & \sin\theta & \dfrac{a}{\rho} \\ a^2 & 0 & 0 & 0 & a^2
\end{pmatrix},
\end{align}
while the corresponding left eigenvectors are the rows of the matrices
\begin{align}\label{adiaroe23}
\mathbf{L_1^\mathbf{P}} = 
\begin{pmatrix}
0 &-\dfrac{\rho}{2a} & 0 & \dfrac{\rho\sin\theta}{2a} & \dfrac{1}{2a^2} \\
1 & 0 & 0 & 0 &-\dfrac{1}{a^2} \\ 0 & 0 & 1 & 0 & 0 \\[1pt] 0 & 0 & 0 & 1 & 0 \\ 0 & \dfrac{\rho}{2a} & 0 &-\dfrac{\rho\sin\theta}{2a} & \dfrac{1}{2a^2}
\end{pmatrix},\quad
\mathbf{L_2^\mathbf{P}} = 
\begin{pmatrix}
0 & 0 &-\dfrac{\rho}{2a} & 0 & \dfrac{1}{2a^2} \\ 1 & 0 & 0 & 0 &-\dfrac{1}{a^2}\\0 & 1 & 0 & 0 & 0 \\[1pt] 0 & 0 & 0 & 1 & 0 \\ 0 & 0 & \dfrac{\rho}{2a} & 0 & \dfrac{1}{2a^2}
\end{pmatrix},\quad
\mathbf{L_3^\mathbf{P}} = 
\begin{pmatrix}
0 & \dfrac{\rho\sin\theta}{2a} & 0 &-\dfrac{\rho}{2a} & \dfrac{1}{2a^2} \\
1 & 0 & 0 & 0 &-\dfrac{1}{a^2} \\
0 & 0 & 1 & 0 & 0 \\[1pt] 0 & 1 & 0 & 0 & 0 \\ 0 &-\dfrac{\rho\sin\theta}{2a} & 0 & \dfrac{\rho}{2a} & \dfrac{1}{2a^2}
\end{pmatrix}.
\end{align}
The explicit forms of the vectors of adiabatic conservative and primitive geometric source terms $\vec{S}^\text{C}$ and $\vec{S}^\text{P}$ are
\begin{align}
\vec{S}^\mathbf{C}=-\frac{1}{R}
\begin{pmatrix}
\rho V^\prime\\[1pt]\rho\vec{V}V^\prime\\[1pt]\rho V^\prime H
\end{pmatrix},\quad
\vec{S}^\mathbf{P} = -\frac{1}{R}
\begin{pmatrix}
\rho V^\prime\\[1pt]\vec{0}\\[1pt]\rho a^2 V^\prime
\end{pmatrix}.
\end{align}

\subsection{Eigensystems of the flaring disk coordinates using the Roe's method: isothermal hydrodynamics}
\label{roeisoflare}
For isothermal hydrodynamics in the \textit{flaring disk coordinate frame} with the \textit{conservative} variables $\vec{U}=(\rho,\Pi_R,\Pi_\phi,\Pi_\theta) = (u_1,\ldots,u_4)$
we write Eq.~\eqref{numappr1} in 3D component form
\citep[see the formalism and terminology given in, e.g.,][]{Toro,2008ApJS..178..137S,2010ApJS..188..290S},
\begin{align}\label{isoroe0}
\frac{\partial\vec{U}}{\partial t}+\xi,\xi^\prime\frac{\partial\vec{F}}{\partial R}+\frac{1}{R}\frac{\partial\vec{G}}{\partial\phi}
+\frac{\cos\theta}{R}\eta,\eta^\prime\frac{\partial\vec{H}}{\partial\theta}=\zeta,\zeta^\prime\vec{S},
\end{align}
where $\vec{F},\,\vec{G},\,\vec{H}$ are the vectors of fluxes in the $R,\,\phi,\,\theta$ directions, respectively, and 
$\vec{S}$ is the vector of geometric source terms. The factors $\xi=1,\xi^\prime=\sin\theta$ stay at particular $R,\theta$ components of the flux $\vec{F}$,
factors $\eta=1,\eta^\prime=\sin\theta$ stay at particular $\theta,R$ components of the flux $\vec{H}$ and
factors $\zeta=1,\zeta^\prime=\sin\theta$ stay at particular $R,\theta$ components of the geometric source term $\vec{S}$. We consistently use in this Section the following substitutions:
$V_{R-\theta}=V_R-V_\theta$, $V_{\theta-R}=V_\theta-V_R$, $V^\prime=V_{R-\theta}\sin\theta$, $V^{\prime\prime}=V_{\theta-R}\sin\theta$, $\mu=\cos\theta$.

The explicit conservative form of the basic isothermal hydrodynamic equations (equation of continuity and three components of momentum equation, omitting the source terms) 
in the \textit{flaring disk coordinate frame} is 
\begin{gather}
\frac{\partial\rho}{\partial t}+\frac{\partial}{\partial R}\left(\rho V^\prime\right)+\frac{1}{R}\frac{\partial}{\partial\phi}\left(\rho V_\phi\right)+
\frac{\mu}{R}\left[\frac{\partial}{\partial\theta}\left(\rho V_\theta\right)-\sin\theta\frac{\partial}{\partial\theta}\left(\rho v_R\right)\right]=-\frac{\rho V^\prime}{R},
\label{isoroe1}\\[1pt]
\frac{\partial}{\partial t}\left(\rho V_R\right)+\frac{\partial}{\partial R}\left[\rho\left(V_RV^\prime+C^2\right)\right]+\frac{1}{R}\frac{\partial}{\partial\phi}\left(\rho V_RV_\phi\right)+
\frac{\mu}{R}\left[\frac{\partial}{\partial\theta}\left(\rho V_RV_\theta\right)-\sin\theta\frac{\partial}{\partial\theta}\left(\rho V_R^2\right)\right]=-\frac{\rho V_RV^\prime}{R},
\label{isoroe2}\\[1pt]
\frac{\partial}{\partial t}\left(\rho V_\phi\right)+\frac{\partial}{\partial R}\left(\rho V_\phi V^\prime\right)+\frac{1}{R}\frac{\partial}{\partial\phi}\left[\rho\left(V_\phi^2+C^2\right)\right]+
\frac{\mu}{R}\left[\frac{\partial}{\partial\theta}\left(\rho V_\phi V_\theta\right)-\sin\theta\frac{\partial}{\partial\theta}\left(\rho V_RV_\phi\right)\right]=
-\frac{\rho V_\phi V^\prime}{R},
\label{isoroe3}\\[1pt]
\frac{\partial}{\partial t}\left(\rho V_\theta\right)+\frac{\partial}{\partial R}\left(\rho V_\theta V^\prime\right)+\frac{1}{R}\frac{\partial}{\partial\phi}\left(\rho V_\phi V_\theta\right)+
\frac{\mu}{R}\left\{\frac{\partial}{\partial\theta}\left[\rho\left(V_\theta^2+C^2\right)\right]-\sin\theta\frac{\partial}{\partial\theta}\left(\rho V_RV_\theta\right)\right\}=
-\frac{\rho V_\theta V^\prime}{R},
\label{isoroe4}
\end{gather}
where we denote the isothermal speed of sound, $C$ , in this Section. 
The explicit components $\vec{F}_\mathbf{iso}$, $\vec{G}_\mathbf{iso}$, $\vec{H}_\mathbf{iso}$, $\vec{S}_\mathbf{iso}$, are
\begin{align}\label{isoroe5}
\vec{F}_\mathbf{iso} = 
\begin{pmatrix}
\rho V_{R-\theta}\\[2pt]\rho V_RV_{R-\theta}+C^2\rho\\[2pt]\rho V_\phi V_{R-\theta}\\[2pt]\rho V_\theta V_{R-\theta}
\end{pmatrix},\quad
\vec{G}_\mathbf{iso} = 
\begin{pmatrix}
\rho V_\phi\\[2pt]\rho V_R V_\phi\\[2pt]\rho V_\phi^2+C^2\rho\\[2pt]\rho V_\phi V_\theta
\end{pmatrix},\quad
\vec{H}_\mathbf{iso} = 
\begin{pmatrix}
\rho V_{\theta-R}\\[2pt]\rho V_RV_{\theta-R}\\[2pt]\rho V_\phi V_{\theta-R}\\[2pt]\rho V_\theta V_{\theta-R}+C^2\rho
\end{pmatrix},\quad
\vec{S}_\mathbf{iso} = -\frac{1}{R}
\begin{pmatrix}
\rho V_{R-\theta}\\[2pt]\rho\vec{V}V_{R-\theta}
\end{pmatrix}.
\end{align}
The vectors $\vec{F}_\mathbf{iso},\vec{G}_\mathbf{iso},\vec{H}_\mathbf{iso}$, written in terms of the four conservative variables $\vec{U} = (u_1,\ldots,u_4)$, 
corresponding to Eq.~\eqref{numappr1} in the flaring disk coordinates, are
\begin{align}\label{isoroe6}
\vec{F}_\mathbf{iso}(\vec{u}) = 
\begin{bmatrix}
u_2-u_4^*\\[1pt]\dfrac{u_2(u_2-u_4^*)}{u_1}+C^2u_1\\[1pt]\dfrac{u_3(u_2-u_4^*)}{u_1}\\[1pt]\dfrac{u_4(u_2-u_4^*)}{u_1}
\end{bmatrix},\quad
\vec{G}_\mathbf{iso}(\vec{u}) = 
\begin{pmatrix}
u_3\\[3pt]\dfrac{u_3u_2}{u_1}\\[1pt]\dfrac{u_3^2}{u_1}+C^2u_1\\[1pt]\dfrac{u_3u_4}{u_1}
\end{pmatrix},\quad
\vec{H}_\mathbf{iso}(\vec{u}) = 
\begin{bmatrix}
u_4-u_2^*\\[1pt]\dfrac{u_2(u_4-u_2^*)}{u_1}\\[1pt]\dfrac{u_3(u_4-u_2^*)}{u_1}\\[1pt]\dfrac{u_4(u_4-u_2^*)}{u_1}+C^2u_1
\end{bmatrix},
\end{align}
where the asterisk denotes the factor $\sin\theta$ staying in front of the derivative of the particular term.

The isothermal conservative Jacobian matrices $\mathbf{A_{1,{\mathbf{iso}}}^\mathbf{C}}=\dfrac{\partial\vec{F}_\mathbf{iso}(\vec{u})}{\partial\vec{U}}$,
$\mathbf{A_{2,{\mathbf{iso}}}^\mathbf{C}}=\dfrac{\partial\vec{G}_\mathbf{iso}(\vec{u})}{\partial\vec{U}}$, 
$\mathbf{A_{3,{\mathbf{iso}}}^\mathbf{C}}=\dfrac{\partial\vec{H}_\mathbf{iso}(\vec{u})}{\partial\vec{U}}$, and the vector 
$\vec{S}_\mathbf{iso}^\mathbf{C}$, are
\begin{align}\label{isoroe7}
\mathbf{A_{1,{\mathbf{iso}}}^\mathbf{C}} = 
\begin{pmatrix}
0&1&0&-\sin\theta\\[1pt]-V_RV^\prime+C^2&V_R+V^\prime&0&-V_R\sin\theta\\[1pt]
-V_\phi V^\prime&V_\phi&V^\prime&-V_\phi\sin\theta\\[1pt]-V_\theta V^\prime&V_\theta&0&V^\prime-V_\theta\sin\theta
\end{pmatrix},\quad
\mathbf{A_{2,{\mathbf{iso}}}^\mathbf{C}} = \frac{1}{R}
\begin{pmatrix}
0&0&1&0\\[1pt]-V_\phi V_R&V_\phi&V_R&0\\[1pt]-V_\phi^2+C^2&0&2V_\phi&0\\[1pt]-V_\phi V_\theta&0&V_\theta&V_\phi
\end{pmatrix},
\end{align}
\begin{align}\label{isoroe8}
\mathbf{A_{3,{\mathbf{iso}}}^\mathbf{C}} = \frac{\mu}{R}
\begin{pmatrix}
0&-\sin\theta&0&1\\[1pt]-V_R V^{\prime\prime}&V^{\prime\prime}-V_R\sin\theta&0&V_R\\[1pt]
-V_\phi V^{\prime\prime}&-V_\phi\sin\theta&V^{\prime\prime}&V_\phi\\[1pt]-V_\theta V^{\prime\prime}+C^2&-V_\theta\sin\theta&0&V_\theta+V^{\prime\prime}
\end{pmatrix},\quad
\vec{S}_\mathbf{iso}^\mathbf{C}=-\frac{1}{R}
\begin{pmatrix}
\rho V^\prime\\[1pt]\rho\vec{V}V^\prime
\end{pmatrix}.
\end{align}
The corresponding eigenvalues $\lambda_{1,{\text{iso}}}^\text{C},\lambda_{2,{\text{iso}}}^\text{C},\lambda_{3,{\text{iso}}}^\text{C}$, 
of the matrices $\mathbf{A_{1,{\mathbf{iso}}}^\mathbf{C}}$, $\mathbf{A_{2,{\mathbf{iso}}}^\mathbf{C}}$, $\mathbf{A_{3,{\mathbf{iso}}}^\mathbf{C}}$, are 
\begin{align}\label{isoroe9}
\lambda_{1,{\text{iso}}}^\text{C} = \left(V^\prime-C,V^\prime,V^\prime,V^\prime,V^\prime+C\right),\,\,
\lambda_{2,{\text{iso}}}^\text{C} = \frac{1}{R}\left(V_\phi-C,V_\phi,V_\phi,V_\phi,V_\phi+C\right),\,\,
\lambda_{3,{\text{iso}}}^\text{C} = \frac{\mu}{R}\left(V^{\prime\prime}-C,V^{\prime\prime},V^{\prime\prime},V^{\prime\prime},V^{\prime\prime}+C\right).
\end{align}
The corresponding right eigenvectors (following the directional swapping of Eqs.~\eqref{isoroe7}-\eqref{isoroe8}) are the columns of the matrices
\begin{align}\label{isoroe10}
\mathbf{R_{1,{\mathbf{iso}}}^\mathbf{C}} = 
\begin{pmatrix}
1&0&0&1\\[1pt] V_R-C&0&\sin\theta&V_R+C\\[1pt] V_\phi&1&0&V_\phi\\[1pt] V_\theta&0&1&V_\theta
\end{pmatrix},\quad
\mathbf{R_{2,{\mathbf{iso}}}^\mathbf{C}} = 
\begin{pmatrix}
1&0&0&1\\[1pt] V_R&0&1&V_R\\[1pt] V_\phi-C&0&0&V_\phi+C\\[1pt] V_\theta&1&0&V_\theta
\end{pmatrix},\quad
\mathbf{R_{3,{\mathbf{iso}}}^\mathbf{C}} = 
\begin{pmatrix}
1&0&0&1\\[1pt] V_R&0&1&V_R\\[1pt] V_\phi&1&0&V_\phi\\[1pt] V_\theta-C&0&\sin\theta&V_\theta+C
\end{pmatrix}.
\end{align}
The corresponding left eigenvectors are the rows of the matrices,
\begin{align}\label{isoroe11}
\mathbf{L_{1,{\mathbf{iso}}}^\mathbf{C}} = 
\begin{pmatrix}
\dfrac{1+V^\prime/C}{2}&-\dfrac{1}{2C}&0&\dfrac{\sin\theta}{2C}\\[7pt]
-V_\phi&0&1&0\\[1pt] -V_\theta&0&0&1\\\dfrac{1-V^\prime/C}{2}&\dfrac{1}{2C}&0&-\dfrac{\sin\theta}{2C}
\end{pmatrix},\quad
\mathbf{L_{2,{\mathbf{iso}}}^\mathbf{C}} = 
\begin{pmatrix}
\dfrac{1+V_\phi/C}{2}&0&-\dfrac{1}{2C}&0\\[7pt] -V_\theta&0&0&1\\[1pt] -V_R&1&0&0\\[1pt]\dfrac{1-V_\phi/C}{2}&0&\dfrac{1}{2C}&0
\end{pmatrix},\quad
\mathbf{L_{3,{\mathbf{iso}}}^\mathbf{C}} = 
\begin{pmatrix}
\dfrac{1+V^{\prime\prime}/C}{2}&\dfrac{\sin\theta}{2C}&0&-\dfrac{1}{2C}\\[7pt]-V_\phi&0&1&0\\[1pt] -V_R&1&0&0\\
\dfrac{1-V^{\prime\prime}/C}{2}&-\dfrac{\sin\theta}{2C}&0&\dfrac{1}{2C}
\end{pmatrix}.
\end{align}

The same set of isothermal hydrodynamic equations \eqref{isoroe1}-\eqref{isoroe4} expressed in terms of primitive variables $\vec{W}=(\rho,V_R,V_\phi,V_\theta)=(\varw_1,\ldots,\varw_4)$ takes the form
\begin{gather}\label{isoroe12}
\frac{\partial\rho}{\partial t}+\rho\left[\frac{\partial V^\prime}{\partial R}+\frac{1}{R}\frac{\partial V_\phi}{\partial\phi}+
\frac{\mu}{R}\left(\frac{\partial V_\theta}{\partial\theta}-\sin\theta\frac{\partial V_R}{\partial\theta}\right)\right]+
V^\prime\frac{\partial\rho}{\partial R}+\frac{V_\phi}{R}\frac{\partial\rho}{\partial\phi}+
\frac{\mu}{R}V^{\prime\prime}\frac{\partial\rho}{\partial\theta}=-\frac{\rho V^\prime}{R},\\[1pt]
\frac{\partial V_R}{\partial t}+V^\prime\frac{\partial V_R}{\partial R}+\frac{V_\phi}{R}\frac{\partial V_R}{\partial\phi}+
\frac{\mu}{R}V^{\prime\prime}\frac{\partial V_R}{\partial\theta}+\frac{C^2}{\rho}\frac{\partial\rho}{\partial R}=0,\\[1pt]
\frac{\partial V_\phi}{\partial t}+V^\prime\frac{\partial V_\phi}{\partial R}+\frac{V_\phi}{R}\frac{\partial V_\phi}{\partial\phi}+
\frac{\mu}{R}V^{\prime\prime}\frac{\partial V_\phi}{\partial\theta}+\frac{C^2}{\rho R}\frac{\partial\rho}{\partial\phi}=0,\\[1pt]
\frac{\partial V_\theta}{\partial t}+V^\prime\frac{\partial V_\theta}{\partial R}+\frac{V_\phi}{R}\frac{\partial V_\theta}{\partial\phi}+
\frac{\mu}{R}V^{\prime\prime}\frac{\partial V_\theta}{\partial\theta}+\frac{\mu}{R}\frac{C^2}{\rho}\frac{\partial\rho}{\partial\theta}=0.\label{skrcouriso4}
\end{gather}
The corresponding matrices $\mathbf{A}_\mathbf{iso}^\mathbf{P}$ and the vector $\vec{S}_\mathbf{iso}^\mathbf{P}$
in terms of the primitive variables $\vec{W}$ become
\begin{align}\label{isoroe13}
\mathbf{A_{1,{\mathbf{iso}}}^\mathbf{P}} = 
\begin{pmatrix}
V^\prime&\rho&0&-\rho\sin\theta\\\dfrac{C^2}{\rho}&V^\prime&0&0\\[7pt] 0&0&V^\prime&0\\[1pt] 0&0&0&V^\prime
\end{pmatrix},\,\,
\mathbf{A_{2,{\mathbf{iso}}}^\mathbf{P}} = \frac{1}{R}
\begin{pmatrix}
V_\phi&0&\rho&0\\0&V_\phi&0&0\\[1pt]\dfrac{C^2}{\rho}&0&V_\phi&0\\[1pt] 0&0&0&V_\phi
\end{pmatrix},\,\,
\mathbf{A_{3,{\mathbf{iso}}}^\mathbf{P}} = \frac{\mu}{R}
\begin{pmatrix}
V^{\prime\prime}&-\rho\sin\theta&0&\rho\\0&V^{\prime\prime}&0&0\\0&0&V^{\prime\prime}&0\\\dfrac{C^2}{\rho}&0&0&V^{\prime\prime}
\end{pmatrix},\,\,
\vec{S}_\mathbf{iso}^\mathbf{P} = -\frac{1}{R}
\begin{pmatrix}
\rho V^\prime\\\vec{0}
\end{pmatrix}.
\end{align}
The corresponding eigenvalues $\lambda_{1,{\text{iso}}}^\text{P},\lambda_{2,{\text{iso}}}^\text{P},\lambda_{3,{\text{iso}}}^\text{P}$, 
of the matrices $\mathbf{A_{1,{\mathbf{iso}}}^\mathbf{P}}$, $\mathbf{A_{2,{\mathbf{iso}}}^\mathbf{P}}$, 
$\mathbf{A_{3,{\mathbf{iso}}}^\mathbf{P}}$, are
\begin{align}\label{isoroe14}
\lambda_{1,{\text{iso}}}^\text{P} = \left(V^\prime-C,V^\prime,V^\prime,V^\prime,V^\prime+C\right),\,\,
\lambda_{2,{\text{iso}}}^\text{P} = \frac{1}{R}\left(V_\phi-C,V_\phi,V_\phi,V_\phi,V_\phi+C\right),\,\,
\lambda_{3,{\text{iso}}}^\text{P} = \frac{\mu}{R}\left(V^{\prime\prime}-C,V^{\prime\prime},V^{\prime\prime},V^{\prime\prime},V^{\prime\prime}+C\right).
\end{align}
The corresponding right eigenvectors (following the directional swapping of Eqs.~\eqref{isoroe7}-\eqref{isoroe8}) are the columns of the matrices
\begin{align}\label{isoroe15}
\mathbf{R_{1,{\mathbf{iso}}}^\mathbf{P}} = 
\begin{pmatrix}
1&0&0&1\\-\dfrac{C}{\rho}&\sin\theta&0&\dfrac{C}{\rho}\\[7pt] 0&0&1&0\\[1pt] 0&1&0&0
\end{pmatrix},\quad
\mathbf{R_{2,{\mathbf{iso}}}^\mathbf{P}} = 
\begin{pmatrix}
1&0&0&1\\[1pt] 0&1&0&0\\-\dfrac{C}{\rho}&0&0&\dfrac{C}{\rho}\\[7pt]0&0&1&0
\end{pmatrix},\quad
\mathbf{R_{3,{\mathbf{iso}}}^\mathbf{P}} = 
\begin{pmatrix}
1&0&0&1\\[1pt] 0&0&1&0\\[1pt] 0&1&0&0\\-\dfrac{C}{\rho}&0&\sin\theta&\dfrac{C}{\rho}
\end{pmatrix}.
\end{align}
The corresponding left eigenvectors are the rows of the matrices
\begin{align}\label{isoroe16}
\mathbf{L_{1,{\mathbf{iso}}}^\mathbf{P}} = 
\begin{pmatrix}
\dfrac{1}{2}&-\dfrac{\rho}{2C}&0&\dfrac{\rho\sin\theta}{2C}\\[7pt] 
0&0&1&0\\[1pt] 0&0&0&1\\\dfrac{1}{2}&\dfrac{\rho}{2C}&0&-\dfrac{\rho\sin\theta}{2C}
\end{pmatrix},\quad
\mathbf{L_{2,{\mathbf{iso}}}^\mathbf{P}} = 
\begin{pmatrix}
\dfrac{1}{2}&0&-\dfrac{\rho}{2C}&0\\[7pt] 0&1&0&0\\[1pt] 0&0&0&1\\\dfrac{1}{2}&0&\dfrac{\rho}{2C}&0
\end{pmatrix},\quad
\mathbf{L_{3,{\mathbf{iso}}}^\mathbf{P}} = 
\begin{pmatrix}
\dfrac{1}{2}&\dfrac{\rho\sin\theta}{2C}&0&-\dfrac{\rho}{2C}\\[7pt] 0&0&1&0\\[1pt] 0&1&0&0\\
\dfrac{1}{2}&-\dfrac{\rho\sin\theta}{2C}&0&\dfrac{\rho}{2C}
\end{pmatrix}.
\end{align}

\subsection{Computation of the time increment of the unsplit algorithm in the flaring coordinates}
\label{roetimeflare}
The principle of the computation of time increment is common for both eigensystems in both types of variables:
we compute a new timestep $\Delta t$ using the pre-defined CFL number $C_0\le 1/2$, satisfying the cell-centered stability condition
\begin{align}\label{Roetimes}
 \Delta t=C_0\,\text{min}\left[\frac{\Delta R}{\text{max}(\lambda_1)},\frac{R\,\Delta\phi}{\text{max}(\lambda_2)},
 \frac{(R/\mu)\,\Delta\theta}{\text{max}(\lambda_3)}\right],
\end{align}
where we however describe the complete 3D form.
The denominators $\max(\lambda_\alpha)$ in Eq.~\eqref{Roetimes} denote either 
the eigenvalues $|V_\alpha|+a$ from Eqs.~\eqref{adiaroe11} and \eqref{adiaroe21} for adiabatic 
hydrodynamics or the eigenvalues $|V_\alpha|+C$ from Eqs.~\eqref{isoroe9} and \eqref{isoroe14} for isothermal
hydrodynamics that are already evaluated using the updated quantities \citep[cf.][]{2008ApJS..178..137S,2010ApJS..188..290S}.

\end{appendix}

\end{document}